\documentclass{aa}  

\usepackage{graphicx}
\usepackage{txfonts}
\usepackage{subeqnarray}
\usepackage{subfigure}
\usepackage{url}
\usepackage[hidelinks]{hyperref}
\hypersetup{
    colorlinks=true,
    citecolor=blue,
    linkcolor=red,
    urlcolor=black
    }  
\makeatletter
\renewcommand*\aa@pageof{, page \thepage{} of \pageref*{LastPage}}
\makeatother

\begin{document}

\title{The Acceleration of Superrotation in Simulated Hot Jupiter Atmospheres }
\titlerunning{Superrotation}
\authorrunning{Debras et al.}

\author{F. Debras \inst{1,2,3} \thanks{corresponding author: florian\_debras@hotmail.com} 
\and N. Mayne\inst{3} 
\and
I. Baraffe\inst{3,2}
\and E. Jaupart\inst{2}
\and P. Mourier\inst{2}
\and G. Laibe \inst{2}
\and T. Goffrey\inst{4}
\and J. Thuburn\inst{5}}

\institute{IRAP, Universit\'e de Toulouse, CNRS, UPS, Toulouse, France 
\and
Ecole normale sup\'erieure de Lyon, CRAL, UMR CNRS 5574, 69364 Lyon Cedex 07,  France
\and
Physics and Astronomy, College of Engineering, Mathematics and Physical Sciences, University of Exeter, Exeter, EX4 4QL, UK 
\and
Centre for Fusion, Space \& Astrophysics, Department of Physics, University of Warwick, Coventry, CV4 7AL, UK
\and
Mathematics, College of Engineering, Mathematics and Physical Sciences, University of Exeter, Exeter, EX4 4QL, UK}

\date{}

\abstract{Atmospheric superrotating flows at the equator are an almost ubiquitous result of simulations of hot Jupiters, and
a theory explaining how this zonally coherent flow reaches an equilibrium has been developed in the literature. However, this understanding relies on the existence of either an initial superrotating or a sheared flow, coupled with a slow evolution such that a linear steady state can be reached.}
{A consistent physical understanding of superrotation is needed for arbitrary drag and radiative timescales, and the relevance of considering linear steady states needs to be assessed.  }
{We obtain an analytical expression for the structure, frequency and decay rate of propagating waves in hot Jupiter atmospheres around a state at rest in the 2D shallow-water $\beta$--plane limit. We solve this expression numerically and confirm the robustness of our results with a 3D linear wave algorithm. We then compare with 3D simulations of hot Jupiter atmospheres and study the non linear momentum fluxes. }
{We show that under strong day--night heating the dynamics does not transit through a linear steady state when starting from an initial atmosphere in solid body rotation.  We further show that non--linear effects favour the initial spin-up of superrotation and that the acceleration due to  the vertical component of the eddy--momentum flux is critical to the initial development of superrotation
. }
{Overall, we describe the initial phases of the acceleration of superrotation, including consideration of differing radiative and drag timescales, 
and conclude that eddy-momentum driven superrotating equatorial jets are robust, physical phenomena in simulations of hot Jupiter atmospheres.}


\maketitle


Accepted in A\&A

\section{Introduction} 
\label{sec:intro}
Understanding the atmospheric dynamics of hot Jupiters, Jovian planets in short period orbits, has been a major challenge since their discovery \citep{Mayor}.
Due to their proximity to their host star hot Jupiters are expected to be tidally--locked \citep[see][for review]{Baraffe2010}, resulting in a permanent day and night side driving atmospheric circulations with no equivalent in our solar system, which in turn likely mix material between the two hemispheres \citep{Drummond2018c,Drummond2018b}.

\citet{Cooper2005} performed the first study of the atmosphere of HD~209458b \citep[e.g.,][]{Charbonneau2002,Sing2008,Snellen2008}  using a General Circulation Model (GCM), and such GCMs have subsequently been used extensively to characterise hot Jupiters \citep[e.g.,][]{Showman2008,Heng2011, Rauscher2012,Dobbs2013,Mayne2014b,Helling2016}. The physical complexity, or completeness, of these GCMs varies greatly, for example treatments of the dynamics and radiative transfer range from those adopting the primitive equations of dynamics, and simple Newtonian cooling, to those solving the full Navier--Stokes equations and more accurate radiative transfer \citep[see, notably,][]{Amundsen2014,Amundsen2016}. Recent advances have also been made in the treatment of chemistry, regarding both the gas phase \citep[see][]{Drummond2016,Tsai2017,Drummond2018,Drummond2018c,Drummond2018b} and the condensates, or clouds \citep{Lee2016,Lines2018,Lines2018b,Roman2019}.

\noindent A common feature has emerged from almost all GCM studies of hot Jupiters: the atmosphere exhibits equatorial superrotation, a prograde atmospheric wind velocity greater than that arising from the rotation of the planet alone, over a range of pressures. Observations have detected an eastward shift of the peak infrared flux from the substellar point in the atmosphere of hot Jupiters \citep{Knutson2007,Zellem2014}, consistent with that found in simulations caused by the advection of heat by the superrotating jet. \citet{Mayne2017} attempted to suppress the formation of the equatorial jet in simulated hot Jupiter atmospheres by forcing the deep atmospheric flow, or altering the model parameters. They found superrotation to be a very robust feature in numerical simulations. However, a recent measurement has inferred an opposite, westward shift for COROT--2b \citep{Dang2018}, and \citet{Armstrong2016} previously obtained variability in the position of the hot spot with time suggesting additional complexity \citep[a potential link to magnetic fields has recently been investigated by][]{Hindle2019}. Superrotation therefore has to be explained with sound physical arguments.

\citet{Showman2011} were the first to study the formation of a superrotating jet in simulated hot Jupiter atmospheres using a simplified two--layer model. Exploring the linear steady state of the atmosphere \citet{Showman2011} highlighted the formation of a Matsuno--Gill \citep[hereafter MG, see][]{Matsuno,Gill80} pattern, where the atmospheric perturbations are `tilted' in the latitude-longitude plane driving momentum transport to the equator and accelerating the jet. \citet{Showman2011} posit that the non--linear equilibrium is reached when the transport of meridional and vertical eddy momentum into the region, acting to accelerate and decelerate the jet, respectively, are balanced by the atmospheric drag. \citet{Tsai2014} extended the study to a full 3D dynamical model including consideration of the resonance of the atmospheric wave response, as well as the 'tilt' of the vertical component which acts to drive the vertical eddy--momentum transport, under the assumption of equal drag and radiative timescales. 
This was followed by \citet{Hammond2018} who explored superrotation in 2D with the addition of a shearing flow. \citet{Perez2013} considered the propagation of waves and the resulting balance for the equilibrated jet and propose diagnostics for predicting the day to night temperature contrast, controlled by the efficiency of the zonal advection. This analysis was later improved upon by \citet{Komacek2016}, across an extensive range of dissipation timescales.

There is however an inherent discrepancy between the works of \citet{Komacek2016} and \citet{Showman2011}: when the atmospheric drag timescale is large, superior to a few $10^5$\ s, the linear steady states obtained in \citet{Komacek2016} tend to decelerate the equator although the associated non linear steady states exhibit equatorial superrotation. This raises the question: is superrotation properly explained through the transition from a linear steady state ? 

Specifically, the study of \citet{Tsai2014} is only valid in the moderate to strong dissipation limit, and that of \citet{Hammond2018} requires an initial sheared superrotation. However, \citet{Komacek2016} showed that superrotation develops only if the dissipation is sufficiently low (see their Figure 4). Current theories are therefore applicable only once an initial flow has been set up, and its evolution is slow compared to the wave propagation time.


In this study, we address the issue of what is driving the initial spin up of superrotation in simulated hot Jupiter atmospheres. In order to do this we develop a description of the time dependent waves supported by our simulations of hot Jupiters atmospheres with arbitrary drag and radiative timescales, and determine which are responsible for driving the evolution of the jet. Firstly, in Section \ref{sec:simple} we state our main assumptions and develop the mathematical framework we adopt throughout this work, before finding the form of the time dependent linear solution to the beta--plane equations. Additionally, in this section we summarise the main results of \citet{Showman2011}, \citet{Tsai2014} and \citet{Komacek2016} upon which we base our study. Obtaining the form of the solution to the time dependent case is not sufficient as the controlling parameters remain unconstrained and are not easily accessible analytically. Therefore, in Section \ref{sec:heating} we numerically explore the sensitivity of the steady linear solution to the shape of the forcing, or heating, showing that the linear steady state requires a day--night heating contrast but is insensitive to the exact shape of the forcing itself. This confirms that the limitations of the current theory do not come from the simplified form of the forcing, but that time--dependent linear considerations must be included. We therefore study numerically the propagating waves, except for the special case of Kelvin waves which have an analytical expression, to build a more complete picture of the physical process of acceleration of superrotation. In Section \ref{sec:waves} we determine the characteristic decay timescale for different waves and arbitrary dissipative timescales, which are used to explain the structure of both the linear steady states presented in \citet{Komacek2016} (their Figure 5) and the time--dependent linear evolution of simulated hot Jupiters. In Section \ref{sec:transition}, we then combine the understanding developed throughout this study to detail the transition to superrotation in 3D GCM simulations through eddy--mean flow interaction under different conditions, revealing the importance of time--dependent linear considerations as well as vertical momentum transport across different drag and forcing regimes. Finally, we summarise our conclusions in Section \ref{sec:conclusions}. Overall, our study shows that
the paradigm of equatorial superrotation in hot Jupiters is robust: superrotation is accelerated by an eddy--mean flow interaction (i.e. atmospheric waves interacting with the background flow), and is strongly influenced by the wave dissipation timescales and vertical momentum convergence.


\section{Solution to 2D Shallow Water Equation}
\label{sec:simple}


\subsection{Theoretical framework}
\label{ssec:framework}

For this study we adopt the 2D shallow water equations under the equatorial $\beta$--plane approximation. As in \citet{Showman2011}, the shallow water approximation consists of considering that the planet can be decomposed into an upper, constant density but dynamically active layer with a free surface at the bottom exchanging heat and momentum with a lower, quiescent layer of much higher density. The equatorial $\beta$--plane simplifies the spherical planet as a local Cartesian plane at the equator, with the Coriolis parameter depending linearly on the Cartesian meridional coordinate, $y$, with a factor of proportionality $\beta = 2 \Omega / R$, where $\Omega$ is the rotation rate of the planet and $R$ its radius. The further away from the equator, the less valid this approximation is but it allows for analytic solutions, particularly suited for the study of equatorial superrotation. \citet{Wu2000} showed that the 3D structure of solutions to the $\beta$--plane system can be decomposed onto an infinite sum of solutions of 2D $\beta$--planes with different characteristic heights. The importance of this decomposition regarding hot Jupiter atmospheric dynamics has been emphasised by \citet{Tsai2014}, where a vertical shift of the wave response is presented when the mean background velocity is changed (their Figure 10). We begin by summarising the main results of \citet{Matsuno}, \citet{Gill80} and \citet{Showman2011}, all of which solve the 2D $\beta$--plane equations. 

Following  \citet{Showman2011}, the non--dimensional, linearised equations of motion for a forced 2D, equatorial $\beta$--plane can be written as
\begin{gather}
\dfrac{\partial u}{\partial t} - yv + \dfrac{\partial h}{\partial x} + \dfrac{u}{\tau_\mathrm{drag}} = 0, \label{eq:sp11_u} \\
\dfrac{\partial v}{\partial t}  + yu + \dfrac{\partial h}{\partial y} + \dfrac{v}{\tau_\mathrm{drag}}= 0,\label{eq:sp11_v} \\
\dfrac{\partial h}{\partial t}  + \dfrac{\partial u}{\partial x} + \dfrac{\partial v}{\partial y} + \dfrac{h}{\tau_\mathrm{rad}}= Q,\label{eq:sp11_h}
\end{gather}
where $x$ and $y$ are the horizontal coordinates, $t$ is time, $u$ is the zonal velocity ($x$ direction), $v$ the meridional ($y$ direction), $h$ the height ($H$) of the shallow water fluid minus the initial, horizontally constant height ($H_0$), i.e. $h = H-H_0$, $\tau_\mathrm{drag}$ the drag timescale, $\tau_\mathrm{rad}$ the radiative timescale and $Q$ the heating function. The characteristic length, speed and time, corresponding to the Rossby deformation radius, the gravity wave speed and the time for a gravity wave to cross a deformation radius in the shallow water system are:
\begin{align}
&L = \left(\beta^{-1}\sqrt{gH_0}\right)^{1/2}, \label{eq:L_dim}\\
&U = \sqrt{gH_0},\label{eq:U_dim} \\
&T = \left(\beta \sqrt{gH_0}\right)^{-1/2}, \label{eq:T_dim}
\end{align}
respectively, where $g$ is the gravitational acceleration assumed constant, and $\beta = 2 \Omega \cos \phi /R$ or the derivative of the Rossby parameter with $\phi$ the latitude of the $\beta$--plane. In the rest of this paper we only consider $\phi = 0$.

Eqs.\eqref{eq:sp11_u} to \eqref{eq:sp11_h} form a linear differential equation of the form $\partial X/\partial t = \mathcal{L} X + \mathcal{Q}$ where $X = (u,v,h)$ is a vector of solutions, $\mathcal{L}$ a linear operator and $\mathcal{Q} = (0,0,Q)$ is the vector form of the forcing. Hence the solution is the sum of a homogeneous and a particular solution. The spatial part of the 3D solution can be expressed as an infinite sum of modes indexed by $m$ with equivalent depth $H_m$ instead of $H_0$. The orthogonal base functions are sinusoidal in $z$, the vertical coordinate, and of the form $\mathrm{e}^{\mathrm{i}(mz)}$ with $m \in \mathbb{N}$ and the heating must be decomposed onto these functions (see \citet{Tsai2014} section 2 for the rescaling of z and \citet{Wu2000} section 2 for a discussion on boundary conditions).

When neither drag nor heating are considered, \citet{Matsuno} expressed the analytic solutions to the homogeneous equations in the form $\{u,v,h\} = \{\tilde{u},\tilde{v},\tilde{h}\} \ \mathrm{exp}(\mathrm{i} \omega t + \mathrm{i}k x)$, where $\omega$ is the complex frequency, $k$ the longitudinal wavenumber and a tilde denotes a function of $y$ only. Dropping the tilde for simplicity, the homogeneous equations can be expressed as:
\begin{align}
&\mathrm{i} \omega u - yv + \mathrm{i}k h = 0, \label{eq:matsuno_u}\\
&\mathrm{i} \omega v + yu + \dfrac{\partial h}{\partial y} = 0, \label{eq:matsuno_v}\\
&\mathrm{i} \omega h +\mathrm{i} k u + \dfrac{\partial v}{\partial y} = 0 \,\, . \label{eq:matsuno_h}
\end{align}
 \citet{Matsuno} showed that this system reduces to a single equation for $v$, namely, 
\begin{equation}
    \dfrac{\partial^2 v}{\partial y^2} + (\omega^2-k^2+\dfrac{k}{\omega}-y^2)v = 0.
    \label{v_matsuno}
\end{equation}
By analogy with the Schr\"odinger equation of a simple harmonic oscillator, the boundary condition $v \rightarrow 0$ when $|y| \rightarrow \infty$ requires
\begin{equation}
    \omega^2-k^2+\dfrac{k}{\omega} = 2n+1,
\end{equation}
with $n \in \mathbb{N}$. As this is a third order equation, the eigenvalues for the frequency are labelled $\omega_{n,l}$ with $l = 0,1,2$, and the corresponding eigenvectors are labelled $\tilde{X}_{n,l} = (u_{n,l},v_{n,l},h_{n,l})$,. Finally, the case where $n=0$ is treated separately, and the important case where $v$ is identically null is similar to a coastal Kelvin wave, with $\omega = -k$ \citep{Matsuno}. The form of the solutions in the $y$ direction are expressed through the use of the parabolic cylinder functions $\psi_n$, given by
\begin{equation}
\psi_n (y) = H_n(y)e^{-y^2/2},
\label{eq:parabolic}
\end{equation}
where $H_n$ is the n$^{\mathrm{th}}$ Hermite polynomial. Finally, simple mathematical arguments show that the eigenvalue $\omega$ is always real: the homogeneous solutions are only neutral modes. \citet{Matsuno} also showed that the eigenvectors of Eqs.\eqref{eq:matsuno_u} to \eqref{eq:matsuno_h} form a complete, orthogonal set of the 2D beta-plane: at a given time, any  
function on the beta plane can be written as a linear combination of the $\psi_n(y) \exp(\mathrm{i}kx)$ functions.

\citet{Matsuno} and \citet{Gill80} obtained the steady state solution to Eqs.\eqref{eq:sp11_u} to \eqref{eq:sp11_h} under the inclusion of heating (and cooling) and drag. The completeness and orthogonality of the above functions allows one to write:
\begin{equation}
\mathcal{Q} = \sum_{n,l} q_{n,l} \tilde{X}_{n,l},
\end{equation}
where $q_{n,l}$ is the projection of $\mathcal{Q}$ onto $\tilde{X}_{n,l}$. \citet{Matsuno} (their Eq.(34)) and \citet{Gill80} showed that a steady solution $X$ to the forced problem with $\tau_\mathrm{drag} = \tau_\mathrm{rad}$ is given by
\begin{equation}
X = \sum_{n,l} \dfrac{q_{n,l}}{\tau_\mathrm{drag}^{-1} - \mathrm{i} \omega_{n,l}}  \tilde{X}_{n,l}.
\label{eq:MG_matsuno}
\end{equation}
\citet{Showman2011} showed that, for a horizontal wavenumber one representing the asymmetric heating of hot Jupiters and $\tau_\mathrm{drag} =\tau_\mathrm{rad} = 10^5$\ s, the steady linear solution exhibits a `chevron' shaped pattern (in pressure, density or temperature), and has been denominated the Matsuno-Gill solution, leading to a net acceleration of the equator at the non linear order. However, it is not clear whether a linear steady state is relevant in  a case where non--linearities are likely dominant, i.e. hot Jupiters where the extreme forcing is likely to trigger non--linear effects over short timescales, and further whether it is appropriate to choose equal values for both dissipation timescales. Therefore, we require the time dependent solution of Eqs.\eqref{eq:sp11_u} to \eqref{eq:sp11_h} in the general case, which are expressed in section \ref{ssec:time_solution}.

\subsection{Non--Linear Accelerations from the Linear Steady State}
\label{ssec:problem_komacek}

Now that we have reviewed the main assumptions and equations for our basic framework, in this section, we move on to summarising the key results of \citet{Showman2011} and \citet{Tsai2014}
A key conclusion of \citet{Showman2011} is that the Matsuno--Gill pattern is a linear steady state, but the non linear accelerations from this circulation trigger an equatorial superrotation. Consider a linear perturbation (a wave or a steady linear circulation) associated with velocities $u',v',w'$ in the longitudinal, latitudinal and vertical directions, respectively. The non linear momentum fluxes per unit mass from this perturbation scale as
\begin{gather}
\phi_l \propto \overline{u'v'}, \\
\phi_v \propto \overline{u'w'},
\end{gather}
where $\phi_l$ is the latitudinal flux of momentum, $\phi_v$ the vertical and an overline denotes a longitudinal average. When $\phi_l$ is positive, there is a net transport of eastward momentum to the North, with a negative value resulting in a net transport to the South. When $\phi_v$ is positive, there is a net transport of eastward momentum upward, or downwards when it is negative.  Therefore, if $\phi_l$ is negative in the mid latitudes of the Northern hemisphere and positive in the Southern hemisphere, there is a net meridional convergence of eastward momentum (a similar argument applies in the vertical coordinate for a 3D systems). For the shallow water $\beta$--plane system used in \citet{Showman2011} the vertical momentum flux is accounted for by the addition of a coupling term between the deeper (high pressure), quiescent atmosphere and the dynamically active (lower pressure) atmosphere \citep[$R$ term in Eqs.(9) and (10) of][]{Showman2011}.

\begin{figure}  
    \begin{center}
    \subfigure[]{\includegraphics[width=8.5cm,angle=0.0,origin=c]{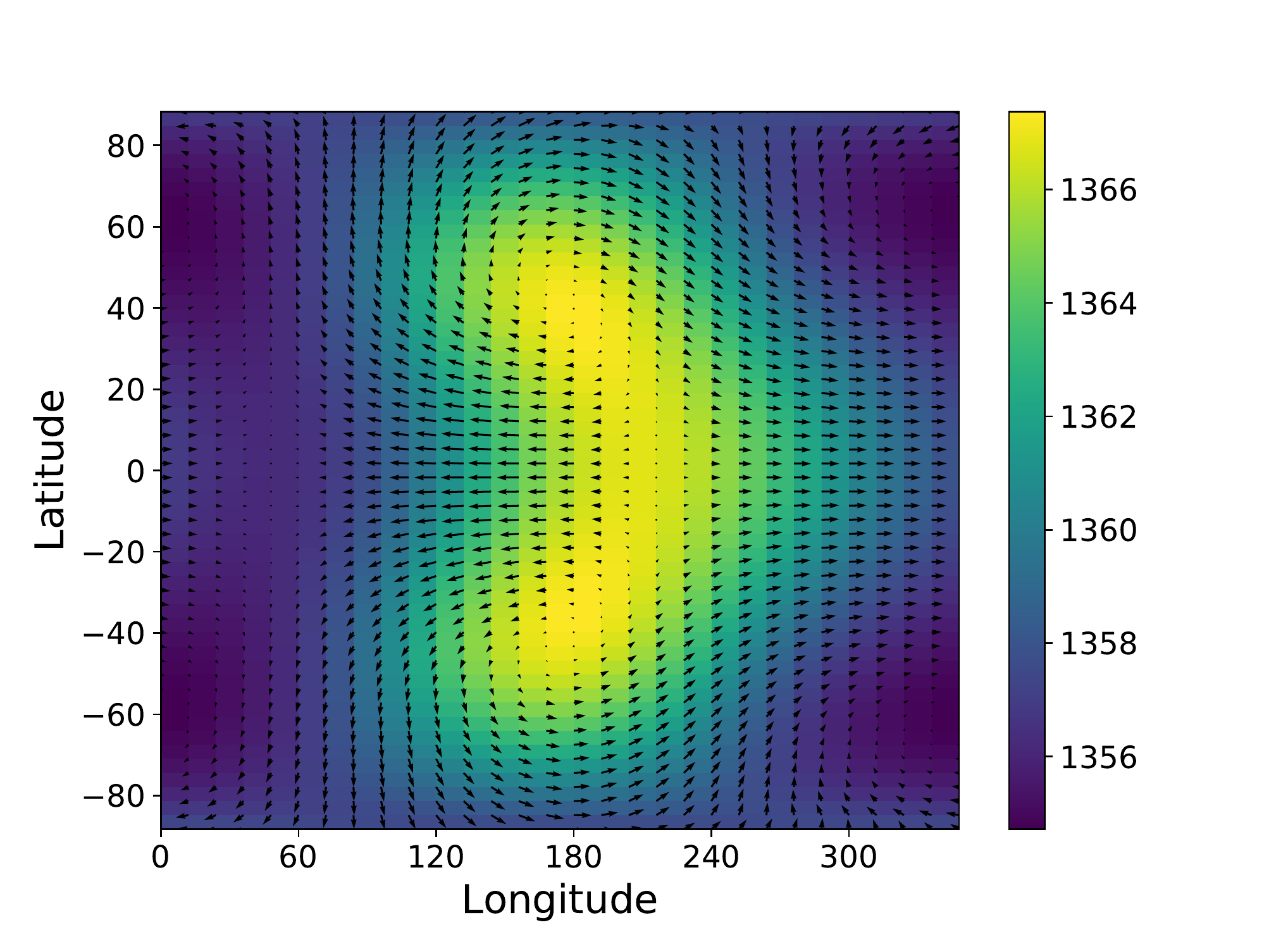}\label{fig:MG_usual}}
    \subfigure[]{\includegraphics[width=8.5cm,angle=0.0,origin=c]{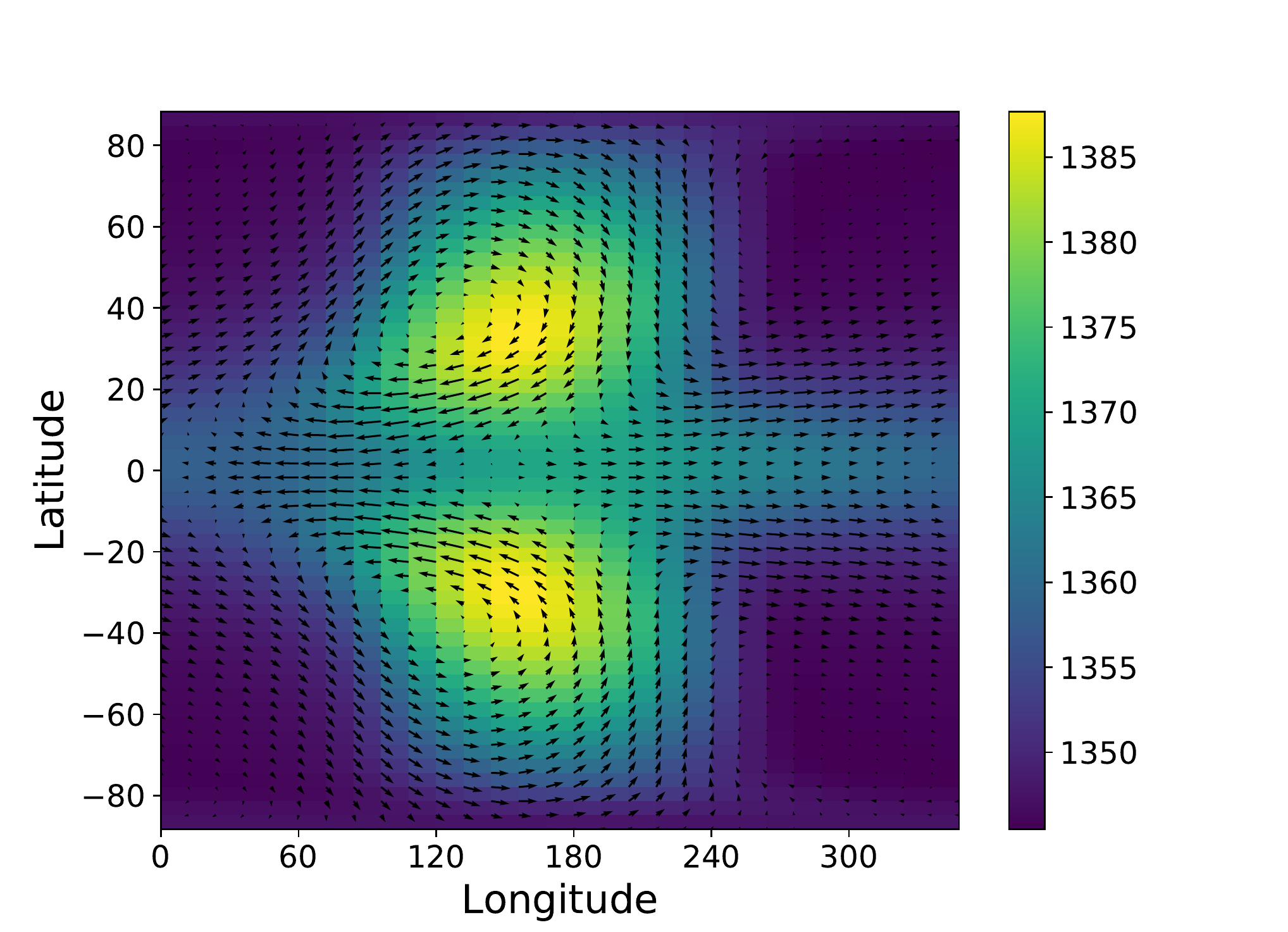}\label{fig:MG_komacek}}
    \caption{Temperature (colourscale in K) and horizontal wind (arrows) as a function of longitude (x axis) and latitude (y axis) at the 40\,mbar pressure level of the linear steady state (denominated Matsuno-Gill circulation)
        obtained using ECLIPS3D \citep{Debras2019} with heating function, drag and radiative timescales following the definitions of \citet{Komacek2016}. Following the notation of \citet{Komacek2016}: (a) $\Delta T_\mathrm{eq,top} = 100$\,K, $\tau_\mathrm{drag,top} = 10^5$\,s and $\tau_\mathrm{rad,top} = 10^5$\,s. The maximum speed at this pressure range is 10$\,\mathrm{m.s^{-1}}$. (b) $\Delta T_\mathrm{eq,top} = 100$\,K, $\tau_\mathrm{drag,top} = 10^6$\,s and $\tau_\mathrm{rad,top} = 10^4$\,s. The maximum speed at this pressure range is 100$\,\mathrm{m.s^{-1}}$. Note that the maximum speed has been multiplied by ten as the drag timescale has been multiplied by ten.}
    \label{fig:MG_usual+komacek}
    \end{center}
\end{figure}

In Figure \ref{fig:MG_usual}, we present the temperature (colour scale, $K$) and wind vectors (vector arrows) as a function of latitude and longitude for a typical Matsuno-Gill circulation. This 3D linear steady state was obtained using ECLIPS3D \citep[see][for details and benchmarking of ECLIPS3D]{Debras2019}, a linear solver for waves, instabilities and linear steady states of an atmosphere under prescribed heating and drag. The initial state around which the equations are linearised is an axisymmetric, hydrostatically balanced state at rest.  The bottom pressure is set to $220$\,bars and the temperature profile follows that of \citet{iro2005}, with the polynomial fit in the log of pressure of \citet{Heng2011}, assuming an ideal gas equation of state. Due to the very high inner boundary pressure, the choice of the inner boundary condition does not impact our results. The physical parameters relevant for HD~209458b used in this setup, namely the radius $R_\mathrm{p}$, the rotation rate $\Omega$, the depth of the atmosphere $R_\mathrm{top}$, the surface gravity $g_\mathrm{p}$, the inner boundary pressure $p_\mathrm{max}$, the specific heat capacity $c_\mathrm{p}$ and the ideal gas constant $R$ are given in Table \ref{tab:basic_par}. Finally, the heating as well as drag and radiative timescales were prescribed as in \citet{Komacek2016} with the addition of an exponential decay of the heating in the upper, lower pressure, part of the atmosphere. The exponential damping acts to mimic the damping of vertical velocities close to the outer boundary, or `sponge layer' applied in 3D GCMs \citep[see for example][]{Mayne2014}. In turn, this damping layer allows the adoption of a `no escape' or reflective outer boundary condition, which would otherwise reflect waves back into the numerical domain. The equilibrium temperature towards which the atmosphere relaxes is a sinusoidal function of latitude and longitude, and $\Delta T_\mathrm{eq}$, the equilibrium day side--night side temperature difference, decreases logarithmically in pressure between $10^{-3}$\,bar where $\Delta T_\mathrm{eq} = \Delta T_\mathrm{eq,top}$ (the value at the top of the atmosphere which is set to 10K in this case) to 10 bars where $\Delta T_\mathrm{eq} = 0$. The drag timescale, $\tau_\mathrm{drag}$, is constant throughout the atmosphere at $10^5 \ $s and the radiative timescale, $\tau_\mathrm{rad}$, is a logarithmically increasing function of pressure \citep[see Eq.(7) of][]{Komacek2016} between $10^{-2}$\,bar where $\tau_\mathrm{rad} = \tau_\mathrm{rad,top} = 10^5$\,s and 10\,bar where $\tau_\mathrm{rad} = 10^7$\,s. 

\begin{table*}
  \caption{Value of the standard parameters for HD~209458b, following \citet{Mayne2014}. }
\label{tab:basic_par}
\centering
\begin{tabular}{lc}
  \hline\hline
  Quantity&Value\\
  \hline
  Radius, $R_{\rm p}$ (m)&$9.44\times10^7$\\
  Rotation rate, $\Omega$ (s$^{-1}$)&$2.06\times 10^{-5}$\\
  Depth of the atmosphere, $R_{\rm top}$ (m)&$1.1\times 10^7$\\
  Surface gravity, $g_{\rm p}$ (ms$^{-2}$)&9.42\\
  Inner boundary pressure, $p_{\rm max}$ (Pascals, Pa)&$220\times 10^5$\\
  Specific heat capacity (constant pressure), $c_{\rm p}$ (Jkg$^{-1}$K$^{-1}$)&14\,308.4\\
  Ideal gas constant, $R$ (Jkg$^{-1}$K$^{-1}$)&4593\\
  \hline
\end{tabular}
\end{table*}

As shown in Figure \ref{fig:MG_usual} the maximum temperature at the equator is shifted eastward from the substellar point (the substellar point is set at a longitude of $180$ degrees) in our results consistent with observations \citep{Knutson2007,Zellem2014}. The meridional circulation exhibits a Rossby wave--type circulation at mid latitudes, with clockwise or anticlockwise rotation around the pressure maxima, and a Kelvin wave type circulation at the equator, with no meridional velocities. The combination of both of these circulations brings eastward momentum to the equator to the East of the substellar point, and advects westward momentum to the mid latitudes to the West of the substellar point. Globally, it is easily shown that $\phi_l$ is indeed negative in the Northern hemisphere and positive in the Southern hemisphere: there is a net convergence of eastward momentum at the equator. According to \citet{Showman2011}, this convergence is associated with divergence of vertical momentum flux, and equilibration occurs when the vertical and meridional terms balance. 

\citet{Tsai2014} have further extended this understanding by expanding to include the vertical transport more completely. \citet{Tsai2014} projected the heating function onto equivalent height $\beta$--plane solutions (see Section \ref{ssec:framework}), showing that the vertical behaviour of the waves can be linked to the equilibration of the jet (a synonym here, and throughout this work for equatorial superrotation). More precisely, \citet{Tsai2014} show that in the limit of slow evolution, or strong dissipation, their linear development around a steady flow with constant background zonal velocity reproduces the wave processes occurring in 3D simulations extremely well. \citet{Tsai2014} show that the wave response of the atmosphere is shifted from West to East when the background zonal velocity is increased (their figure 10): this is interpreted as a convergence towards a single equilibrium state, where the non linear acceleration from the linear processes cancel. Although very detailed and physically relevant, the results of \citet{Tsai2014} are, as they state, only applicable to the strong or modest damping scenario, dictated by the fact that the waves must have the time to reach a stationary state before non linearities become significant. Throughout this work we define the drag regime relative to the {\it initial} acceleration of the jet: in the 'weak' drag regime, non-linear terms become non negligible before a linear steady state (MG) could have been reached, in the 'modest' drag regime the time to reach the MG state is comparable with the time to depart from this linear steady state, and in the `strong' drag regime we can decouple the linear and the non linear evolution of the planet, as considered by \citet{Showman2011}.  
Once an initial jet has been accelerated, the evolution of the atmosphere is much slower than its initial acceleration and the results of \citet{Tsai2014} therefore apply, even in a weak drag regime, explaining the consistency of their work for the evolution of the jet towards an equilibrated state.

\citet{Komacek2016} compare the steady states from various 3D GCM simulations across a range of $\tau_\mathrm{rad}$ and $\tau_\mathrm{drag}$ values (their figures 4 and 5). The simulations of \citet{Komacek2016} extend from low forcing, hence a linear steady state, to strong forcing, hence a non linear steady state. Contrary to the conclusions of \citet{Showman2011}, \citet{Komacek2016} also show that when the linear steady state resembles that of Figure \ref{fig:MG_usual}, the associated non linear steady state is {\it not} (or weakly) superrotating. This can be understood by the fact that $\tau_\mathrm{drag}$ is smaller than the characteristic timescale of advection by the superrotating jet over the whole planet, hence the jet is dissipated before it can reach a steady state. In Figure \ref{fig:MG_komacek}, we present the results from ECLIPS3D obtained when reproducing a particular setup of \citet{Komacek2016}, namely with $\tau_\mathrm{drag} = 10^6$\,s and $\tau_\mathrm{rad,top} = 10^4$\,s. According to the analysis of \citet{Komacek2016} the non linear steady state associated with the linear steady state of Figure \ref{fig:MG_komacek} {\it does} exhibit strong superrotation, although the tilt of the wave in Figure \ref{fig:MG_komacek} would lead to a removal of momentum at the equator. \citet{Komacek2016} acknowledge this: "these phase tilts are the exact opposite of those that are needed to drive superrotation". For the explanation of \citet{Showman2011} the stationary wave pattern obtained from the heating is postulated to accelerate superrotation, but \citet{Komacek2016} present results which oppose this scenario: when the linear steady state accelerates the equator (Figure \ref{fig:MG_usual}, with $\tau_\mathrm{drag} = 10^5$\,s) the associated non linear steady state is not superrotating. However, when the linear steady states takes momentum away from the equator (Figure \ref{fig:MG_komacek}, with $\tau_\mathrm{drag} = 10^6$\,s), the non linear steady state is superrotating. Thus, there is an inherent discrepancy between \citet{Showman2011} and \citet{Komacek2016}. In order to understand this discrepancy, we need to go a step beyond the sole consideration of a linear steady state, and study the evolution of the linear solution with time. This is the objective of the sections \ref{ssec:time_solution} and \ref{sec:waves}. 

\subsection{Time dependent solutions}
\label{ssec:time_solution}

Now that we have established the basic mathematical system, and summarised the current picture of superrotation in hot Jupiter atmospheres, we move to expressing the time dependent solution to the forced problem which provides us with the shape of the atmospheric wave response. Our main assumption is that the heating function can be decomposed onto the homogeneous solutions of Eqs.\eqref{eq:dissip_u} to \eqref{eq:dissip_uh}. When $\tau_\mathrm{rad} =\tau_\mathrm{drag}$, the horizontal part (defined as ${X}_{n,l} (x,y,t=0)$ in Eq.\eqref{eq:eigen}) of the eigenvectors still form a complete set of the equatorial $\beta$--plane because of the orthogonality and completeness of the Hermite functions, as in \citet{Matsuno}. However, when $\tau_\mathrm{rad} \neq \tau_\mathrm{drag}$ the eigenvectors are no longer orthogonal, as shown in Appendix \ref{app:orthogonality}, and a rigorous proof would be needed to show that they still form a complete set of solutions\footnote{Although the eigenvectors remain linearly independent, hence a Gram-Schmidt method could ensure creation of an orthogonal set of these eigenvectors.}. From a physical perspective, it is expected that the heating function will trigger linear waves which are solutions of the homogeneous equation, and such a decomposition of the heating function onto these waves therefore probably exists, although it is no longer simply given by a scalar product. Finally, it is worth stating that solving $\partial X/\partial t = a X + \mathcal{Q}$ is straightforward (except that we don't know the eigenvalues and eigenvectors of the homogeneous equation), however, employing a Green's function to solve this equation provides a more physically intuitive result in terms of wave propagation and dissipation.

With the addition of a drag timescale, $\tau_\mathrm{drag}$, 
and a radiative timescale, $\tau_\mathrm{rad}$, Eqs.\eqref{eq:matsuno_u} to 
\eqref{eq:matsuno_h} can be modified to yield
\begin{gather}
\left(\mathrm{i} \omega+ \dfrac{1}{\tau_\mathrm{drag}}\right) u - yv + \mathrm{i}k h = 0, \label{eq:dissip_u} \\
\left(\mathrm{i} \omega+ \dfrac{1}{\tau_\mathrm{drag}}\right)  v + yu + \dfrac{\partial h}{\partial y} = 0,\label{eq:dissip_v} \\
\left(\mathrm{i} \omega+ \dfrac{1}{\tau_\mathrm{rad}}\right)  h +\mathrm{i} k u + \dfrac{\partial v}{\partial y} = 0
\,\, .\label{eq:dissip_uh}
\end{gather}
Indexing again the solutions by $n$ and $l$ as in \citet{Matsuno}, we define
\begin{equation}
X_{n,l} = (u_{n,l},v_{n,l},h_{n,l})=\tilde{X}_{n,l}(y)e^{\mathrm{i} k x+(\mathrm{i}\nu_{n,l}-\sigma_{n,l})t}
\label{eq:eigen}
\end{equation} as an eigenvector of Eqs.\eqref{eq:dissip_u}, \eqref{eq:dissip_v} and \eqref{eq:dissip_uh}, 
with $\tilde{X}_{n,l} (y)$ the amplitude of the wave, $k$ the
horizontal wave number, $\nu_{n,l}$ its frequency and $\sigma_{n,l}$ its damping
(or growing/growth) rate (note that $\omega_{n,l} = \nu_{n,l}+\mathrm{i} \sigma_{n,l}$). We also define $\mathcal{L}$ as the operator of the same equations, such that $\mathcal{L}X_{n,l} = 0$ for all $n$ and $l$. The general equation can then be written as $\mathcal{L}X_\mathrm{F} = \mathcal{Q}$, where $\mathcal{Q}$ is the forcing which, as it is only present in the third individual equation is given by $\mathcal{Q} = (0,0,Q)$, and $X_\mathrm{F}$ is the forced, time dependent solution. A homogeneous solution $X_\mathrm{H}$ can be written in its general form as 
\begin{equation}
X_\mathrm{H} = \sum_{n,l} \alpha_{n,l} X_{n,l},
\end{equation} 
where $\alpha_{n,l}$ are scalars.

When $\tau_\mathrm{drag} = \tau_\mathrm{rad}$, $\nu_{n,l}$ are similar to the $\omega_{n,l}$ of \citet{Matsuno} and $\sigma_{n,l} = \tau_\mathrm{drag}^{-1}$ for all $(n,l)$. When $\tau_\mathrm{drag} \neq \tau_\mathrm{rad}$, the analytical expressions for $\nu_{n,l}$ and $\sigma_{n,l}$ are not known a priori. In order to solve the general equation, we seek the causal Green function $X_\mathrm{G}$ that represents the solution at time $t$ due to switching on the forcing at time $t'$ only. Therefore, for all $t$ and $t^{\prime}$:
\begin{equation}
\mathcal{L}X_\mathrm{G}(x,y,t,t^{\prime})= \delta(t-t^{\prime}) F(x,y,t'),
\label{eq:def_green}
\end{equation} 
where $\delta(t)$ is the Dirac distribution and $F$ is the heating function. In the case of simulated hot Jupiter atmospheres,
the star is effectively `switched on' at $t=0$ after which the heating is constant with time (in the linear limit).  $F$ can then simply be expressed as $F(x,y,t^{\prime}) = \Theta(t^{\prime}) \mathcal{Q}(x,y)$, where $\Theta(t)$ is the Heavyside function (null when $t<0$ and equal to $1$ otherwise). The forced solution of Eqs.\eqref{eq:dissip_u} to \eqref{eq:dissip_uh} is simply the integral over $t^{\prime}$ of the causal Green function, hence the sum of the responses of the atmosphere excited by a Dirac function of the forcing at time $t^{\prime}$:
\begin{equation}
X_\mathrm{F} = \displaystyle \int_{-\infty}^{\infty} X_\mathrm{G}(t-t^{\prime}) \Theta(t^{\prime}) dt^{\prime} = 
\displaystyle \int_{0}^{t} X_\mathrm{G}(t-t^{\prime}) \Theta(t^{\prime}) dt^{\prime} ,
\label{eq:green_forced}
\end{equation}
where the change in the upper limit of integration can be made due to the fact that the Green function is causal, and is therefore zero when $t-t^{\prime}$ is negative 
. 
From the definition of the Green function (Eq.\eqref{eq:def_green}), for $(t-t^{\prime})>0$
we have $\mathcal{L}X_\mathrm{G} = 0$. Hence $X_G$ is a homogeneous solution of Eqs.\eqref{eq:dissip_u} to \eqref{eq:dissip_uh} when $t > t^{\prime}$. It is then logical to choose for $X_\mathrm{G}$:
\begin{equation}
X_\mathrm{G}(t,t^{\prime})= \Theta(t-t^{\prime}) \sum_{n,l} \alpha_{n,l} X_{n,l}(t-t^{\prime}),
\label{eq:green_choice}
\end{equation}
and it is easily verified that
\begin{gather}
\mathcal{L}X_\mathrm{G}= \delta(t-t^{\prime}) \sum_{n,l} \alpha_{n,l} X_{n,l}(t-t^{\prime}) \nonumber \\
+\Theta(t-t^{\prime}) \dfrac{\partial}{\partial t}\sum_{n,l} \alpha_{n,l} X_{n,l}(t-t^{\prime})\nonumber \\
+\mathcal{L}_h\Theta(t-t^{\prime}) \sum_{n,l} \alpha_{n,l} X_{n,l}(t-t^{\prime}),
 \nonumber \\
 = \delta(t-t^{\prime}) \sum_{n,l} \alpha_{n,l} X_{n,l}(t-t^{\prime}) + 
\Theta(t-t^{\prime}) \mathcal{L}\sum_{n,l} \alpha_{n,l} X_{n,l}(t-t^{\prime}), \nonumber \\
 = \delta(t-t^{\prime}) \sum_{n,l} \alpha_{n,l} X_{n,l}(t-t^{\prime}),
\end{gather}
where we have used the fact that the derivative of the Heavyside function is the Dirac
distribution, $\mathcal{L} = \partial/\partial t+\mathcal{L}_h$ where $\mathcal{L}_h$ is an operator 
acting on the horizontal coordinates only and $\mathcal{L} X_{n,l} = 0$. 
Therefore, in order to solve the forced problem we can project the forcing onto the homogeneous solutions and write for $t=t^{\prime}$: 
\begin{gather}
\sum_{n,l} \alpha_{n,l} X_{n,l}(t=t^{\prime}) = \sum_{n,l} \alpha_{n,l} \tilde{X}_{n,l} = \mathcal{Q}(x,y). \,\,
\label{eq:scalar}
\end{gather}
 The first equality simply arises from the definition
of $\tilde{X}_{n,l}$, whereas the second uses Eq.\eqref{eq:def_green}. Hence if we know the projection of $\mathcal{Q}$ onto the $\tilde{X}_{n,l}$, the $\alpha_{n,l}$ quantities, the final solution can be obtained. By setting $\alpha_{n,l} = q_{n,l}$, we recover the results of the previous section
(these are termed $b_{m}$ and $b_{m,n,l}$ in \citet{Matsuno} and \citet{Tsai2014}, respectively where the latter is in 3D). The solution to the forced problem is given by injecting the Green function (Eq.\eqref{eq:green_choice}) into Eq.\eqref{eq:green_forced}: 
\begin{align}
X_\mathrm{F} &= \displaystyle \int_{-\infty}^{\infty} \sum_{n,l} q_{n,l} X_{n,l}(t-t^{\prime}) \Theta(t-t^{\prime}) \Theta(t^{\prime}) dt^{\prime}, \nonumber \\
&= \sum_{n,l}\displaystyle \int_{0}^{t}  q_{n,l} \tilde{X}_{n,l}(x,y) 
e^{(i\nu_{n,l}-\sigma_{n,l}) (t-t^{\prime})} \Theta(t^{\prime}) dt^{\prime}.
\label{eq:green_integral}
\end{align}
Under this integral form, it is clear that the solution is the continual excitation (the $\Theta$ term)
of waves with a characteristic frequency $\nu_{n,l}$ and time of decay of $\sigma_{n,l}^{-1}$. The amplitude of the excited waves is proportional to their projection onto the forcing function $Q$, as explained by \citet{Gill80} and  \citet{Tsai2014}. Solving this integral yields:
\begin{equation}
X_\mathrm{F} = \sum_{n,l} \dfrac{q_{n,l} \tilde{X}_{n,l}}{\sigma_{n,l}-i\nu_{n,l}} 
\left( 1-e^{(i\nu_{n,l}-\sigma_{n,l}) (t)} \right).
\label{eq:linear_evolution}
\end{equation}
In this form we simply recover the results of \citet{Matsuno}, and notably their Eq.(34) or our Eq.\eqref{eq:MG_matsuno},
\begin{equation}
X_\mathrm{MG} = \sum_{n,l} \dfrac{q_{n,l} \tilde{X}_{n,l}}{\sigma_{n,l}-i\nu_{n,l}}, 
\label{eq:MG_green}
\end{equation}
where $X_\mathrm{MG}$ is the Matsuno-Gill solution, hence the steady solution to the forced problem. The time dependent part of the solution could have been obtained from a simple first order equation solution. However, the Green function formalism allows us to determine that the solution consists of permanently forced waves that are damped with time, and that the shape of the atmosphere is given by the interactions between these waves. Notably, with Eq.\eqref{eq:green_integral}, we can write
\begin{equation}
X_\mathrm{MG} = \lim_{t\to\infty} \sum_{n,l}\displaystyle \int_{0}^{t}  q_{n,l} \tilde{X}_{n,l}
e^{(i\nu_{n,l}-\sigma_{n,l}) (t-t^{\prime})}
\Theta(t^{\prime}) dt^{\prime},
\label{eq:MG_integral}
\end{equation}
the form of which confirms the interpretation of the stationary solution as an infinite interaction of waves. Additionally, it shows that for a given heating function, changing the value of $\tau_\mathrm{drag}$ (but keeping $\tau_\mathrm{rad} = \tau_\mathrm{drag}$), hence not altering the $\tilde{X_{n,l}}$ and $\nu_{n,l}$ but only $\sigma_{n,l} = \tau_\mathrm{drag}^{-1}$, will change the linear steady solution. This is as the excited waves will not propagate to the same length before being damped. This was first realised by \citet{Wu2001}, where they show that the zonal wave decay length is of the order of $\sqrt{\tau_\mathrm{rad}^{-1}\tau_\mathrm{drag}^{-1}}$ for arbitrary
$\tau_\mathrm{rad}$ and $\tau_\mathrm{drag}$.  

From Eq.\eqref{eq:green_integral}, we see that the linear solution is controlled by three main parameters: the shape of the forcing function ($q_{n,l}$), the global behaviour of the waves (horizontal shape of $\tilde{X}_{n,l}$ and $\nu_{n,l}$) and the dissipation of the waves ($\sigma_{n,l}$). Although Eq.\eqref{eq:green_integral} provides the form of the solution to the time dependent problem, the three main parameters we have detailed remain unknown. Therefore, we move to quantifying the sensitivity of the solution to the forcing function, hence the influence of the $q_{n,l}$ in the next section. 

\section{Insensitivity of the Matsuno-Gill solution to the differential heating}
\label{sec:heating}

The interpretation of \citet{Showman2011} relies on a simplified treatment of the forcing, the impact of which we must first assess before discussing the impact of the linear evolution of the atmosphere.
Firstly, in order to derive analytical results, \citet{Showman2011} impose an antisymmetric (sinusoidal) heating where the night side of the planet is cooled as much as the day side is heated. In this case, the linear steady solution gives rise to the chevron shaped pattern they denominate the `Matsuno-Gill' circulation. However, the actual structure of the heating is not just a sinusoidal function. Moreover, from \citet{Mayne2017} and \citet{Amundsen2016} we know that there are qualitative differences between the steady state of GCM simulations of hot Jupiters calculated using either a Newtonian heating or with a more sophisticated radiative transfer scheme. In that regard, the first intuitive idea to test is whether the MG pattern is robust when the heating function is changed. With the addition of the vertical dimension, \citet{Tsai2014} have shown that the linear solution strongly resembles the MG circulation at low pressures in the atmosphere. It is not clear whether this holds with realistic, three dimensional heating functions.

From Eq.\eqref{eq:MG_matsuno} or \eqref{eq:MG_integral}, as ${X}_{n,l}(x,y) = \tilde{X}_{n,l}(y)\exp(\mathrm{i}kx)$, we know that the projection of the heating function onto different wavenumbers, $k$, can alter the resulting Matsuno-Gill circulation from that obtained at wavenumber one. To verify this point, using ECLIPS3D we solve for linear steady circulations across a range of prescribed heating rates, adopting the drag and radiative timescales of \citet{Komacek2016}, as used previously for the data presented in Figure \ref{fig:MG_usual}. We use three forms of heating, the first two modified from that used for Figure \ref{fig:MG_usual}, and a third one, inspired from 3D GCM simulations:
\begin{itemize}
    \item The same day-side forcing as Figure \ref{fig:MG_usual}, but no night-side cooling.
    \item The same forcing as Figure \ref{fig:MG_usual} but with the night-side cooling enhanced by a factor of 3.
    \item A heating profile matching that obtained from 3D GCM simulations \citep[taken from][]{Amundsen2016}, including net day side heating and night side cooling. 
\end{itemize}
The first two cases allow us to test the robustness of the pattern under extreme situations, with the third mimicking the GCM simulations. ECLIPS3D calculates the linear steady states by inverting the linear matrix obtained with the full 3D equations \citep[see description in][]{Debras2019}. The outer boundary condition is a `no escape' condition (zero vertical velocity $w$), and we have applied an exponential decay of the forcing in the high atmosphere (low pressure) to mimic the damping of vertical velocities, or sponge layer, employed in the UM and other GCMs. For the inner boundary, we adopt a solid boundary condition for the vertical velocity ($w=0$), and a free slip (no vertical derivatives of the horizontal velocities, $u$ and $v$) or no slip (no horizontal velocities) condition on $u$ and $v$ give qualitatively similar results because of the very high inner pressure. These assumptions are obviously simplifications of the real physical situation, but we have assessed their impact by changing the range of pressures at the inner and outer boundaries, and find no significant change in the qualitative results. The physical parameters adopted for HD~209458b are the same as in \citet{Mayne2017} and given earlier in Table \ref{tab:basic_par}.

\begin{figure*} 
    \subfigure[]{\includegraphics[width=8.5cm,angle=0.0,origin=c]{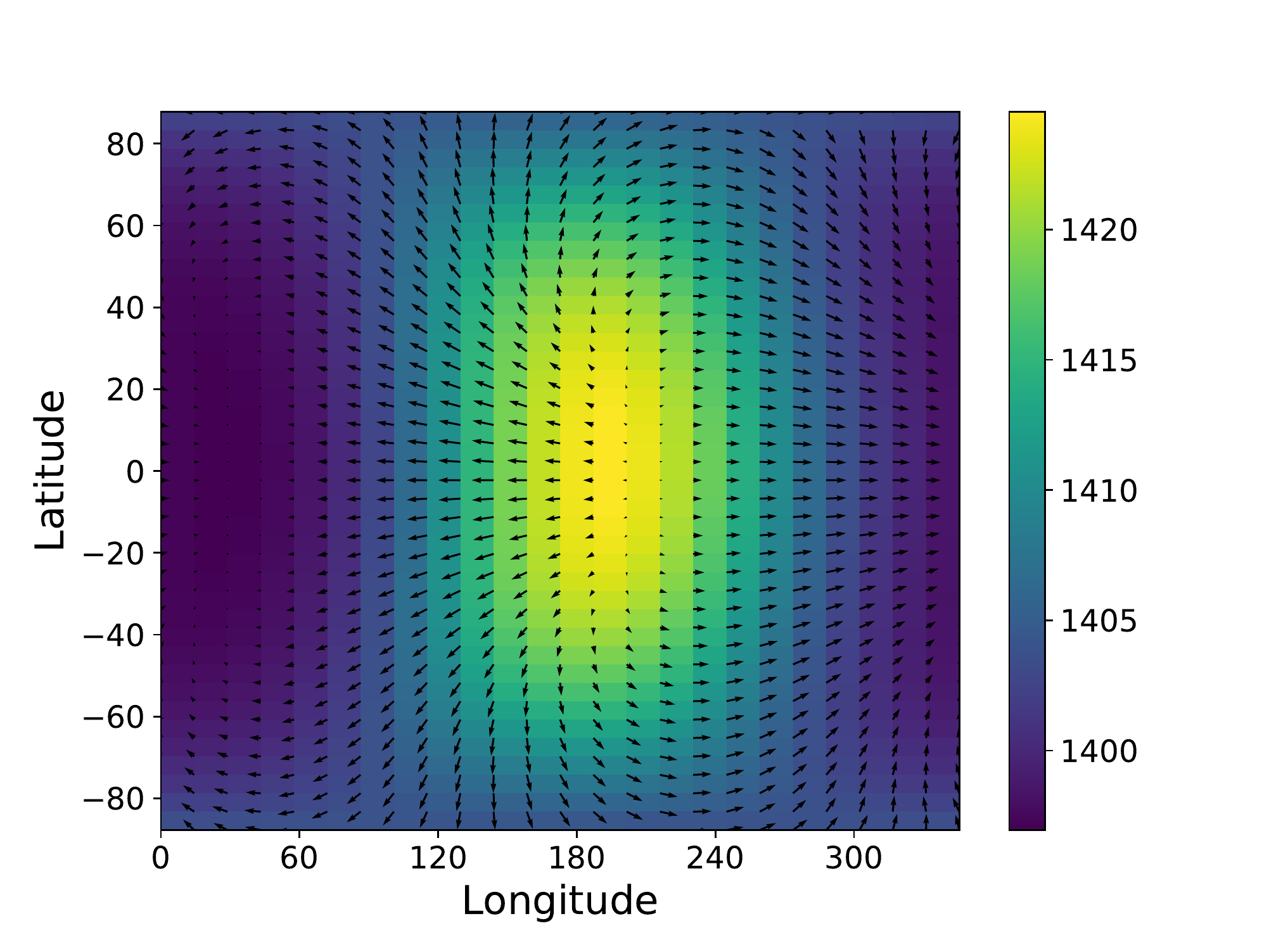}\label{fig:normal}}
    \subfigure[]{\includegraphics[width=8.5cm,angle=0.0,origin=c]{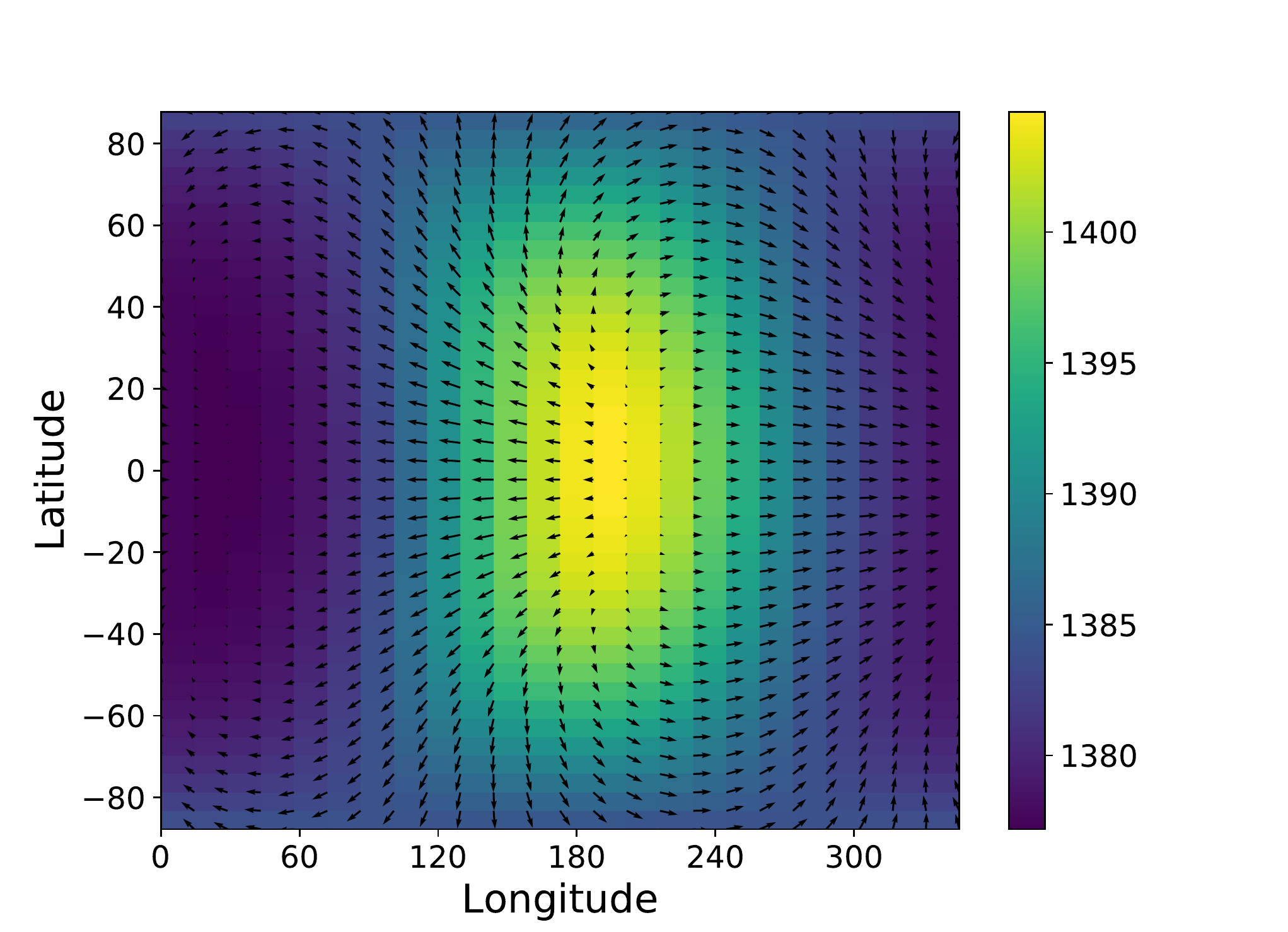}\label{fig:low_heating}} 
     \vspace{-1\baselineskip} \\
    \subfigure[]{\includegraphics[width=8.5cm,angle=0.0,origin=c]{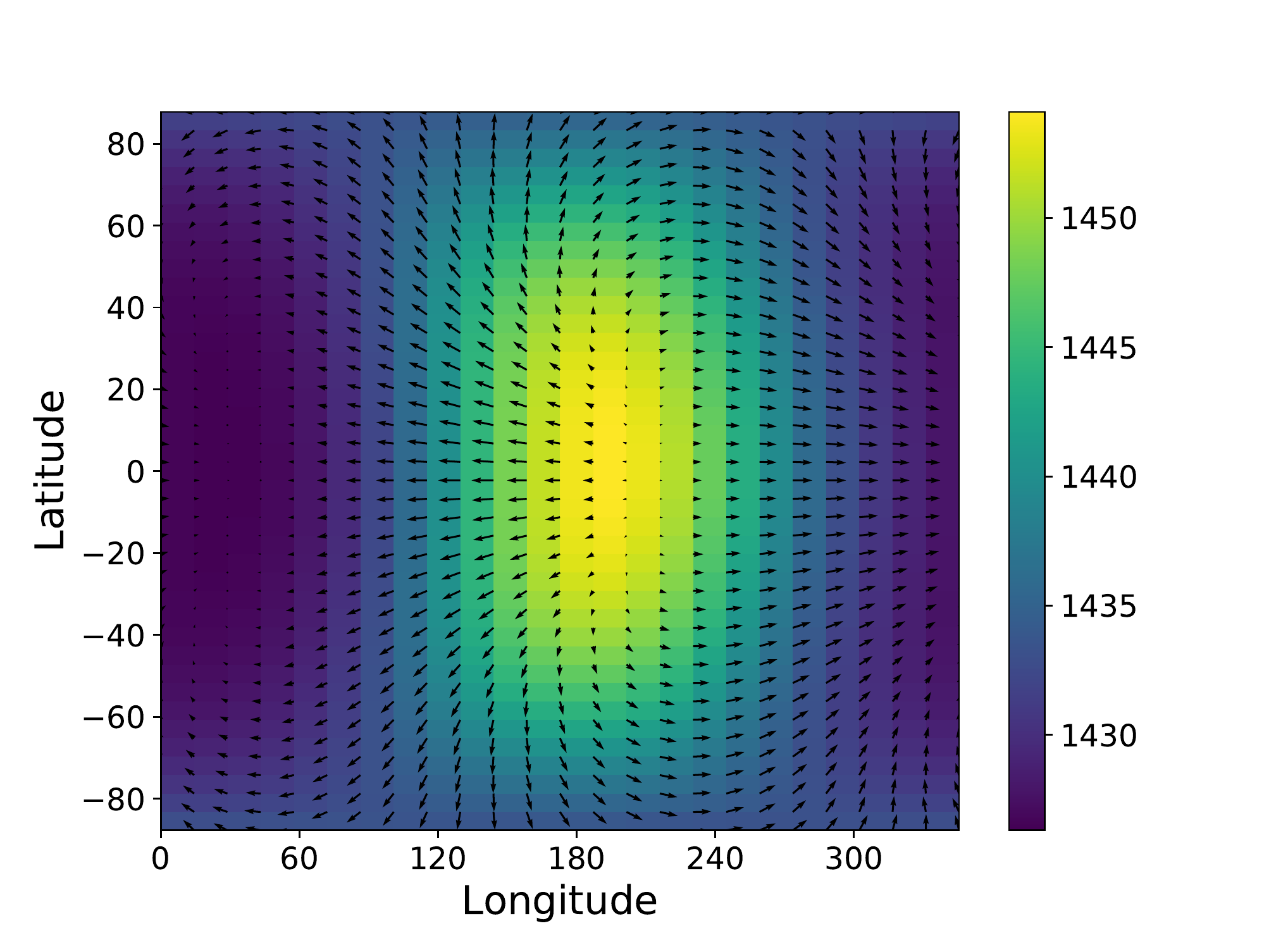}\label{fig:heating_nosym}}
    \subfigure[]{\includegraphics[width=8.5cm,angle=0.0,origin=c]{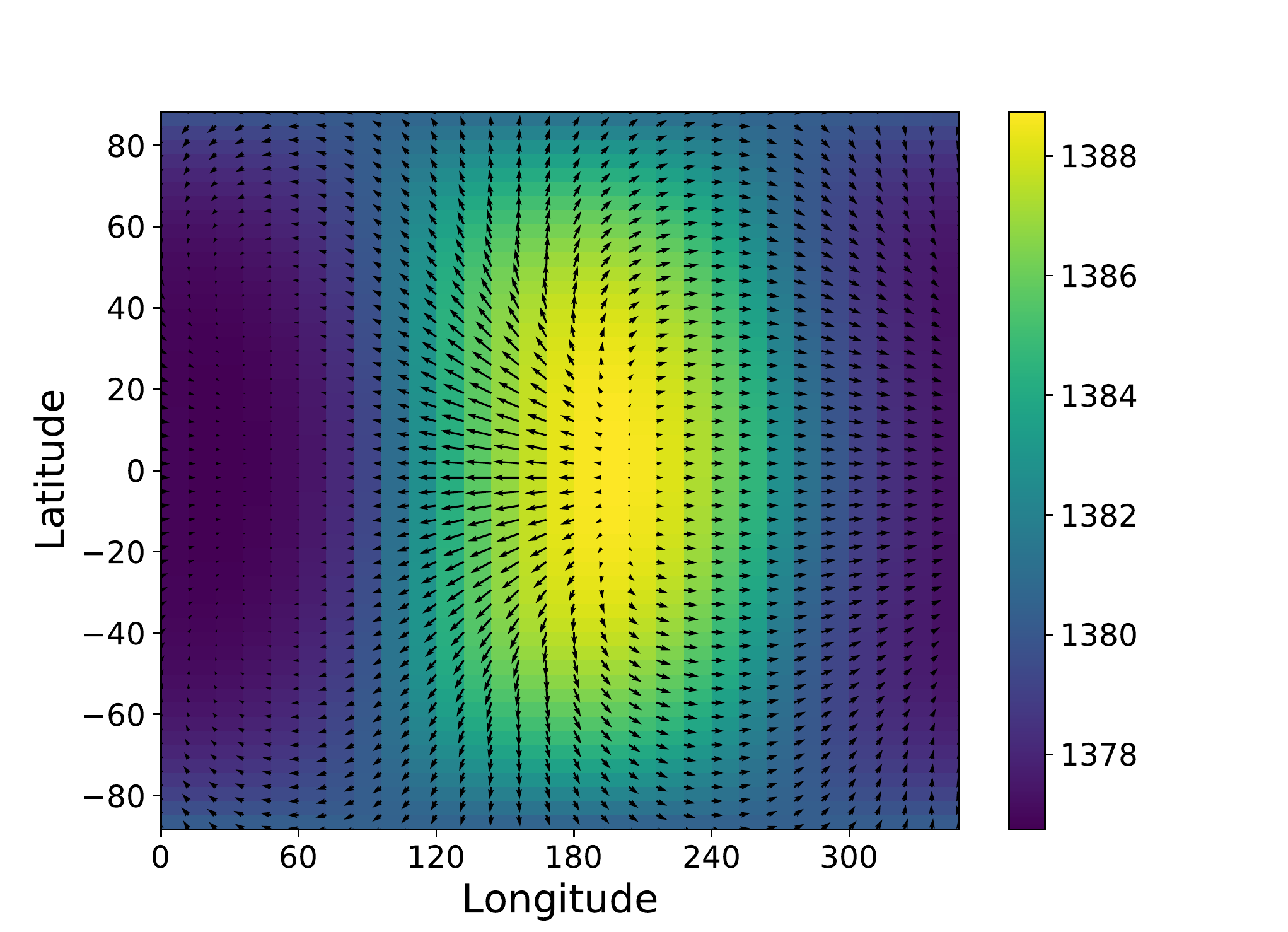}\label{fig:heating_RT}}
    \caption{Temperature (colorscale in $\mathrm{K}$) and winds (arrows) as a function of longitude and latitude with different heating functions, with drag and radiative timescales defined as in \citet{Komacek2016} with non-dimensional values of 1 (almost half an Earth day). For figures (a), (b) and (c) the initial pressure at this height is 50mbar which corresponds to an initial temperature of 1326K and the forcing is $\Delta T_\mathrm{eq,top}= 50K$. For figure (d), the initial pressure is 40mbar and initial temperature 1322K. (a):  Heating as in \citet{Komacek2016} (b): Same as \citet{Komacek2016} but with a cooling at the night side 3 times more efficient than the heating on the day side. (c): Same as \citet{Komacek2016} with no cooling at the night side. (d): Heating rate extracted from the full radiative transfer calculations of the GCM (divided by ten to obtain comparable values).}
    \label{fig:heating_change}
\end{figure*}

Figure \ref{fig:heating_change} shows the perturbed temperature (steady temperature minus initial {temperature}, colorscale) and winds (arrows) as a function of longitude and latitude of the MG solutions for the three different heating profiles described previously, at a height where the initial pressure is 50\ mbar.  Figure \ref{fig:heating_change} shows that the shape of the linear steady circulation does not qualitatively depend on the forcing: we always recover an eastward shift of the hot spot, associated with a tilt of the eddy patterns leading to a net acceleration of the equator for the non linear order. As long as there is a differential heating between the day and the night side, the linear steady solution of the atmosphere exhibits the "chevron-shaped" pattern of the MG circulation (we have verified that a constant heating across the whole planet or just a cooling on the night side does not lead to solutions of a similar form to the MG solutions). There is however a change in the quantitative values, without affecting the qualitative aspects of the non linear momentum transfer (see Section \ref{sec:transition}). Therefore, relaxing the approximation of a wavenumber one (e.g., day--night antisymmetric forcing) heating function has no influence on the non linear acceleration around the linear steady state: the projection onto different zonal wavenumbers changes the zonal amplitude of the MG circulation, but not its qualitative shape. We only presented results with a characteristic drag and radiative timescale of $10^5$s, but we have verified that changing these values affects the shape of the solutions in all cases in a similar way. 

Globally, even for simulations with a proper treatment of the radiative transfer, as long as the planet is tidally locked we expect the linear steady circulation to be MG-like. In the paradigm proposed by \citet{Showman2011} and \citet{Tsai2014}, the propagation of planetary waves impose this global MG circulation, exhibiting no superrotation, with the acceleration on to superrotation being due to eddy mean flow interactions around this primordial state. Therefore, the time to reach the linear steady state, which is by no means a non linear steady state, must be small relatively to other dynamical times in the system. In the case where the drag and radiative timescales are equal, this led \citet{Tsai2014} to conclude that their work could not apply to long diffusion timescales. We now seek to assess if similar conclusions can be made in cases where the drag and radiative timescales differ, in order to determine how the atmosphere reacts at first order. To develop this argument, we  analytically restrict ourselves to the quasi geostrophic set of equations (2D Cartesian, shallow water $\beta$--plane), as detailed in section \ref{ssec:framework}.

\section{Wave propagation and dissipation}
\label{sec:waves}

We have established, using a Green's function, that the linear solution to the 2D shallow water, $\beta$--plane equations, can be expressed as the continual excitation of waves with a characteristic frequency and decay timescale (Eq.\eqref{eq:green_integral} Section \ref{ssec:time_solution}). The decay timescales themselves are crucial as they can be used to determine the qualitative form of the linear steady state itself, and provide insight into the response of the atmosphere over short timescales. In this section we focus on the mathematical derivation of the characteristic decay timescales reaching an expression (Section \ref{sssec:decay}), which we then solve numerically for different types of atmospheric waves (gravity, Rossby and Kelvin waves). A short summary is then provided in Section \ref{sssec:waves_summary}. Following this, we extend our arguments to 3D in Section \ref{ssec:eclips3d_confirm}. The entirety of this section is focused on the mathematical nature of the supported waves, with a physical interpretation provided later in Section \ref{sec:transition}.

\subsection{Characteristics of waves in 2D}
\label{ssec:2d}

\subsubsection{Decay time of damped waves}
\label{sssec:decay}

The first task is to derive an equation for the decay timescale, or growth rate, for a general damped wave in our framework. If $\tau_\mathrm{drag} = \tau_\mathrm{rad}$, using the complex frequency $\omega \in \mathbb{C}$ (contrary to \citet{Matsuno} where $\omega \in \mathbb{R}$), we can define $\mathrm{i}\lambda = \mathrm{i} \omega+ {1}/{\tau_\mathrm{drag}} $  to transform Eqs.\eqref{eq:dissip_u} to \eqref{eq:dissip_uh} into Eqs.\eqref{eq:matsuno_u} to 
\eqref{eq:matsuno_h}, equations of which we know the eigenvalues from \citet{Matsuno}. Therefore, $\lambda$ is real which implies that the real part of $\omega$, the frequency, is unaltered from the original equations of \citet{Matsuno}, but the imaginary part of $\omega$ becomes,
\begin{equation}
    \Im(\omega) = \dfrac{1}{\tau_\mathrm{drag}}\,\, .
\end{equation}
One can then express e.g., $u_{n,l}$ as: 
\begin{equation}
    u_{n,l} = \tilde{u}_{n,l} \mathrm{e}^{-t/\tau_\mathrm{drag}} \mathrm{e}^{\mathrm{i}(\nu_{n,l}t+kx)}.
\end{equation}
This shows that all modes decay over a characteristic timescale, namely, the drag (or radiative) timescale. For the case where $\tau_\mathrm{rad} = \tau_\mathrm{drag}$
the time to converge to the Matsuno-Gill circulation is the decay timescale, as one would naively expect, and all waves have the same exponential decay in time. 

If $\tau_\mathrm{drag} \ne \tau_\mathrm{rad}$, Eq.(\ref{v_matsuno}) must be modified to obtain: \begin{align}
    \dfrac{\partial^2 v}{\partial y^2} - \bigg((\mathrm{i} \omega+ \tau_\mathrm{drag}^{-1})(\mathrm{i} \omega+ \tau_\mathrm{rad}^{-1})+k^2 \nonumber \\
    +y^2\dfrac{\mathrm{i} \omega+ \tau_\mathrm{rad}^{-1}}{\mathrm{i} \omega+ \tau_\mathrm{drag}^{-1}}
    - \dfrac{ik}{\mathrm{i} \omega+ \tau_\mathrm{drag}^{-1}} \bigg)v = 0 \,\, .
    \label{v_tot}
\end{align}
It is easy to verify that setting $\tau_\mathrm{drag}^{-1} = \tau_\mathrm{rad}^{-1}=0$ gives back Eq.\eqref{v_matsuno}.


To go one step further, we define a complex number $c$ such that:
\begin{gather}
    c^4 = \dfrac{\mathrm{i} \omega+ \tau_\mathrm{rad}^{-1}}{\mathrm{i} \omega+ \tau_\mathrm{drag}^{-1}}, \label{eq:defc4} 
\end{gather}
and choose for $c$ the only root with positive real and imaginary part. 
This definition ensures that the real part of $c^2$ is always positive, which allows for the solutions to decay at infinity, see Appendix \ref{app:orthogonality}. We then choose as a variable $z = cy$ (the
cases $c = 0$ and $c = \infty$ are of no physical interest). Using this new variable, ${\partial^2 v}/{\partial y^2} = c^2 {\partial^2 v}/{\partial z^2}$, Eq.\eqref{v_tot} simplifies to
\begin{align}
    c^2\dfrac{\partial^2 v}{\partial z^2} - \bigg((\mathrm{i} \omega+ \tau_\mathrm{drag}^{-1})(\mathrm{i} \omega+ \tau_\mathrm{rad}^{-1})+k^2
     +y^2 c^4- \dfrac{ik}{\mathrm{i} \omega+ \tau_\mathrm{drag}^{-1}} \bigg)v = 0 \,\, .
    \label{v_bis}
\end{align}
Dividing this equation by $c^2$ (and recalling the definition of $z$, $z = cy$), we obtain 
\begin{align}
    \dfrac{\partial^2 v}{\partial z^2} -  \dfrac{(\mathrm{i} \omega+ \tau_\mathrm{drag}^{-1})(\mathrm{i} \omega+ \tau_\mathrm{rad}^{-1})+k^2 
    - ik/(\mathrm{i} \omega+ \tau_\mathrm{drag}^{-1})}{c^2} v 
    - z^2 v = 0.
\end{align}
Defining the multiplier of $v$ in the second term as $m$, the equation can be expressed as,
\begin{align}
    \dfrac{\partial^2 v}{\partial z^2} +  (m-z^2) v = 0 \,\, .
    \label{v_fin}
\end{align}
The major difference with the Matsuno case, Eq.\eqref{v_matsuno} is that now $z \in \mathbb{C}$, and so the boundary conditions are altered. As $|z| \rightarrow \infty$ when $y \rightarrow \pm \infty$, we have to solve this equation with the following boundary condition: 
\begin{align}
v \rightarrow 0 \ \ \ \text{when} \ \ \ |z| \rightarrow \infty \,\, .
\end{align}
As in the case where $m$ is real, one could show \citep[see, for example,][]{abramowitz1965} that the only solutions are the parabolic cylinder functions, Eq.\eqref{eq:parabolic}: $H_n(cy) \mathrm{e}^{-c^2y^2/2}$ where $n \in \mathbb{N}$, hence the need for $\Re(c^2) > 0$ so that the solutions decay when $y \rightarrow \pm \infty$, and provided that:
\begin{equation}
    \dfrac{-(\mathrm{i} \omega+ \tau_\mathrm{drag}^{-1})(\mathrm{i} \omega+ \tau_\mathrm{rad}^{-1})-k^2 
    + ik/(\mathrm{i} \omega+ \tau_\mathrm{drag}^{-1})}{c^2} = 2n+1,
    \label{eq:odd}
\end{equation}
 Defining $x = \mathrm{i} \omega + \tau_\mathrm{drag}^{-1}$ and $\gamma = \tau_\mathrm{rad}^{-1}-\tau_\mathrm{drag}^{-1}$, followed by taking the square of Eq.\eqref{eq:odd} in order to obtain $c^4$ yields, 
\begin{flalign}
    x^4(x+\gamma)^2 + 2k^2 x^3 (x+\gamma) -2ikx^2(x+\gamma)
    \nonumber \\- (2n+1)^2 x(x+\gamma)  
    + k^4x^2 -2ik^3 x - k^2 = 0 \,\,. 
    \label{eq:polynomial_6}
\end{flalign}
Eq.\eqref{eq:polynomial_6} has already been obtained by \citet{Heng2014} (their Eq.(121) with different notations: their $F_0$ is $ \tau_\mathrm{rad}^{-1}$ in our work, $\omega_R $ is our $\omega$ and $\omega_I$ is $- \tau_\mathrm{drag}^{-1}$), in order to derive steady state solutions as performed in \citet{Wu2001} and \citet{Showman2011}.
Eq.\eqref{eq:polynomial_6} is a polynomial of order six, but a thorough study of Eq.\eqref{eq:odd} reported in Appendix \ref{app:pol} shows that only three different waves propagate, and two for n=0, as in \citet{Matsuno} . 

The horizontal shape of the solutions to Eqs.\eqref{eq:dissip_u} to \eqref{eq:dissip_uh} in the general case are given by Eq.\eqref{eq:shape_degueu}, but we would also require the solutions of Eq.\eqref{eq:polynomial_6} to obtain a fully analytic expression for the waves. Therefore, we have solved Eqs.\eqref{eq:dissip_u} to \eqref{eq:dissip_uh} numerically over an extensive range of $n$, $k$, $\tau_\mathrm{drag}$ and $\tau_\mathrm{rad}$ values (we have verified that our numerical solutions recover the limits $\tau_\mathrm{drag}^{-1} = \tau_\mathrm{rad}^{-1} = 0$ and $\tau_\mathrm{drag}^{-1} = \tau_\mathrm{rad}^{-1}$).  First, as expected no mode
can exponentially grow given a background state at rest. The cases of $k \sim 1$ and $n = 1, 3, 5$ and $7$ are the most important for hot Jupiters, as the heating function is dominated by wavenumber 1, i.e. a day and night side (non--dimensional value around 0.7, see Section \ref{sec:transition}). Here the $n$ number represents the order of the Hermite polynomial, hence the number of zero nodes in latitude (note that as $c^2$ can be complex, the number of zeros in latitude is no more solely defined by the Hermite polynomials as in the $c=1$ case). If the heating function is a Gaussian function, as chosen by \citet{Matsuno,Showman2011,Tsai2014}, then the projection of $\mathcal{Q}$ on to the parabolic cylinder function stops at the third order (this is not necessarily true when $\tau_\mathrm{drag} \ne \tau_\mathrm{rad}$ but we don't expect large amplitude in the $n>3$ waves, as the forcing exhibits no zeros in latitude).

Using our numerical solutions we explore the behaviour of gravity (Section \ref{sssec:gravity}), Rossby (Section \ref{sssec:rossby}) and Kelvin waves (Section \ref{sssec:kelvin}) before summarising our results (Section \ref{sssec:waves_summary}). For all cases the frequency of the waves remains within an order of magnitude of the free wave frequency (when $\tau_\mathrm{drag}^{-1} = \tau_\mathrm{rad}^{-1} =0$). Hence the major changes, between cases, are obtained for the decay rate and horizontal shapes. The shapes of the waves with non zero $\tau_\mathrm{drag}$ and $\tau_\mathrm{rad}$ are shown in Appendix \ref{app:waves}.Regarding the decay rates, we express them as power laws fitting reasonably well the numerical values in the next section. We have also implemented the 2D--shallow water equations in ECLIPS3D \citep[detailed in][]{Debras2019} to verify our numerical results discussed in this section. For all cases for matching parameters (characteristic values, $\tau_\mathrm{rad}$, $\tau_\mathrm{drag}$), the agreement between ECLIPS3D and the results discussed here (obtained using the Mathematica software) for both the decay timescales or growth rates and frequency of waves is excellent. Furthermore, inserting the numerical values obtained here in Eq.\eqref{eq:shape_degueu} recovers the exact modes as obtained using the shallow water version of ECLIPS3D.

\subsubsection{Gravity waves}
\label{sssec:gravity}

In this work we define "gravity waves" as the solutions to Eqs.\eqref{eq:dissip_u} to \eqref{eq:dissip_uh} which tend to the standard definition of a gravity wave in the limit $\tau_\mathrm{drag} \rightarrow \tau_\mathrm{rad}$  (we have verified that the identification is unchanged for $\tau_\mathrm{rad} \rightarrow \tau_\mathrm{drag}$). As there are only three solutions to these equations, see e.g., \citet{Matsuno}, the Rossby wave is therefore the last mode.

Our numerical results gave a characteristic time of decay for gravity waves of $\sim \tau_\mathrm{drag}$, when $\tau_\mathrm{drag} \sim \tau_\mathrm{rad}$, as expected. For cases where the drag is dominant over the radiative forcing, $\tau_\mathrm{drag} \ll \tau_\mathrm{rad}$, the decay timescale obtained numerically is $\sim \tau_\mathrm{drag}$ i.e., the drag controls the timescale of the convergence of the atmosphere.
Physically, this is expected as drag will prevent the wave from propagating and damp the perturbed velocity efficiently, preventing the temperature and pressure to depart significantly from the forced equilibrium profile.

However, when the radiative forcing is dominant over the drag, $\tau_\mathrm{drag} \gg \tau_\mathrm{rad}$, we find two cases. Firstly, when $\tau_\mathrm{rad} \ll 1$ the numerically obtained decay {\it rate} for the gravity waves ($\sigma_\mathrm{g1}$) is given by

\begin{equation}
\sigma_\mathrm{g1} \approx \tau_\mathrm{drag}^{-1}+ \tau_\mathrm{rad}^{1/(n+2)} \,\,. 
\end{equation}

For this case, if $\tau_\mathrm{rad} \rightarrow 0$, $\sigma_\mathrm{g1}^{-1}$ (the decay timescale) is given by the drag timescale, as for the case of dominant drag. However, if $\tau_\mathrm{rad}$ is larger the behaviour is more complex and includes a dependence on the order of the Hermite polynomial $n$, although this still results in the decay timescale being the same order of magnitude as the drag timescale.
We interpret it as the fact that, although the temperature and pressure perturbation will be dictated by the forcing, the time to damp the wave is still governed by the time for the velocities to be damped, hence the drag timescale.

Secondly, for the case where $\tau_\mathrm{rad} \gg 1$  the numerically obtained limit for the decay rate ($\sigma_\mathrm{g2}$) is given by
\begin{equation}
\sigma_\mathrm{g2} \approx \dfrac{\tau_\mathrm{rad}^{-1}}{3} \,\,. 
\end{equation}
In this case, the radiative timescale is long enough to be imposed as the characteristic time of decay even for the velocities, and the decay of the wave is only controlled by this timescale.
\subsubsection{Rossby waves}
\label{sssec:rossby}

The behaviour of the Rossby wave decay timescale is more complex than that of gravity waves. When $|\tau_\mathrm{drag}^{-1} - \tau_\mathrm{rad}^{-1}| \gtrsim 0.5$, for all individual values of $\tau_\mathrm{drag}^{-1}$ or $\tau_\mathrm{rad}^{-1}$, the absolute value of the imaginary part of $c^2$ is much larger than that of the real component. This means that the Rossby wave modes oscillate in the $y$ direction several times before being damped, in these conditions. Additionally, for increasing values of $|\tau_\mathrm{drag}^{-1} - \tau_\mathrm{rad}^{-1}|$, the amplitude at the equator becomes negligible, and the mode's peak amplitude moves to higher latitudes, where the equatorial $\beta$--plane approximation begins to break down. Therefore, our numerical results show that the simple shallow--water, equatorial $\beta$--plane framework adopted in this work is not usefully applicable to the case of Rossby waves where $|\tau_\mathrm{drag}^{-1} - \tau_\mathrm{rad}^{-1}| \gtrsim 0.5$. This is confirmed by the graphical representation of these waves in Appendix \ref{app:waves}. We will therefore rely on numerical results of section \ref{ssec:eclips3d_confirm} for this region of the parameter space. However, for $|\tau_\mathrm{drag}^{-1} - \tau_\mathrm{rad}^{-1}| \lesssim 0.5$ (which is the case for all $\tau_\mathrm{drag}^{-1}, \tau_\mathrm{rad}^{-1} < 0.5$), 
the decay rate for Rossby waves ($\sigma_\mathrm{R}$) we have derived numerically can then be approximated by,
\begin{equation}
\sigma_\mathrm{R} \approx \dfrac{1}{2} \left(\tau_\mathrm{rad}^{-1} +
\tau_\mathrm{drag}^{-1} \right) \, \,.
\label{eq:rossby_sigma}
\end{equation}

Therefore, for long radiative and drag timescales, Rossby waves are equally sensitive to the damping of velocities and temperature. Such a result is expected from the conservation of potential vorticity, which gives rise to Rossby waves, and is defined in the shallow water system as $(\xi+f)/h$ where $\xi = \vec{\nabla} \wedge \vec{u}$ is the vorticity and $f = 2 \Omega \cos \phi$ the Coriolis parameter. When neither $\xi$ and $h$ are strongly damped, we might therefore expect a combination of the drag and radiative damping to return to equilibrium. Comparing the decay timescale for Rossby waves with that obtained for gravity waves in Section \ref{sssec:gravity} shows that the decay rates differ between these two cases.

\subsubsection{Kelvin waves}
\label{sssec:kelvin}

Kelvin waves are a particular solution of the homogeneous Eqs.\eqref{eq:dissip_u} to \eqref{eq:dissip_uh}, as first pointed out by \citet{Matsuno}, where the meridional velocity is zero, and can be characterised analytically. Combining Eqs.\eqref{eq:dissip_u} and \eqref{eq:dissip_uh} with $v=0$ yields,
\begin{gather}
\dfrac{\partial u}{\partial y} = \dfrac{\mathrm{i}k}{\mathrm{i} \omega + 1/\tau_\mathrm{drag}} y u,\\
\, \, \text{ and hence} \nonumber \\
u = A \mathrm{exp}\left(\dfrac{\mathrm{i}k}{\mathrm{i} \omega + 1/\tau_\mathrm{drag}} \dfrac{y^2}{2}\right),
\end{gather}
where $A$ is a constant. If the boundary condition $u = 0$ for $y \rightarrow \pm \infty$ is to be satisfied, the factor of $y^2/2$ must have a negative real component. Additionally, in order for $u$ and $h$ not to be identically zero, $\omega$ must be a root of a second order polynomial given by
\begin{equation}
\omega^2 - \mathrm{i} \omega \left(\dfrac{1}{\tau_\mathrm{rad}} +\dfrac{1}{\tau_\mathrm{drag}} \right)
- k^2 - \dfrac{1}{\tau_\mathrm{rad} \tau_\mathrm{drag}} = 0\ ,
\end{equation}
that is,
\begin{equation}
\omega = \dfrac{1}{2} \left(\mathrm{i}\left(\dfrac{1}{\tau_\mathrm{rad}} +\dfrac{1}{\tau_\mathrm{drag}} \right)
 \pm \sqrt{4k^2 - \left(\dfrac{1}{\tau_\mathrm{rad}} -\dfrac{1}{\tau_\mathrm{drag}} \right)^2} \right),
 \end{equation}
 where the term under the square root can be negative, and therefore provide an imaginary component. Further algebraic manipulation then yields,
 \begin{equation}
\dfrac{\mathrm{i}k}{\mathrm{i} \omega + \dfrac{1}{\tau_\mathrm{drag}}} = 
\dfrac{2 \mathrm{i} k} {\left(\dfrac{1}{\tau_\mathrm{drag}} -\dfrac{1}{\tau_\mathrm{rad}} \right)
\pm \mathrm{i}\sqrt{4k^2 - \left(\dfrac{1}{\tau_\mathrm{rad}} -\dfrac{1}{\tau_\mathrm{drag}} \right)^2} } \,\, .
\end{equation}
In order to satisfy the boundary conditions the term under the square root in this equation must be positive, or the result is a pure imaginary number. In other words, Kelvin waves are able to propagate only when the condition,
\begin{equation}
4k^2 > \left(\dfrac{1}{\tau_\mathrm{rad}} -\dfrac{1}{\tau_\mathrm{drag}} \right)^2\,\,,
\label{eq:Kelvin_k}
\end{equation}
is met. 
Additionally, this simple estimation of the regimes where Kelvin waves can be supported in the atmosphere may well be an over estimate for the 3D, spherical coordinate case as the characteristic scale of the damping of the Kelvin wave must be smaller than the scale of the planet's atmosphere itself. The real part of ${\mathrm{i}k}/({\mathrm{i} \omega + {\tau_\mathrm{drag}}^{-1}})$ must therefore be large enough (and negative). Finally,  when Kelvin waves propagate their characteristic decay rate ($\sigma_{K}$) is given by,
\begin{equation}
\sigma_{K} = \dfrac{1}{2}\left(\dfrac{1}{\tau_\mathrm{rad}} +\dfrac{1}{\tau_\mathrm{drag}} \right).
\label{eq:kelvin_sigma}
\end{equation}
This result is similar to the decay timescale for Rossby waves (compare Eqs.\eqref{eq:rossby_sigma} and \eqref{eq:kelvin_sigma}). $\tau_\mathrm{drag}$ and $\tau_\mathrm{rad}$ therefore have a symmetric contribution for Kelvin waves, as expected when considering Eqs.\eqref{eq:dissip_u}  and \eqref{eq:dissip_uh} for $v=0$: they have a symmetric effect on $u$ and $h$.


\subsubsection{Decay Timescale Summary}
\label{sssec:waves_summary}

We have now obtained expressions for the asymptotic values of the decay timescales for damped waves under the 2D shallow--water, $\beta$--plane framework (see Section \ref{ssec:framework}). In particular, for the case of Kelvin waves we obtained an analytical expression for the decay rate, Eq.\eqref{eq:kelvin_sigma}. We have also shown that for the regime where the analytical calculations are valid, Rossby waves exhibit the same decay rate as found for Kelvin waves. For the more general case, aside from considerations of whether the waves can be supported by the atmosphere we have two limits:

\begin{enumerate}
    \item For $\tau_\mathrm{drag} \sim \tau_\mathrm{rad}$ and $\tau_\mathrm{drag} \ll \tau_\mathrm{rad}$, simply, $\sigma_\mathrm{R} \sim \sigma_\mathrm{K} \sim \sigma_\mathrm{g} \sim \tau_\mathrm{drag}^{-1} $ within a factor of $\sim2$. 
    
    \item For $\tau_\mathrm{drag} \gg \tau_\mathrm{rad}$, the Kelvin decay rate is the invert of the radiative timescale. The Rossby decay rate is the invert of the drag timescale if $|\tau_\mathrm{drag}^{-1} - \tau_\mathrm{rad}^{-1}| \lesssim 0.5$ and we obtained no semi-analytical solution if $|\tau_\mathrm{drag}^{-1} - \tau_\mathrm{rad}^{-1}| \gtrsim 0.5$. Finally, the gravity waves decay rate becomes:
    \begin{itemize}
        \item For $\tau_\mathrm{rad} \ll 1$: $\sigma_g \sim  \tau_\mathrm{drag}^{-1} + \tau_\mathrm{rad}^{1/(n+2)}$ 
        \item For $\tau_\mathrm{rad} \gg 1$: $\sigma_g \sim \tau_\mathrm{rad}^{-1}/3$ 
    \end{itemize}
\end{enumerate}

Some of the numerical values we used to derive these asymptotic evaluations are reported in plain lines on Figure \ref{fig:decay}. As our results have been derived for the 2D shallow--water, $\beta$--plane system, and for the case of Rossby waves in particular, the behaviour of the waves may not be captured correctly. Therefore, we next move to verifying and extending our approach into 3D using ECLIPS3D.

\subsection{Extension to 3D with ECLIPS3D}
\label{ssec:eclips3d_confirm}

So far we have determined the characteristic frequencies and decay timescales for various atmospheric waves in our 2D framework, introduced in Section \ref{ssec:framework}. In this section we extend our analysis to full 3D spherical coordinates using ECLIPS3D. In 3D, spherical coordinates, the dependency of the waves on the stratification and the value of the drag and radiative timescales is difficult to predict theoretically. Therefore, we approach the problem numerically using ECLIPS3D, studying the modes which propagate in a stratified atmosphere at rest, as is used for the initial condition when simulating hot Jupiter atmospheres. The background temperature--pressure profile is set to that of \citet{iro2005}, employing the polynomial fit of \citet{Heng2011}. The pressure at the bottom of the atmosphere is set to 10\,bars (the depth of the atmosphere is now $7 \times 10^6$\ m), capturing the dynamically active region of the atmosphere, driven by the forcing in the first phase of simulation but without detailing the innermost regions where the density is much higher. As before, we have varied the inner boundary condition to test its impact on our results, and find our conclusions to be robust to this choice. The selection of the modes of interest is performed by first excluding modes with unrealistic amplitudes at the pole or boundary, and then selecting modes with only one oscillation in longitude i.e. wavenumber 1, matching the heating function. Additionally, we restrict to modes with at most two nodes in the latitude direction which are the dominant modes (see discussion previously in this section). This selection process is also described in \citet{Debras2019}.

For the first study we have verified that all modes supported when $\tau_\mathrm{drag} = \tau_\mathrm{rad} = 10^6$\,s is adopted throughout the atmosphere are similar in form to when $\tau_\mathrm{drag} = \tau_\mathrm{rad} = 0$\,s but exhibit a characteristic decay frequency of $10^{-6} \,\mathrm{s}^{-1}$. Although Rossby and gravity waves are supported with characteristic heights of order the height of the atmosphere itself, Kelvin waves seem only to be supported at the pressure scale height, or smaller. 

We have subsequently applied ECLIPS3D with $\tau_\mathrm{rad}$ prescribed as a function of pressure following \citet{iro2005}, and $\tau_\mathrm{drag}$ set as a constant between $10^5$\,s and $10^6$\,s. For this setup, we recover the usual \citep[see e.g.,][]{Thuburn} Rossby and gravity modes over different vertical wavelengths. Therefore, although our 2D analysis breaks for Rossby modes with large $\tau_\mathrm{drag}$ (see Section \ref{sssec:rossby}), we recover them in the full 3D spherical coordinate treatment using ECLIPS3D, {\it with their decay rate always comparable to the reciprocal of the drag timescale}. However, in this setup, with a pressure dependent radiative timescale, we do not find Kelvin modes with atmospheric-scale characteristic heights (we use 40 points in the $z$ coordinates, therefore we are unable to resolve modes with $H \lesssim  10^5$\,m). However, we obtain Kelvin modes with smaller characteristic heights in the deepest, highest pressure, region of the atmosphere where the radiative timescale is long, in agreement with our previous estimations (see Section \ref{sssec:kelvin}, Eq.\eqref{eq:Kelvin_k}). For this setup we also obtain mixed Kelvin--gravity modes, concentrated at the equator, as well as mixed Rossby--gravity modes, with the Rossby component dominating in the high latitudes and gravity component component near the equator. In the case of the pure gravity modes the resulting frequencies and decay rates, from ECLIPS3D in 3D spherical coordinates, are in good agreement with the 2D estimations (Section \ref{sssec:gravity}). However, for Kelvin modes although the decay rates obtained from ECLIPS3D are in good agreement with our 2D analytical expressions (Section \ref{sssec:kelvin}), the obtained frequencies are a little larger than our analytical analysis would suggest. Finally, for pure Rossby modes, for the range where our 2D analysis is valid, the decay timescales are again in good agreement between our 2D estimations (Section \ref{sssec:rossby}) and the numerical results in 3D, but similarly to the Kelvin modes the frequencies are slightly underestimated. 

From our ECLIPS3D outputs we have calculated the value of $\sqrt{\sigma^2+\nu^2}$ for all modes supported in the simulated atmosphere (keeping $\tau_\mathrm{rad}$ prescribed following \citet{iro2005} and $\tau_\mathrm{drag}=10^6$\,s), as the amplitude of a given mode in the linear steady state of the atmosphere is inversely proportional to $\sqrt{\sigma^2+\nu^2}$ (Eq.\eqref{eq:linear_evolution}). These results show that the value of $\sqrt{\sigma^2+\nu^2}$ is an order of magnitude smaller for Rossby modes, compared with Kelvin or gravity modes. As discussed in Section \ref{sssec:rossby} the frequencies of the modes are not significantly altered by the drag timescale, and \citet{Matsuno} has shown that Rossby waves have frequencies $\sim 10$ times smaller than gravity waves and $\sim 3$ times smaller than Kelvin waves in this regime. Additionally, when $\tau_\mathrm{drag}$ is large but $\tau_\mathrm{rad}$ is modest, the decay rate of Kelvin waves is controlled by the radiative timescale, whereas the decay rate of the gravity waves and Rossby waves is conversely controlled by the drag timescale. Therefore, the value of $\sqrt{\sigma^2+\nu^2}$ will be bigger for Kelvin waves, over that obtained for Rossby and gravity waves, for both of which this quantity will be of order the frequency, which is ten times smaller for Rossby waves compared to gravity waves. This demonstrates that the Rossby waves will propagate over greater timescales and lengths, and dissipate globally more energy in the linear steady state (see Eq.\eqref{eq:green_integral}). The influence of the Rossby waves in the steady linear circulation regime will be dominant over Kelvin and gravity modes, explaining the qualitative shape of Figure \ref{fig:MG_komacek}.  

\begin{figure}  
    \begin{center}
    \includegraphics[width=8.5cm,angle=0.0,origin=c]{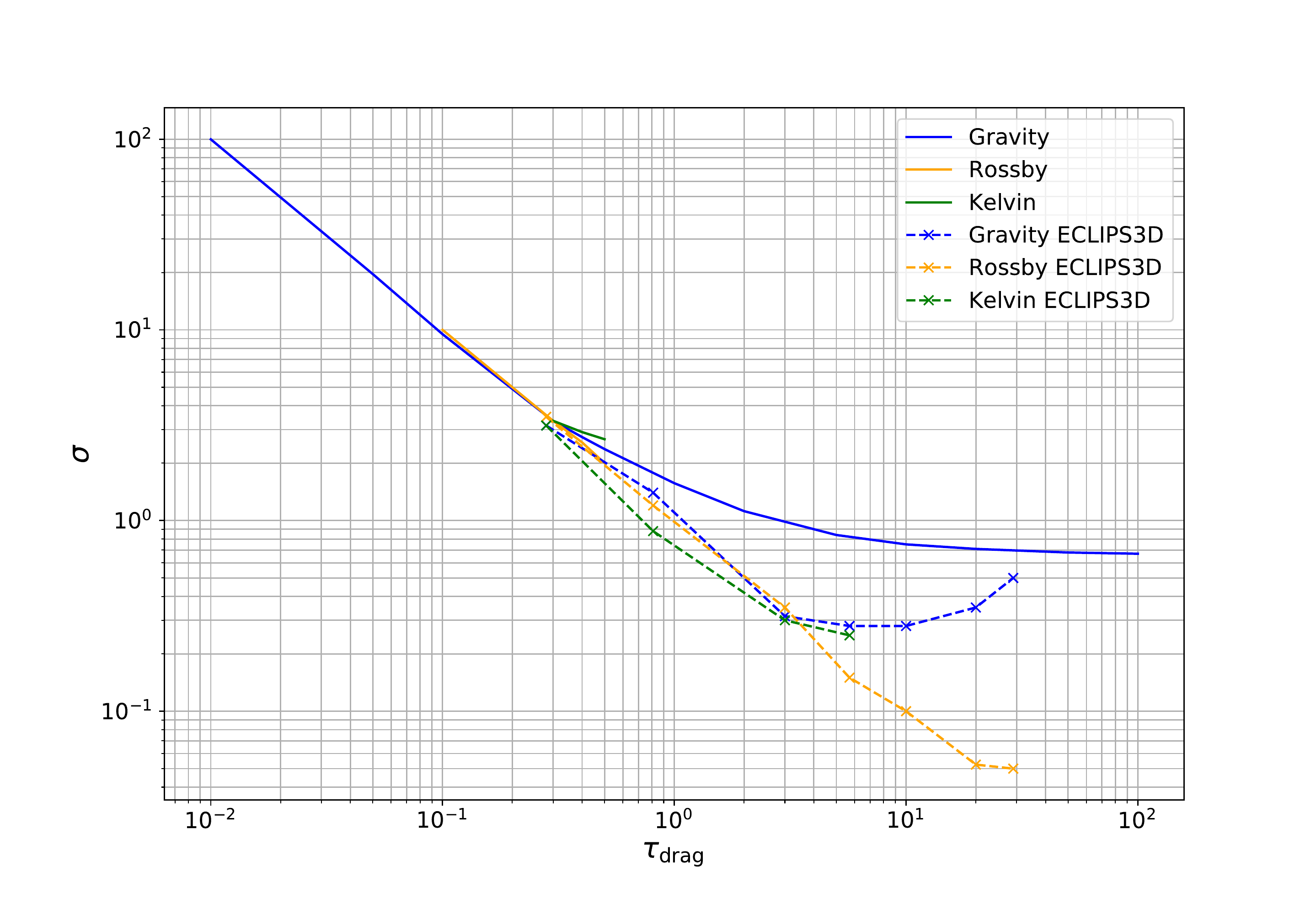} \\
    \includegraphics[width=8.5cm,angle=0.0,origin=c]{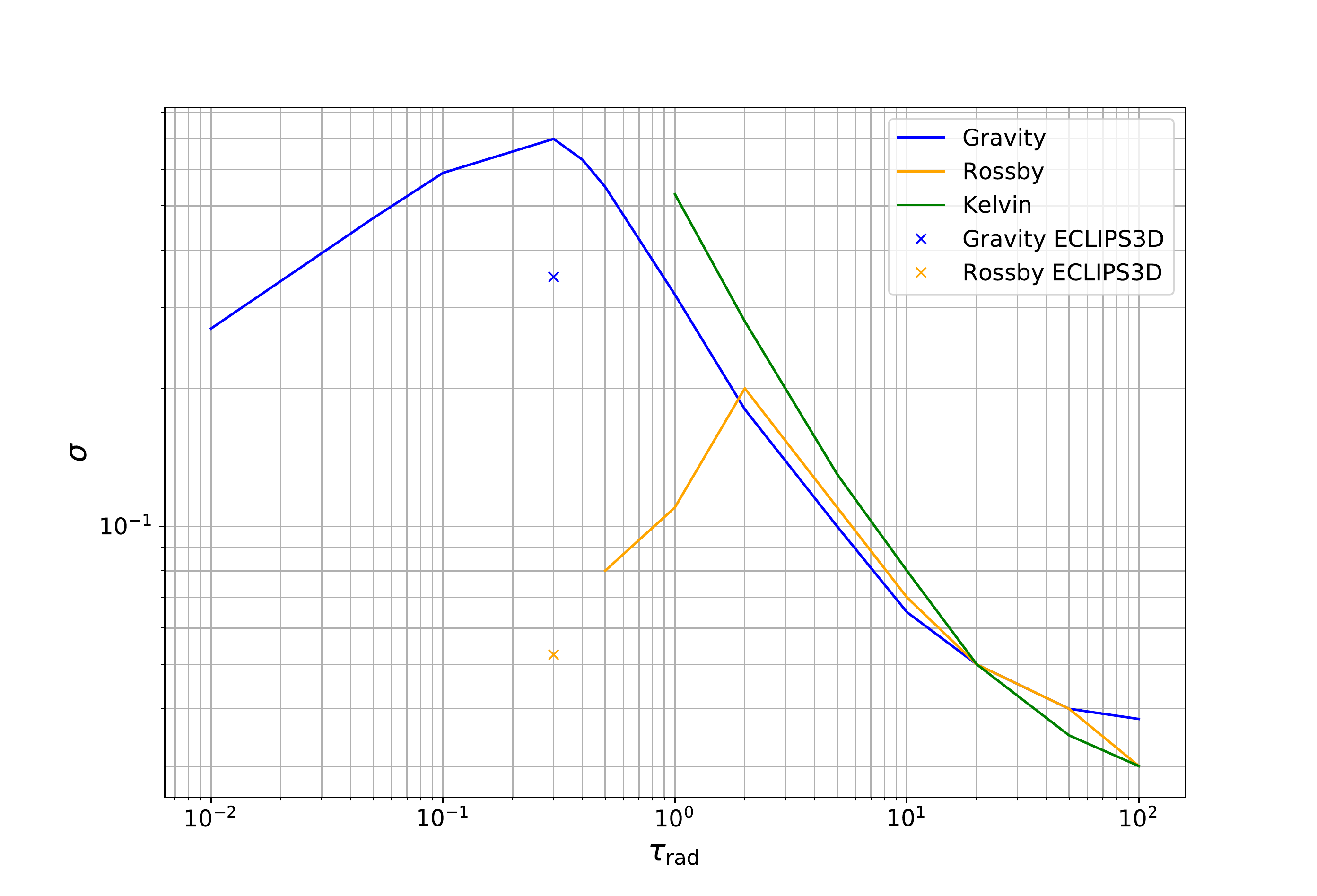}
    \caption{Typical decay rate $\sigma$ for gravity, Rossby and Kelvin waves as a function of (top) $\tau_\mathrm{drag}$ for $\tau_\mathrm{rad} = 0.3 $ and (bottom) $\tau_\mathrm{rad}$ for $\tau_\mathrm{drag} = 20$. The lines are values obtained using Eqs.\eqref{eq:polynomial_6}, \eqref{eq:kelvin_sigma} while crosses are results from seven ECLIPS3D calculations. As $\tau_\mathrm{rad}$ is not constant with depth in the ECLIPS3D results, we have chosen to use an arbitrary value of $\tau_\mathrm{rad} = 0.3$ for comparison. However when $\tau_\mathrm{drag}$ increases the location where the wave exhibits its maximum perturbation moves to higher pressures which should correspond to an increase in the equivalent $\tau_\mathrm{rad}$.
    For the Rossby waves, the low $\tau_\mathrm{drag}$ limit has not been studied numerically as it is irrelevant for superrotation. Kelvin waves of comparable height with Rossby and gravity waves are only clearly identified in four ECLIPS3D calculations. }
    \label{fig:decay}
    \end{center}
\end{figure}

As a summary, Figure \ref{fig:decay}, shows the decay timescales for gravity, Rossby and Kelvin waves, obtained from Eqs.\eqref{eq:polynomial_6}, \eqref{eq:kelvin_sigma} as a function of the drag timescale for $\tau_\mathrm{rad} = 0.3$ (in dimensional units, a few $10^4$\ s), a characteristic value in the superrotating regions of HD~209458b, or as a function of the radiative timescale for $\tau_\mathrm{drag} = 20$ (about $10^6$\ s), a value which allows for superrotation in the non linear limit. We also plot values extracted from numerical exploration with ECLIPS3D, although the radiative timescale is not constant in these numerical results and the characteristic height might differ.
We recover the fact that Kelvin waves are more damped than other waves for the timescales used in this section, thought to be representative of hot Jupiter atmospheres, but not for all timescales (notably when $\tau_\mathrm{drag} =\tau_\mathrm{rad}$). Additionally,  there are many regions of the parameter space where Kelvin waves don't actually propagate, as evidenced in section \ref{sssec:kelvin}. This highlights the need to constrain the timescales to understand the spin up of superrotation, and to know the wave behaviour across different timescales.

\begin{figure*}  
    \begin{center}
    \subfigure[]{\includegraphics[width=8.5cm,angle=0.0]{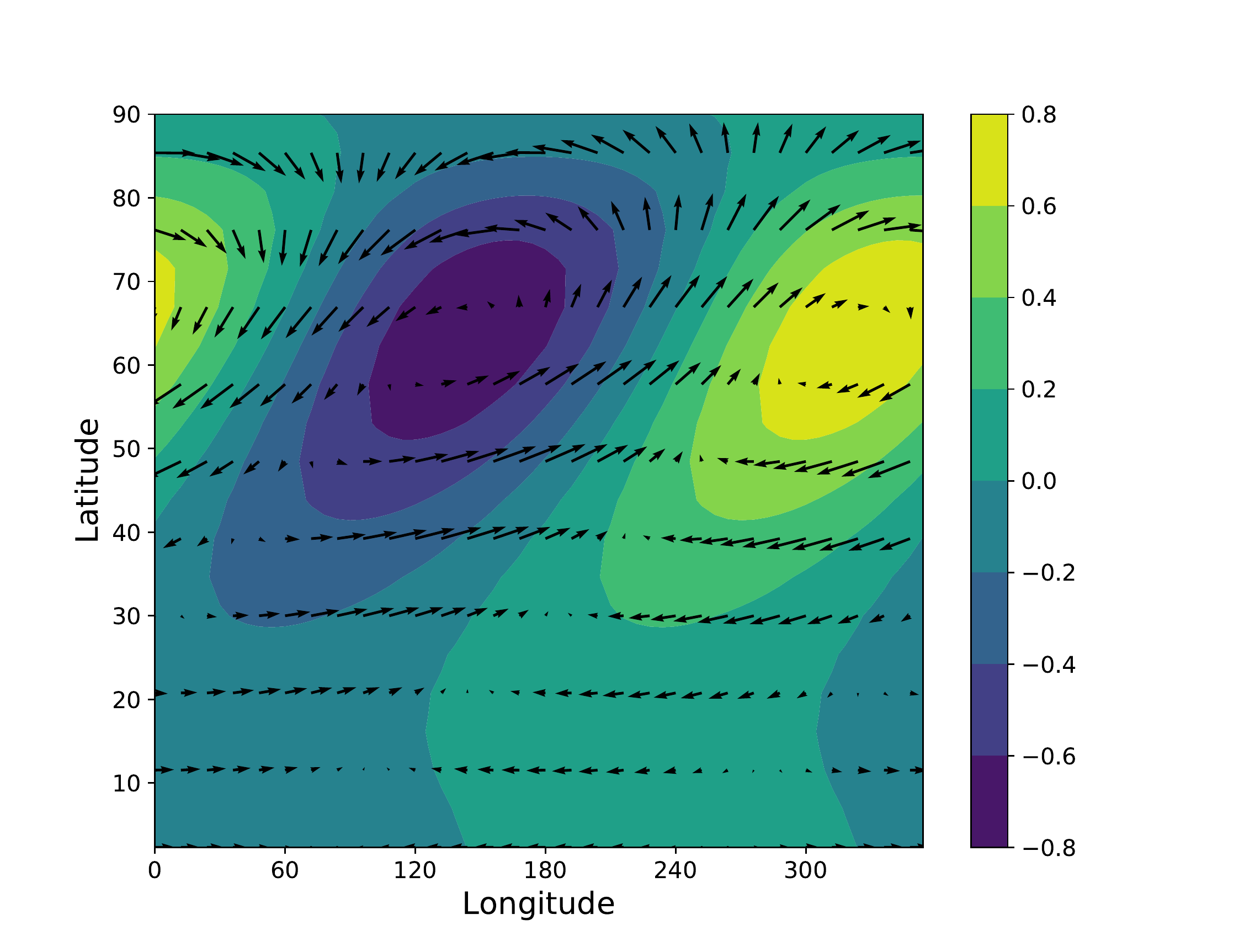}\label{fig:Rossby_1}}
    \subfigure[]{\includegraphics[width=8.5cm,angle=0.0]{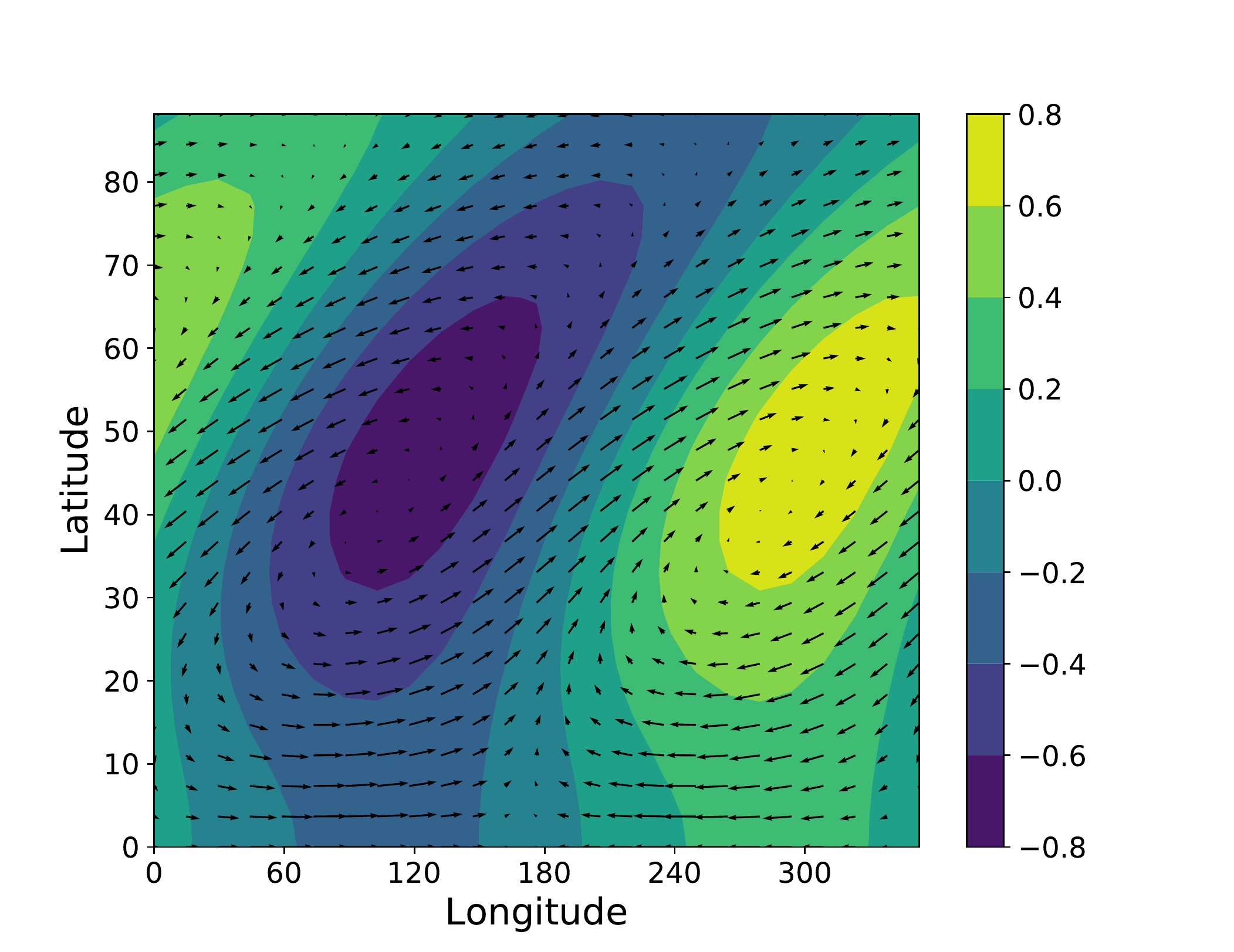}\label{fig:Rossby_2}}
    \subfigure[]{\includegraphics[width=8.5cm,angle=0.0]{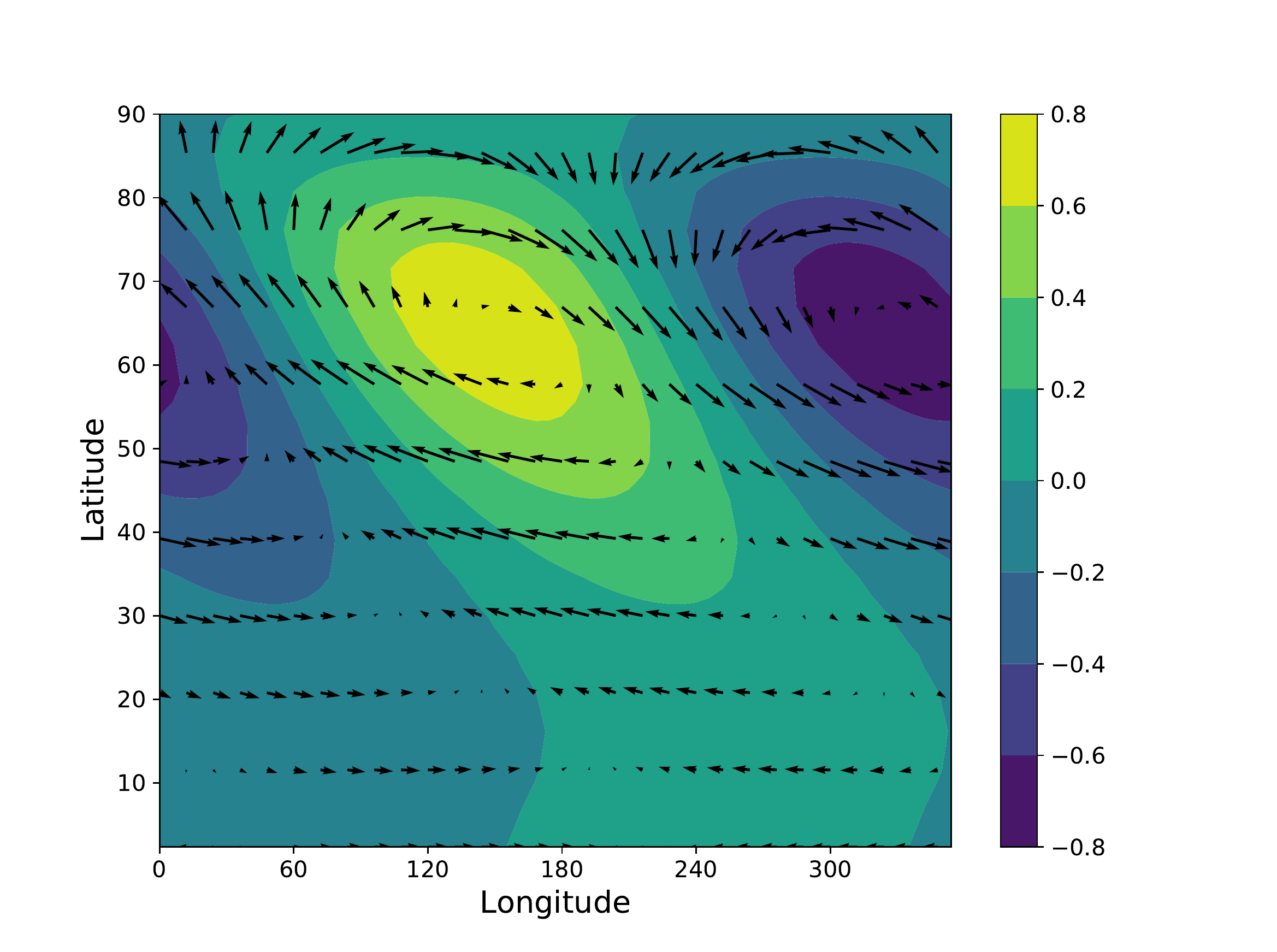}\label{fig:Rossby_3}}
    \subfigure[]{\includegraphics[width=8.5cm,angle=0.0]{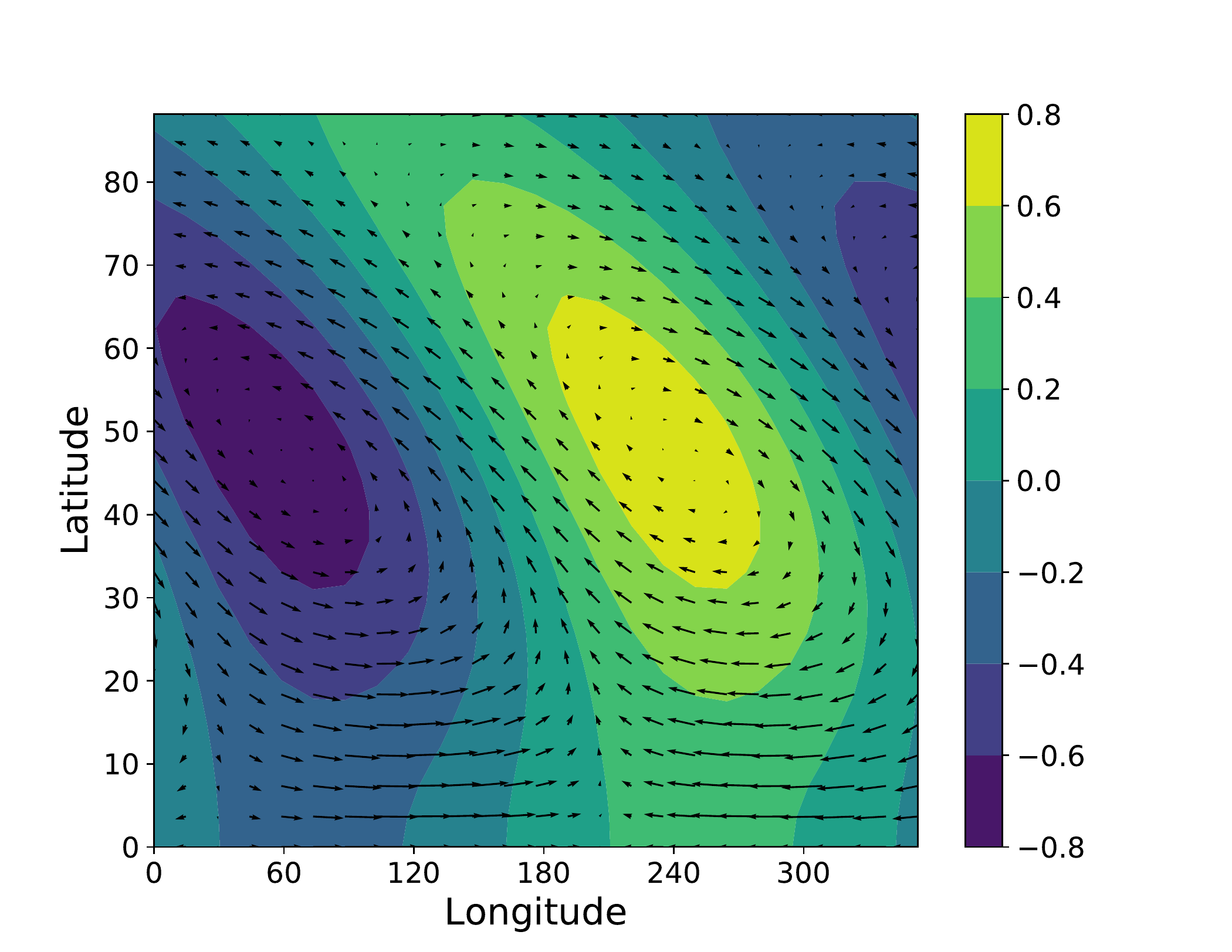}\label{fig:Rossby_4}}
    \caption{Pressure perturbations (colorscale) and winds (arrows) as a function of longitude and latitude for four Rossby waves for an axisymmetric hot Jupiter atmosphere at rest. Units are arbitrary. The values of the drag and radiative timescales are described in Section \ref{ssec:eclips3d_confirm}. The left column shows the results from ECLIPS3D 3D spherical coordinate calculations and the right column for the analytical results to the equation developed in appendix \ref{app:orthogonality}. For the top row, the drag is dominant, and the bottom row radiation is dominant. Note: as the initial state is at rest and axisymmetric, there is an uncontrolled phase in longitude, meaning the longitudinal location is abitrary.}
    \label{fig:Rossbys}
    \end{center}
\end{figure*}

Figure \ref{fig:Rossbys} shows the pressure perturbations (colour scale, total pressure minus initial pressure) and horizontal winds (vectors) as a function of longitude and latitude, for four Rossby modes. Two of the modes in Figure \ref{fig:Rossbys} are from ECLIPS3D, 3D spherical coordinate calcuations, and two from the analytical studies (i.e. derived using equations in Appendix \ref{app:orthogonality};  with \citet{Matsuno} notations, they both have $n=1$ and $l=3$), shown as the left and right columns, respectively. The locations in longitude are arbitrary as the initial state is axisymmetric and at rest. The Rossby modes shown in Figure \ref{fig:Rossbys} from ECLIPS3D have been chosen such that the maximum amplitude was present in the deeper, high pressure, regions where drag is dominant (top panel), in one case, and for the other case the amplitude was maximum in the upper, low pressure, part of the atmosphere, where the radiation timescale is shorter than that of the drag (bottom panel). Figure \ref{fig:Rossbys} shows that the ECLIPS3D results and those from our 2D analytical treatments are in good agreement. Specifically, the `tilt' of the modes in the latitude--longitude plane, and the location of the maximum perturbation in pressure are broadly consistent between the analytical 2D and numerical 3D results. This agreement is comforting given that one case is a simplification of an atmosphere on a 2D shallow water $\beta$--plane and the other one a restriction onto one height of a fully 3D, spherical mode. There are however discrepancies, notably at mid and high latitudes. 

Interestingly, with ECLIPS3D we also recover Rossby waves with the opposite tilt in latitude than the results of the shallow water equations (right panel of Figure \ref{fig:Rossbys}), as well as Rossby waves with no tilt (their horizontal shape being similar to the Rossby waves from \citet{Matsuno}). All of these waves have comparable frequencies and decay rates, no matter the tilt. We attribute the existence of these waves to the density stratification and the dependence of the radiative timescale with height, not considered in the shallow water equations. Therefore, it is no easy task to predict what will be the shape of the linear steady state as it depends on the projection of the heating function on all these waves with different tilts, but also different characteristic heights. The fact that the waves are not orthogonal anymore (Appendix \ref{app:orthogonality}) further prevents an easy evaluation of the projection of the heating function on each wave.
However, it is primordial to note that Rossby waves always exhibit a much lower amplitude in the pressure/temperature perturbation at the equator rather than in mid latitudes. This provides additional confidence in the assertion that Figure \ref{fig:MG_komacek} is really dominated by the Rossby wave component.

It is, therefore, difficult to conclude on the behaviour of the linear solutions solely from the results of the ECLIPS3D calculations in 3D, as we also require the projection of the heating function onto the waves. However, clearly we recover tilted Rossby waves, and an absence of Kelvin waves in the upper part of the atmosphere where superrotation develops (or they must have a very small characteristic height). This is in contradiction with \citet{Showman2011}, where they link the acceleration of superrotation to the interaction between a standing equatorial Kelvin wave and mid latitude Rossby wave, but in agreement with our analytical estimations for the domain of existence of Kelvin waves. As already stated in this paper, this does not refute the theories of \citet{Showman2011} and \citet{Tsai2014} regarding the equilibration of superrotation, but shows that the spin-up of an {\it initial} jet is not due to a linear, chevron-shaped steady state and time dependent linear considerations must be taken into account. Globally, our 2D semi-analytical arguments seem to be a good approximation of the 3D linear evolution of the atmosphere of hot Jupiters, and we devote the next section to an application of these estimations to provide a physical understanding to the acceleration of superrotation.

\section{Transition to superrotation}
\label{sec:transition}

In Section \ref{sec:heating}, we have shown that the linear steady state of our atmosphere was not significantly altered when moving to a more realistic heating profile taken from a 3D GCM simulation. Therefore, to gain further insight into the acceleration of the superrotation we turn to the time--dependent, linear effects. This led us to develop expressions for the time--dependent linear solution to the problem in Section \ref{sec:waves}. In this current section we use these solutions to understand the transition to superrotation in simulated hot Jupiter atmospheres. We first assess the validity of our own results by comparing them with the simulations of \citet{Komacek2016} in Section \ref{ssec:shapes}. This is followed, in Section \ref{ssec:magnitude}, by an order of magnitude analysis, which allows us to conclude that we can not explain the acceleration to superrotation under the simplifications employed for the linear steady state \citep[although such simplifications are well suited for studying the equilibration of the superrotation, see][for example]{Tsai2014}. Finally, we test our interpretation against the results of 3D GCM simulations in Section \ref{ssec:superrot_3D}, revealing that vertical accelerations are vital to the process.

\subsection{Qualitative structure of linear steady state}
\label{ssec:shapes}

Before discussing the transition to superrotation in the non linear limit, we first apply our understanding from Section \ref{sec:waves} to interpret the form of the various linear steady states presented in \citet{Komacek2016}, their Figure 5. As shown by \citet{Wu2001} the zonal damping rate is proportional to $\sqrt{\tau_\mathrm{drag}^{-1} \tau_\mathrm{rad}^{-1}}$. Therefore, if the two timescales (drag and radiative) are small, or one of them is vanishingly small, the zonal propagation of waves will be extremely limited in longitude. This is clearly seen in \citet{Komacek2016}, as when one or two of the timescales are short the temperature gradient is huge between the day and night side, and the zonal flows restricted largely to the day side. Essentially, in this case the waves excited on the day side, have been damped before reaching the night side and therefore can not lead to efficient wind generation and heat redistribution (in the linear limit). This is also discussed in  \citet{Komacek2016}.

Due to the strong asymmetric, steady forcing in the atmospheres of hot Jupiters, the dominant wavenumber $k$ ensures one oscillation around the planet, i.e. $2 \pi r k = 2 \pi$. 
Typical conditions for hot Jupiter atmospheres yield $gH \sim 4 \times 10^6\,m^2s^{-2}$ \citep{Showman2011} and therefore $L \sim 7 \times 10^7\,\mathrm{m}$ and $T \sim 3.5 \times 10^4 $\,s (see Eqs.\eqref{eq:T_dim}, \eqref{eq:L_dim}). 

Let us suppose that one of the timescales (i.e. radiative or drag) is much shorter than the other one, hence is the dominant timescale, which we denote simply as $\tau$. In order for Kelvin waves to propagate we require, from the dimensional form of Eq.\eqref{eq:Kelvin_k} :
\begin{equation}
\tau \gtrsim \dfrac{r}{2 \sqrt{gH}} \sim 2.5 \times 10^4 \ s \ .
\label{eq:Kelvin_adimensional}
\end{equation}
Therefore, if the dominant (shorter) timescale is inferior to $ \sim 2.5 \times 10^4 $\ s, Kelvin waves cannot propagate unless both timescales are equal. However, if both these timescales are $\lesssim 2.5 \times 10^4$s, the dissipation time for Kelvin waves will be very short. Assuming a simple characteristic speed of waves to be $\sqrt{gH}$, which in our case is roughly $2\times10^3\mathrm{m.s^{-1}}$, the time for a wave to travel around the whole planet is $\sim 2\times 10^5$s. Therefore, for drag or radiative timescales of $\lesssim 2.5 \times 10^4$s, even in the cases where Kelvin waves exist they can not propagate around the whole planet and, thereby, generate the stationary chevron shaped MG pattern of \citet{Showman2011}. The chevron shaped pattern of the linear steady state can therefore only exist when both timescales are $\gtrsim 10^5$\ s and comparable in value. This is seen in Figure 5 of \citet{Komacek2016}, where this pattern is clearly not ubiquitous, and this restricts strongly the cases where the explanation of \citet{Showman2011} is applicable. In other words, acceleration of superrotation in hot Jupiters from the MG or chevron shaped, tilted linear steady state is only possible over a restricted parameter space, which is therefore not likely to provide an explanation for all exoplanet cases. 


Additionally, Eq.\eqref{eq:Kelvin_adimensional} shows that for a given $\tau$ but for a varying $H$, there is a minimum height for the propagation of Kelvin waves. As shown by \citet{Wu2000} and later by \citet{Tsai2014}, the three dimensional structure of the propagating waves can be decomposed onto waves in a  shallow water system but with differing equivalent depths. The modes are solutions to the homogeneous equations but with a characteristic height which varies between modes. \citet{Tsai2014} also show, in their Figure 2 that the projection of the heating function of the vertical modes has a high amplitude for modes with equivalent height between $5H_\mathrm{P}$ and $0.2 H_\mathrm{P}$,
where $H_\mathrm{P}$ is the pressure scale height, roughly of the order of $4 \times 10^5m$ in hot Jupiter atmospheres. Therefore, adopting these limits we have $H = 5 H_\mathrm{P} \sim 2 \times 10^6$m and 
$H = 0.2 H_\mathrm{P} \sim 8 \times 10^4 $m, and obtain
\begin{equation}
10^4 \mathrm{s} < \tau < 6 \times 10^4 \mathrm{s}\,\,.
\label{eq:Kelvin_dimensional}
\end{equation}
Therefore, in our case, this implies that if $\tau < 10^4$s, Kelvin waves are unable to propagate with characteristic height less than $5H_\mathrm{P}$: the linear steady state will have an almost null projection onto Kelvin waves. However, if $\tau > 6 \times 10^4$s, the majority of the Kelvin waves excited by the forcing can propagate (we recall that if both timescale are equal all the wave can propagate, as in the neutral case, but we expect the radiative timescale to be at least an order of magnitude smaller in superrotating regions). Additionally, as introduced in Section \ref{sec:waves}, our estimates for the regimes where Kelvin waves are supported by the atmosphere is likely to be wider than the real situation, meaning that the criteria for Kelvin wave propagation are also likely to be stricter.

The behaviour outlined in this section is readily apparent in Figure 5 of \citet{Komacek2016}. In the limit where the drag is strong, the waves are damped efficiently, the thermal structure strongly resembles the thermal forcing, and there is no planetary Kelvin wave structure evident at the equator. However, in the case of weak drag ($\tau_\mathrm{drag} > 10^5$\ s) the Kelvin wave component is visible in the temperature structure only when the radiative timescale is comparable (i.e. $>$ a few $10^4$\ s, in other cases the temperature is almost uniform at the equator). Finally, in the limit of short radiative timescale the Kelvin waves do not propagate, the dynamical shape of the atmosphere is dominated by other components and the linear steady state strongly resembles Figure \ref{fig:MG_komacek}. Interestingly, it appears that the cases that superrotate in the non linear limit all have a negligible Kelvin wave contribution in their linear steady state, in other words, either the equator is not dominated by Kelvin--type circulation or the high latitudes are dominant. 

To conclude this section we detail further the case $\tau_\mathrm{drag} = 10^5$s and $\tau_\mathrm{rad,top} = 10^4$s. Following Eq.(9) of \citet{Komacek2016}, this choice for $\tau_\mathrm{rad,top}$ implies a radiative timescale of $8 \times 10^4$s at $P=80$mbar (the isobaric surface presented in their Figure 5) and the drag timescale $10^5$s. Using (Eq.\eqref{eq:Kelvin_k}), Kelvin waves are able to propagate in this scenario. Additionally, the difference between the drag and radiative timescales enables us to properly consider the behaviour of the Rossby wave component. Solving Eq.\eqref{eq:polynomial_6} for these prescribed radiative and drag timescales yields a decay timescale of $\sim 0.38$ (in non--dimensional units) for both Rossby and Kelvin waves, while the decay timescale of the gravity waves is $\sim 0.36$. Therefore, when $\tau_\mathrm{drag} = 10^5$s and $\tau_\mathrm{rad,top} = 10^4$s we have Kelvin waves propagation, with a decay timescale long enough for the waves to traverse the entire planet and comparable lifetimes for all three wave types considered (Kelvin, Rossby and gravity). In this instance we expect the steady state of the atmosphere to be a combination of planetary waves all with roughly the same magnitude (depending on the projection of the heating function), which leads to the the chevron shaped pattern predicted by \citet{Showman2011} in the Matsuno-Gill circulation. This is confirmed by Figure 5 of \citet{Komacek2016}, where, in the limit discussed, the linear steady state is similar to that shown in our Figure \ref{fig:MG_usual}.

These comparisons of our estimations with results from this work and previous studies show that our semi--analytical analysis is actually rather powerful in understanding the resulting linear steady state response of a hot Jupiter like atmosphere. The natural next step is to explore the implications for numerical simulations of a hot Jupiter atmosphere from the initial condition to the final superrotating state.

\subsection{Order of magnitude analysis}
\label{ssec:magnitude}

In this section, we estimate the  maximum forcing under which the consideration of a linear steady state is relevant, then we estimate the time dependent wave response in the weak drag regime.

As we are performing a linear study, for constant $\tau_\mathrm{rad}$ and $\tau_\mathrm{drag}$ the value of the maximum velocity is proportional to the amplitude of the forcing, which is well represented by the dayside---nightside equilibrium temperature difference we apply
at the top of the atmosphere, $\Delta T_\mathrm{eq,top}$. Assuming that the radiative timescale as a function of pressure within the atmosphere is appropriately represented by the polynomial fit of \citet{Heng2011}, adapted from \citet{iro2005}, the amplitude of the MG--circulation will then only depend on $\Delta T_\mathrm{eq,top}$ and the drag timescale.

Denoting the zonal velocity of the linear steady state as $u_\mathrm{MG}$, the linear steady state can only be reached if the non--linear terms can be neglected when $u = u_\mathrm{MG}$, i.e., once the linear steady state is formed. The non--linear terms scale with the zonal advection, $u_\mathrm{MG} \partial u_\mathrm{MG}/\partial x
\sim u_\mathrm{MG}^2/L$, where L is a characteristic length. Whereas, the linear terms are of the order of $u_\mathrm{MG}/\tau_\mathrm{drag}$. Therefore, equating these two estimates provides a maximum zonal speed for which the non--linear terms may be accurately neglected, $u_\mathrm{max}$, where
\begin{equation}
u_\mathrm{max} \sim \dfrac{L}{\tau_\mathrm{drag}} \,\, ,
\label{eq:MG_max_1}
\end{equation}
and above which a linear steady state will not be reached by the 3D GCM. For the case of hot Jupiters, $L$ ranges from half the planetary circumference in the MG case to the full circumference in the superrotating case i.e., $L \sim \pi R$. If we denote the maximum zonal, equatorial wind by $u_{\mathrm{MG},1}$, for the MG solution with $\Delta T_\mathrm{eq,top} = 1$K, using 
the linear relationship of $u_\mathrm{MG}$ to the forcing we have
\begin{equation}
u_{\mathrm{MG}} \approx u_{\mathrm{MG},1}\times \left(\dfrac{\Delta T_\mathrm{eq,top}}{1\mathrm{K}}\right)\,\, ,
\label{eq:MG_max_2}
\end{equation}
for any $\Delta T_\mathrm{eq,top}$ at a constant $\tau_\mathrm{drag}$. Combining Eqs.\eqref{eq:MG_max_1} and \eqref{eq:MG_max_2} then yields a maximum forcing temperature difference value for which the linear steady state can be reached:
\begin{equation}
\left(\dfrac{\Delta T_\mathrm{max}}{1K}\right) \approx \dfrac{1}{u_{\mathrm{MG},1}}\dfrac{L}{\tau_\mathrm{drag}} \,\, .
\label{eq:T_max}
\end{equation}
For equilibrium temperature contrast at the top of the atmosphere greater than the value in Eq.\eqref{eq:T_max} non--linear effects can no longer be neglected during the acceleration to the linear steady state, which would not be reached by a GCM. This has already been noted in Section 3.3.2 of \citet{Tsai2014}, where they acknowledge that  their analysis is strictly valid only in the strong or moderate damping scenario. As we see in Eq.\eqref{eq:T_max}, if $\tau_\mathrm{drag}$ is too large,  i.e. a low damping scenario, the maximum forcing will be very small, and the linear approximation becomes invalid for forcing amplitudes relevant to hot Jupiters. This analysis allows us to more rigorously define the weak, modest and strong damping regimes we had previously mentioned in Section \ref{sec:simple}.

In the strong or moderate damping scenario, $\Delta T_\mathrm{eq,top} \lesssim \Delta T_\mathrm{max} $, the atmosphere first reaches the linear steady solution and then the subsequent evolution is controlled by non linear acceleration. In this regime, the non linear evolution from the Matsuno-Gill linear steady state is given by $\partial u_{\mathrm{MG}}/\partial t \sim u_{\mathrm{MG}}^2/L$, where the $u_{\mathrm{MG}}^2/L$ term comes from advection, then the characteristic time $\tau_\mathrm{evol}$ for the atmosphere to significantly depart from the MG state is 
\begin{equation}
\tau_\mathrm{evol} = \dfrac{L}{u_\mathrm{MG}} \, \, .
\label{eq:t_evol}
\end{equation}
This limit is that studied in \citet{Tsai2014}, where the waves change the mean flow in a quasi--static way leading to the emergence and equilibration of superrotation. 

\begin{figure}  
    \begin{center}
    \includegraphics[width=8.5cm,angle=0.0,origin=c]{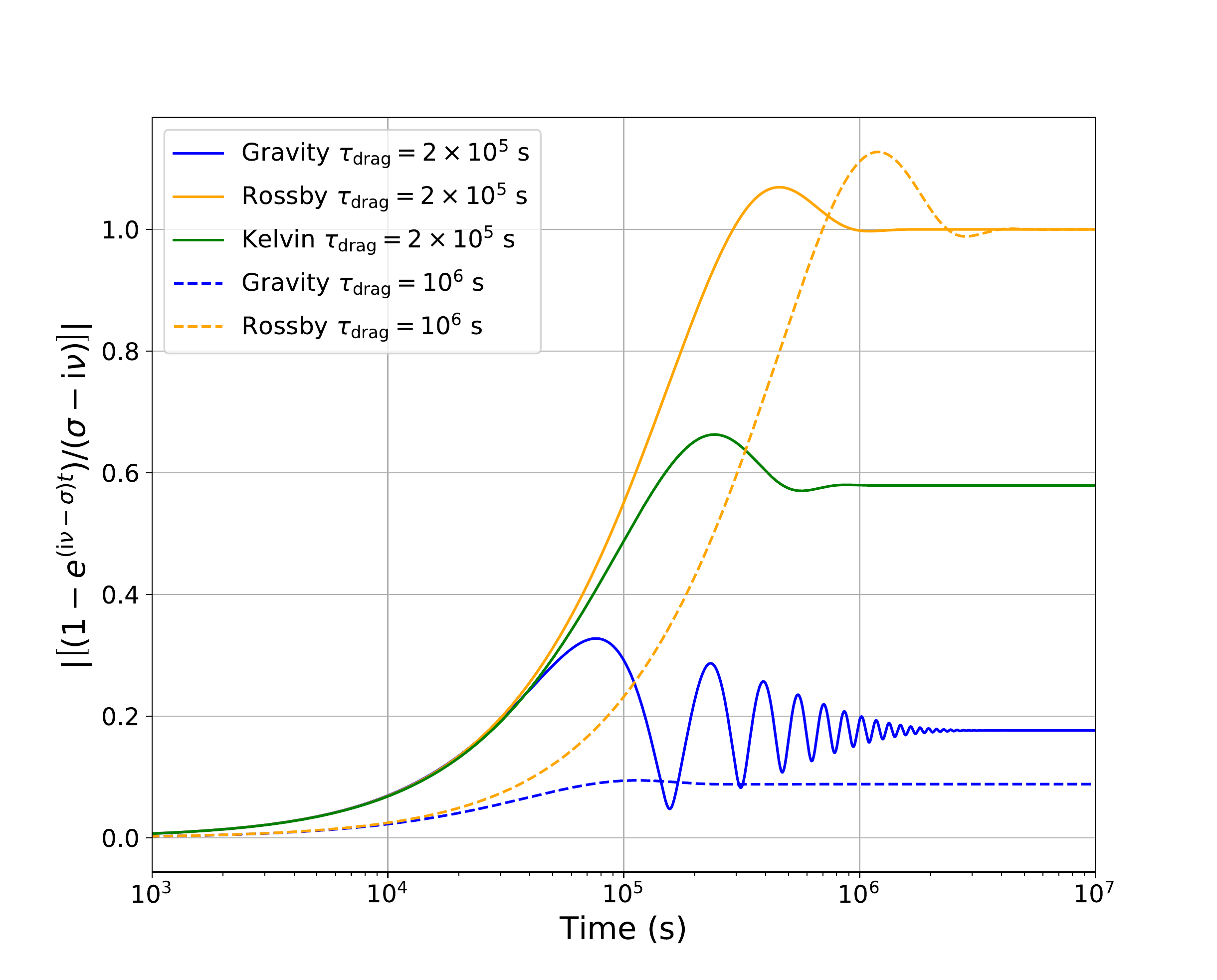}
    \caption{Modulus of $\left(1-e^{(\mathrm{i}\nu-\sigma) (t)}\right)/(\sigma-\mathrm{i}\nu)$ normalized by the final value for the Rossby wave as a function of time for gravity, Rossby and Kelvin waves with the numerical frequencies and decay rates obtained with ECLIPS3D. $\tau_\mathrm{rad}$ follows the prescription of \citet{iro2005} whereas $\tau_\mathrm{drag}$ is set constant either to $2 \times 10^5$ s (plain lines) or $10^6$ s (dashed lines). }
    \label{fig:amplitude}
    \end{center}
\end{figure}

In the low damping scenario however, the atmosphere never reaches the MG steady state, and non linear considerations must be taken into account when the characteristic speed exceeds $u_\mathrm{max}$ (hence Eq.\eqref{eq:MG_green} is irrelevant and only Eq.\eqref{eq:green_integral}-\eqref{eq:linear_evolution} can be used to understand the transition to superrotation). Let us suppose for example, that the atmosphere is composed of two waves: a slowly oscillating, slowly decaying Rossby wave, hence $\nu_1, \sigma_1 \ll 1$ and a quickly oscillating, strongly damped Kelvin or gravity wave with $\nu_2, \sigma_2 \gg 1$. Our analysis of Section \ref{sec:waves} shows that when $\tau_\mathrm{rad} \sim 10^4 $\ s and $\tau_\mathrm{drag} \sim 10^6 $\ s Rossby waves indeed have small frequency and decay rate whereas Kelvin and gravity waves have order of magnitude higher frequencies and decay rates (provided they can propagate). In this simplified case, Eq.\eqref{eq:linear_evolution} simplifies to 
\begin{align}
X_\mathrm{F} = &\dfrac{q_{1}}{\sigma_{1}-\mathrm{i}\nu_1} \tilde{X}_{1}(x,y) 
\left(1-e^{(\mathrm{i}\nu_1-\sigma_{1}) (t)}\right) \nonumber \\
+& \dfrac{q_{2}}{\sigma_{2}-\mathrm{i}\nu_{2}} \tilde{X}_{2}(x,y) 
\left(1-e^{(\mathrm{i}\nu_{2}-\sigma_{2}) (t)}\right),
\label{eq:XF_magnitude}
\end{align}
where $X_\mathrm{F}$ is the time dependent solution vector. We know that in the linear steady state, assuming $q_1 \sim q_2$, the Rossby wave component will hold much more power than the Kelvin or gravity wave as $|\mathrm{i}\nu_1-\sigma_1| \ll |\mathrm{i}\nu_2-\sigma_2|$. However, if we select a time $t$ such that $|\mathrm{i}\nu_{2}-\sigma_{2}|t \ll 1$ and (necessarily) $|\mathrm{i}\nu_1-\sigma_{1}|t \ll 1$, Eq.\eqref{eq:XF_magnitude} can be linearised to first order, yielding
\begin{equation}
X_\mathrm{F}(\sigma_1 t,\sigma_2 t \ll 1) \approx q_1 \tilde{X}_1 t +  q_2 \tilde{X}_2 t \,\, .
\label{eq:first_order}
\end{equation}
Therefore, in the limit of very early times in the evolution the two wave components in the time dependent solution of the atmosphere are comparable (provided $q_1 \sim q_2$). The wave components remain comparable even when $|\mathrm{i}\nu_{2}-\sigma_{2}|t \sim 1$ but $|\mathrm{i}\nu_1-\sigma_{1}|t \ll 1$, where the exponential term for the Kelvin or gravity wave is almost zero, but the linearisation holds for the Rossby wave, hence,
\begin{equation}
X_\mathrm{F}(\sigma_1 t \ll 1,\sigma_2 t \gtrsim 1) \approx q_1 \tilde{X}_1 t +  
\dfrac{q_{2}}{\sigma_{2}-\mathrm{i}\nu_{2}} \tilde{X}_2 \,\, .
\label{eq:sigmat_1}
\end{equation}
Dividing the amplitude of our first, Rossby wave, in this linear {\it time dependent} state, $\alpha_1$, by the amplitude of wave 2, $\alpha_2$ yields,
\begin{equation}
\left|\dfrac{\alpha_1}{\alpha_2}\right| = \left| \dfrac{q_1}{q_2} (\sigma_{2}-\mathrm{i}\nu_{2}) t \right| \sim \dfrac{q_1}{q_2} \,\,.
\label{eq:contribution}
\end{equation} 
Provided that the projection of the heating function on the two wave components are similar i.e. $q_1 \sim q_2$, the time dependent solution will exhibit comparable amplitudes for both waves before the (slowest) Rossby wave has grown much larger than the asymptotic amplitude of the Kelvin or gravity wave. However, the steady state will be dominated by the Rossby wave component. Therefore, although the linear steady state strongly depends on $\tau_\mathrm{rad}$ and $\tau_\mathrm{drag}$, in the limits of short timescales the structure of the atmosphere only depends on the projection of the heating function on the propagating waves. On Figure \ref{fig:amplitude},
we show the modulus of $\left(1-e^{(\mathrm{i}\nu-\sigma) (t)}\right)/(\sigma-\mathrm{i}\nu)$ for a Rossby , Kelvin and gravity waves obtained by ECLIPS3D (see Figure \ref{fig:decay}) when $\tau_\mathrm{drag} = 2 \times 10^5 $\ s and for a Rossby and gravity waves when $\tau_\mathrm{drag} = 10^6 $\ s. In both cases, after a few $10^4$ s (a day being $\sim 9 \times 10^4$ s), the amplitude of the different waves is of the same order of magnitude whereas Rossby wave contribution dominates at later times.

This analysis suggests that the linear steady state is not responsible for accelerating superrotation in hot Jupiter atmospheres for the low drag limit, and allows us to resolve the problem explained in Section \ref{ssec:problem_komacek}. As discussed in Section \ref{ssec:problem_komacek}, superrotating atmospheres were found by \citet{Komacek2016} despite structures implying negative convergence of momentum at the equator in the linear limit (a 'reverse' MG structure), in contradiction with the mechanism of \citet{Showman2011} which invokes the linear MG steady state solution. 

This problem persists for physically plausible choices on the equilibrium temperature contrast and the drag and radiative timescale, typically, $\Delta T_\mathrm{eq,top} = 500$\ K, $\tau_\mathrm{drag} \sim 5 = 10^5$\ s and $\tau_\mathrm{rad} = 5 \times 10^4$\ s which yields $u_\mathrm{max} \sim 200\mathrm{m.s}^{-1}$. When solved numerically, we obtain $u_\mathrm{MG,1} \sim 5 \mathrm{m.s}^{-1}$ hence $\Delta T_\mathrm{max} \sim 150$\ K: the linear steady state cannot be reached. Specifically, after one day of simulation, our analysis shows that the linear steady state will not have been reached {\it but} numerically we already have $u > u_\mathrm{max}$, hence non linear effects can not be neglected anymore. Regarding the dissipation timescales for the waves (Section \ref{sec:waves}), after one day we have $\sigma_K t, \nu_K t \sim 1$,$\sigma_g t, \nu_g t \sim 1$ and $\sigma_R t, \nu_R t \ll 1$, where $\nu_K$, $\nu_g$, $\nu_R$ are the Kelvin, gravity and Rossby waves frequencies, respectively, and $\sigma_K$, $\sigma_g$ $\sigma_R$ the Kelvin, gravity and Rossby dissipation rate, respectively. Our estimates of this section, Eq.\eqref{eq:contribution}, show that this leads to an equivalent contribution of  Rossby, gravity and Kelvin waves in the circulation.

In summary, after 1 day of simulation:
\begin{itemize}
\item the structure of the atmosphere exhibits similar contribution in Rossby, gravity and Kelvin waves, which is characteristic of the chevron-shaped pattern of Figure \ref{fig:MG_usual}. As shown by \citet{Showman2011}, the eddies from such a circulation favour the meridional convergence of eastward momentum at the equator. 
\item Additionally, simple orders of magnitude show that the non linear terms (hence the contribution from the eddies) can no longer be neglected as they are of the same order of magnitude of the linear terms.
\end{itemize}
Therefore, when the non linear terms become dominant, they lead to a net acceleration of the equator. Even though the eddy acceleration from the linear steady state (our Figure \ref{fig:MG_komacek}) would tend to decelerate the equator, the equator is accelerated non linearly after one day of simulation because of this chevron shaped, {\it non steady} linear circulation. This initiates superrotation, and the later, slower evolution is well described by \citet{Tsai2014}. We study this further in the next section using our own GCM \citep[the Unified Model, UM, presented in ][]{Mayne2014,Mayne2017}.

We must underline that this separation of scales between linear and non linear behaviour is obviously very simplified. As already noted by \citet{Showman2011}, the non linear accelerations are mostly due to the wave--mean flow interactions (rather than wave--wave or mean flow--mean flow, as we confirm in the next section). Therefore, a quasi linear analysis or statistical studies of momentum transfer might allow more rigorous insight into the jet acceleration (see e.g., \citet{Srivinasan2012,Tobias2013,Bakas2013,Bouchet2013,Bakas2015,Herbert2019}).




\subsection{3D GCM simulations}
\label{ssec:superrot_3D}

In order to assess the applicability of the linear shallow water result developed in this work to a full 3D calculation, we have performed simulations using the UM across a range of forcing scenarios. The simulations employ Newtonian heating as discussed in \citet{Mayne2014}, and adopt the baseline hot Jupiter setup presented in \citet{Mayne2014b} which follows that of \citet{Heng2011} and for the radiative timescale, \citet{iro2005}. For our simulations we have then varied the day to night temperature contrast, $\Delta T$ from 1 to 100K \citep[see Eqs.(B2) and (B3) of][]{Heng2011}. Obviously, this is only a toy model as we do not expect to find a tidally locked planet with an effective temperature of $1300$K and a day night contrast of $0.1$K, but it allows us to study the physical mechanisms controlling the atmospheric structure. Atmospheric drag has not been explicitly implemented in the UM, but is provided by a diffusion scheme as detailed in \citet{Mayne2014b}. We have verified that all of our simulations conserve mass and angular momentum to an order of $10^{-6}$. 
We use these simulations to first explore the resulting, qualitative structure of the atmosphere and then the accelerations within it, in Sections \ref{sssec:gcm_qual} and \ref{sssec:gcm_accel}, respectively.
\subsubsection{Qualitative structure of the atmosphere}
\label{sssec:gcm_qual}

As long as the linear effects are dominant, we expect all our simulations to have qualitatively similar features but with quantitative values that scale with the forcing. After one day of simulation this is indeed the case, where all of our simulations have the same qualitative structure matching Figure \ref{fig:MG_usual}, although the magnitude of the temperature differences and wind velocities vary between simulations (increasing with larger temperature contrast). The structure matches the `chevron' shaped pattern of \citet{Showman2011}, but we again state that this is {\it not} the steady Matsuno-Gill solution. It is a specific time in the evolution of the atmosphere when all waves have the same projection in the circulation, as discussed in the previous section.

\begin{figure}  
    \begin{center}
    \subfigure[]{\includegraphics[width=9.5cm,angle=0.0,origin=c]{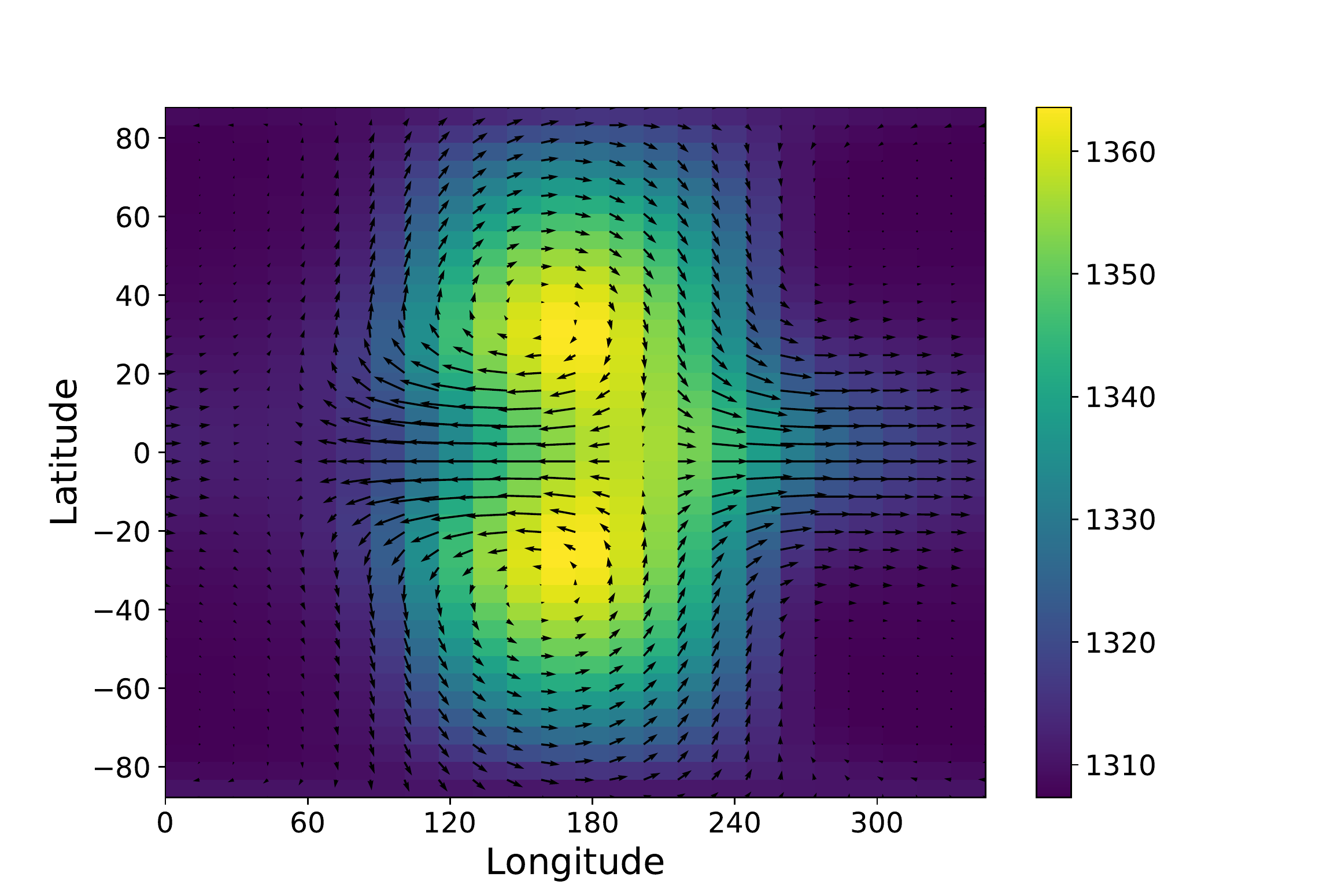}\label{fig:dt_100_r45.pdf}}
    \subfigure[]{\includegraphics[width=9.5cm,angle=0.0,origin=c]{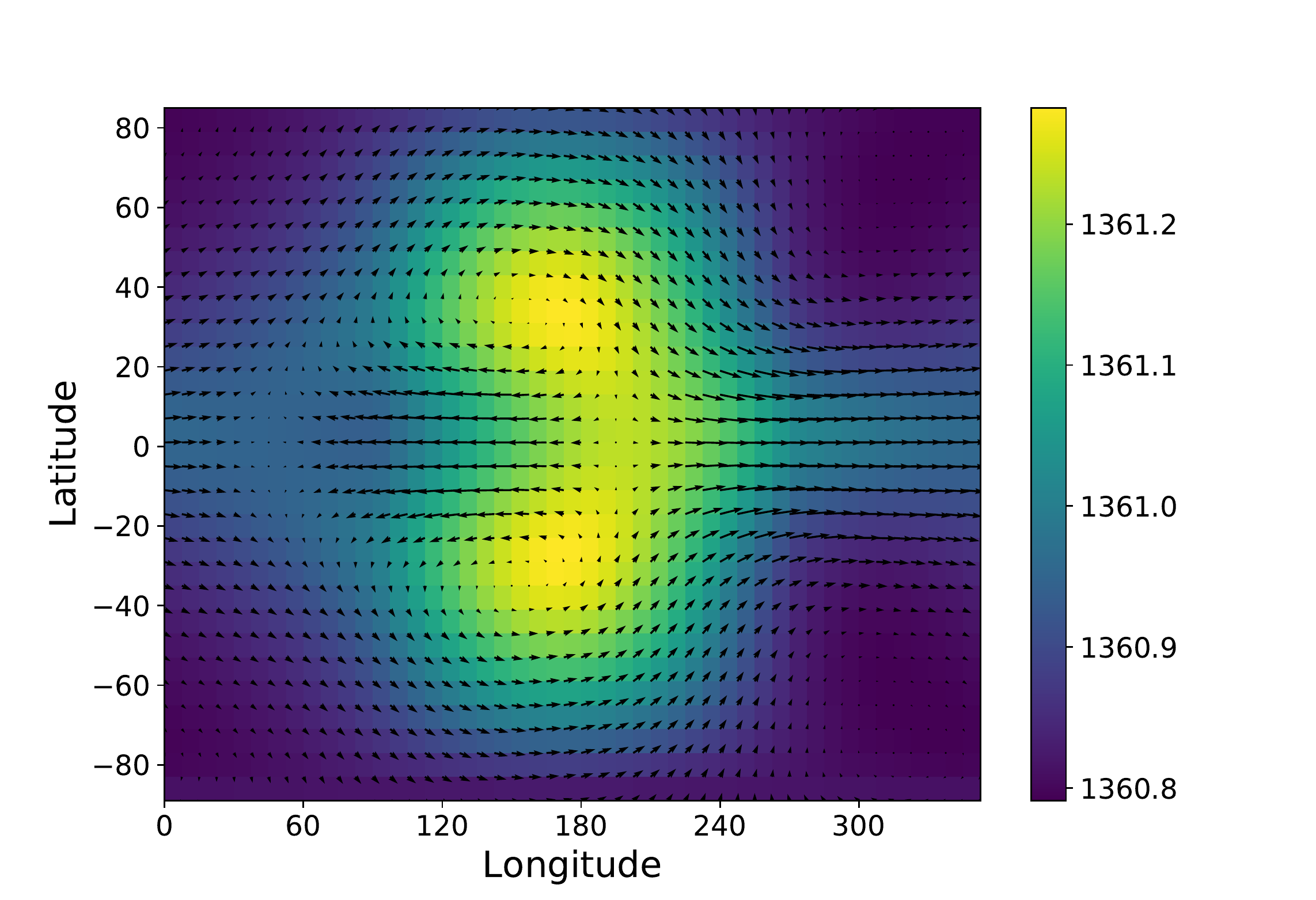}\label{fig:dt_1_10day}}
    \subfigure[]{\includegraphics[width=9.5cm,angle=0.0,origin=c]{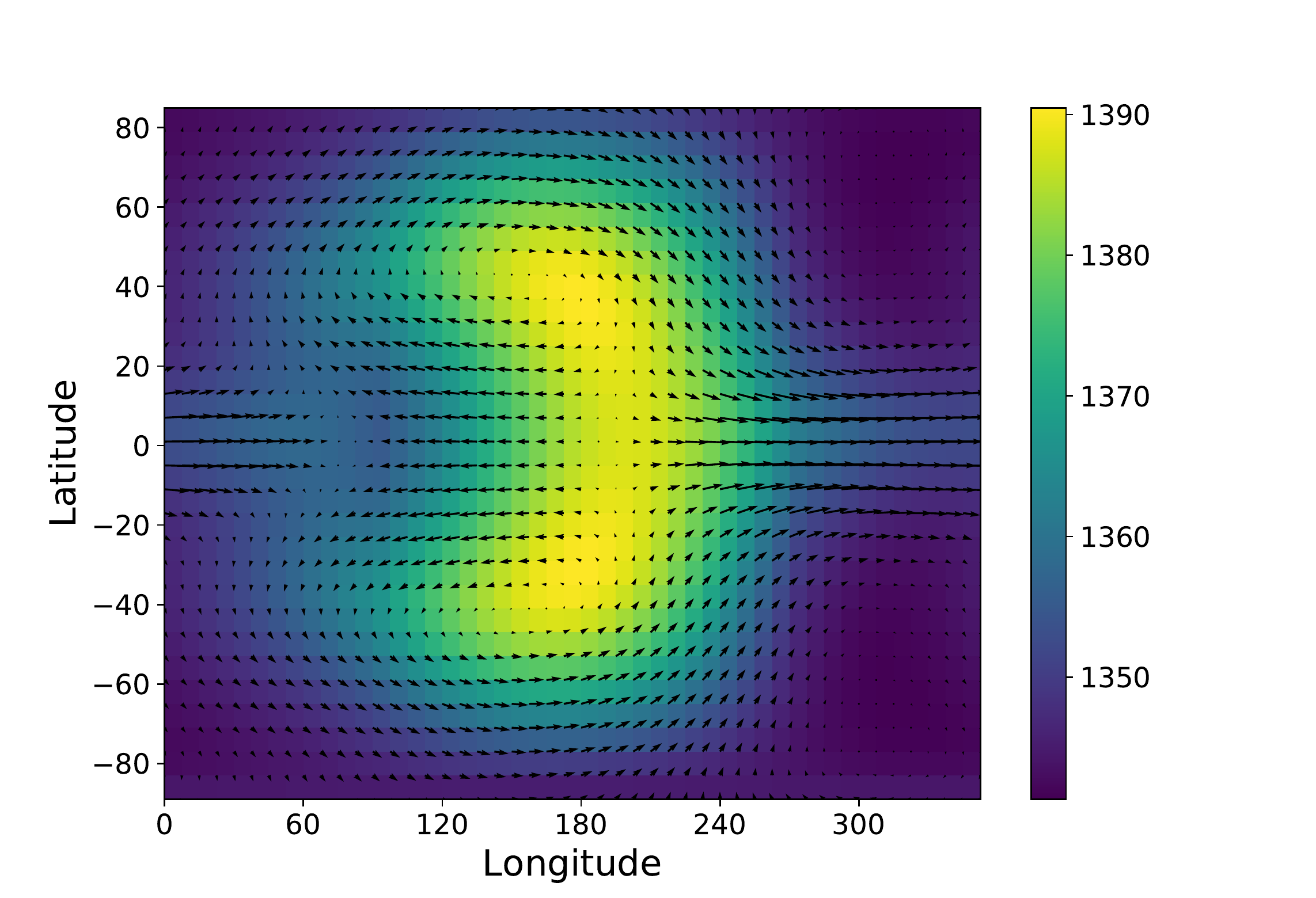}\label{fig:dt_100_10day}}
    \caption{Temperature (color) and winds (arrows) as a function of longitude and latitude for (a) linear steady state calculated with ECLIPS3D with $\Delta T_\mathrm{eq,top} = 100$\,K, $\tau_\mathrm{drag} = 2 \times 10^5$\,s and $\tau_\mathrm{rad}$ following \citet{iro2005}. Maximum speed is $400\,\mathrm{m.s^{-1}}$ (b): 3D GCM result at the 80\,mbar level after 10 days of simulation, with $\Delta T_\mathrm{eq,top} = 1$\,K and the radiative timescale of \citet{iro2005}. The maximum speed is $2.5\, \mathrm{m.s^{-1}}$; (c): 3D GCM result at the 80\,mbar level after 10 days of simulation, with $\Delta T_\mathrm{eq,top} = 100$\,K and the radiative timescale of \citet{iro2005}. The maximum speed is $350\,\mathrm{m.s^{-1}}$}
    \label{fig:atmo_10day}
    \end{center}
\end{figure}

Comparing the very low temperature contrast case with linear steady states from ECLIPS3D, the dissipation within our simulations is equivalent to $\tau_\mathrm{drag} \sim \, \, \text{a few} \, \, 10^5$s. At $P \sim 80$\ mbar the radiative timescale of \citet{Heng2011}, adapted from \citet{iro2005}, is of the order of $2.5 \times 10^4$s. For these parameters the linear MG state can only be considered as reached after $\sim 10$ days, the time for the gravity and Rossby waves (which are the most long lived components) to be completely dissipated (see Section \ref{sssec:waves_summary}). Figure \ref{fig:atmo_10day} shows the temperature and wind structure for three calculations, the first one using ECLIPS3D with parameters set to those matching the GCM simulations (Figure \ref{fig:dt_100_r45.pdf}), and the next two from GCM simulations after 10 days of simulations, at 80\,mbar, for a small and large temperature contrast at the top of the atmosphere $\Delta T_\mathrm{eq,top} = 1$\,K (Figure \ref{fig:dt_1_10day}) and  $\Delta T_\mathrm{eq,top} = 100$\,K, (Figure \ref{fig:dt_100_10day}).

The ECLIPS3D calculation, hence the linear steady state, Figure \ref{fig:dt_100_r45.pdf}, recovers the dominant mid--latitude Rossby gyres, associated with equatorial winds but it clearly differs from Figure \ref{fig:MG_usual} in the sense that the equatorial circulation is not impacting the Rossby gyres significantly. We therefore have an intermediate between Figures \ref{fig:MG_usual} and \ref{fig:MG_komacek}. For completeness, we also show the evolution of the atmosphere for the same simulations but with $\tau_\mathrm{drag} = 10^6$s in Appendix \ref{app:figure}, which leads to a linear steady circulation equivalent to Figure \ref{fig:MG_komacek}. For the GCM simulations adopting $\Delta T_\mathrm{eq,top} = 1K$, Figure \ref{fig:dt_1_10day}, the circulation of Figure \ref{fig:dt_100_r45.pdf} is recovered after 10 days of simulation. For the GCM simulations adopting $\Delta T_\mathrm{eq,top} = 100K$,  Figure \ref{fig:dt_100_10day}, respectively, the longitudinal extent of the westward wind is reduced after 10 days compared to the weak forcing regime. Hence, for the simulation with the larger temperature contrast, after 10 days, the atmosphere has already diverged from the linear evolution of the atmosphere. Although both simulations were qualitatively identical after 1 day of simulation, the low forcing simulations then reaches the linear steady state whereas higher forcing simulations never go through this state. Using the steady state wind velocities for the smaller temperature contrast simulation, $\sim 2 \mathrm{ms^{-1}}$ and Eq.\eqref{eq:t_evol}, alongside the fact that we expect the linear steady wind to scale with the forcing, we can estimate the timescale to depart from the MG state in the stronger forcing simulation: $\sim 10^6$s, which is about 10 days. This timescale matches the timescale estimated above, from the analysis of the atmospheric waves, for convergence to the MG state and hence the MG state is never actually reached.

In our simulations, the case of not reaching the linear steady state occurs for day--night temperature contrasts of $\sim 100$K or greater,  at drag timescales of $\tau_\mathrm{drag} \sim \, \, \text{a few} \, \, 10^5$\ s (as presented in Figure \ref{fig:atmo_10day}). If the drag timescale is further increased, the temperature contrast for which the linear steady state could not be reached would decrease further\footnote{and it is not possible to reduce the drag timescale below $\sim 10^5$s as it leads to a suppression of superrotation (in this situation,  the drag timescale is lower than the advective timescale of a superrotating jet, explaining the breaking of superrotation).},  and HD~209458b is rather expected to experience a $\sim 500$\ K temperature contrast.
Hence it does not seem possible to reconciliate the initial acceleration of superrotation with the consideration of steady linear effect. The linear considerations can only be used in the first day or so of simulation, and the linear steady state is irrelevant.

Therefore if superrotation does exist in hot Jupiter atmospheres, the steady linear considerations are not likely sufficient to explain the initial acceleration as non linear effects quickly dominate. Notably, the study of both \citet{Tsai2014} and \citet{Showman2011} \citep[and more recently][]{Hammond2018} would only apply in the limit of slow evolution, hence once the atmosphere is already close to a non linear steady state. This explains why Figure 16 of \citet{Tsai2014} which represents their linear consideration compares so well with their Figure 15, taken from 3D numerical simulations: when an initial superrotation is already settled, the further evolution is slow and can be understood in the linear limit. However, \citet{Tsai2014} provide no comparison between the linear expectation and the 3D simulations during the original acceleration phase of superrotation. 

As we have seen in this section, linear considerations apply as long as we consider the time dependent solution. Although we cannot use our linear prescriptions to predict the evolution of the atmosphere in the non linear phase (as stressed in section \ref{ssec:magnitude}), we can estimate the duration of validity of the linear approximation and show that the atmosphere does not go through a linear steady state during the acceleration phase. Therefore, it is worth noting that the westward shift of the hot spot in the steady linear limit studied by \citet{Hindle2019}, with the addition of a magnetic field, is not a robust enough diagnostic to predict whether the atmosphere is superrotating.

Interestingly, our low forcing simulation also converged to a superrotating state after a much longer time (scaling as one would expect as the inverse of the forcing). This was unexpected as Figure \ref{fig:dt_1_10day} is not associated with a strong deposition of eastward momentum at the equator. More surprisingly, we also recover a superrotating jet for low forcing simulations when we further increase the drag timescale, and the atmosphere goes through a linear steady state resembling Figure \ref{fig:MG_komacek}. To understand this phenomenon, we conclude our study with the considerations of 3D accelerations in the spin up and equilibration of superrotation for GCM results.

\subsubsection{Accelerations}
\label{sssec:gcm_accel}

The final step is to study the acceleration of the mean flow in the initial stages of the acceleration of superrotation within the 3D GCM simulations and assess the relevance of the 2D studies. As discussed, once the jet is settled and evolving slowly, the studies of \citet{Tsai2014} and \citet{Hammond2018} describe the evolution of superrotation, but the initial acceleration is less clear as we have explained through this paper. For this purpose, we study the acceleration of the jet in a simulation with $\Delta T_\mathrm{eq,top} = 100$K and $\tau_\mathrm{drag} \sim 10^5$s, where the first phases of the development of superrotation can be captured over about 30\,days. Following \citet{Mayne2017} \citep[who adapted the treatment of][]{Hardiman2010}, the acceleration of the zonal mean flow 
can be written as
\begin{align}
&(\overline{\rho}\,\overline{u})_{, t}=-\frac{(\overline{\rho v}\,\overline{u}\cos^2\phi)_{, \phi}}{r\cos^2\phi}-\frac{(\overline{\rho w}\,\overline{u}r^3)_{, r}}{r^3}
+2\Omega\overline{\rho v}\sin\phi \nonumber \\
&-2\Omega\overline{\rho w}\cos\phi -(\overline{\rho^{\prime}u^{\prime}})_{, t}-\frac{\left[ \overline{(\rho v)^{\prime}u^{\prime}}\cos^2\phi\right]_{, \phi}}{r\cos^2\phi} \nonumber \\&-
\frac{\left[ \overline{(\rho w)^{\prime}u^{\prime}}r^3\right]_{, r}}{r^3}+
\overline{\rho G_{\lambda}},
\label{eq:acceleration}
\end{align}
where $G_\lambda$ denotes the body forces acting in the longitudinal direction (not considered here), the subscripts denote partial derivatives, and every quantity $X$ is defined as  $X = \bar{X} + X'$ where a bar denotes an average on longitude. In this section, we do not consider the mean flow-mean flow accelerations as they are negligible during the initial acceleration within our simulations. However, once the superrotating jet has formed, these accelerations should be taken into account as they balance the eddy accelerations and eventually lead to a non linear steady state \citep[see notably the conclusions of][]{Tsai2014}.

\begin{figure}[ht!]  
    \begin{center}
    \subfigure[]{\includegraphics[width=8.5cm,angle=0.0,origin=c]{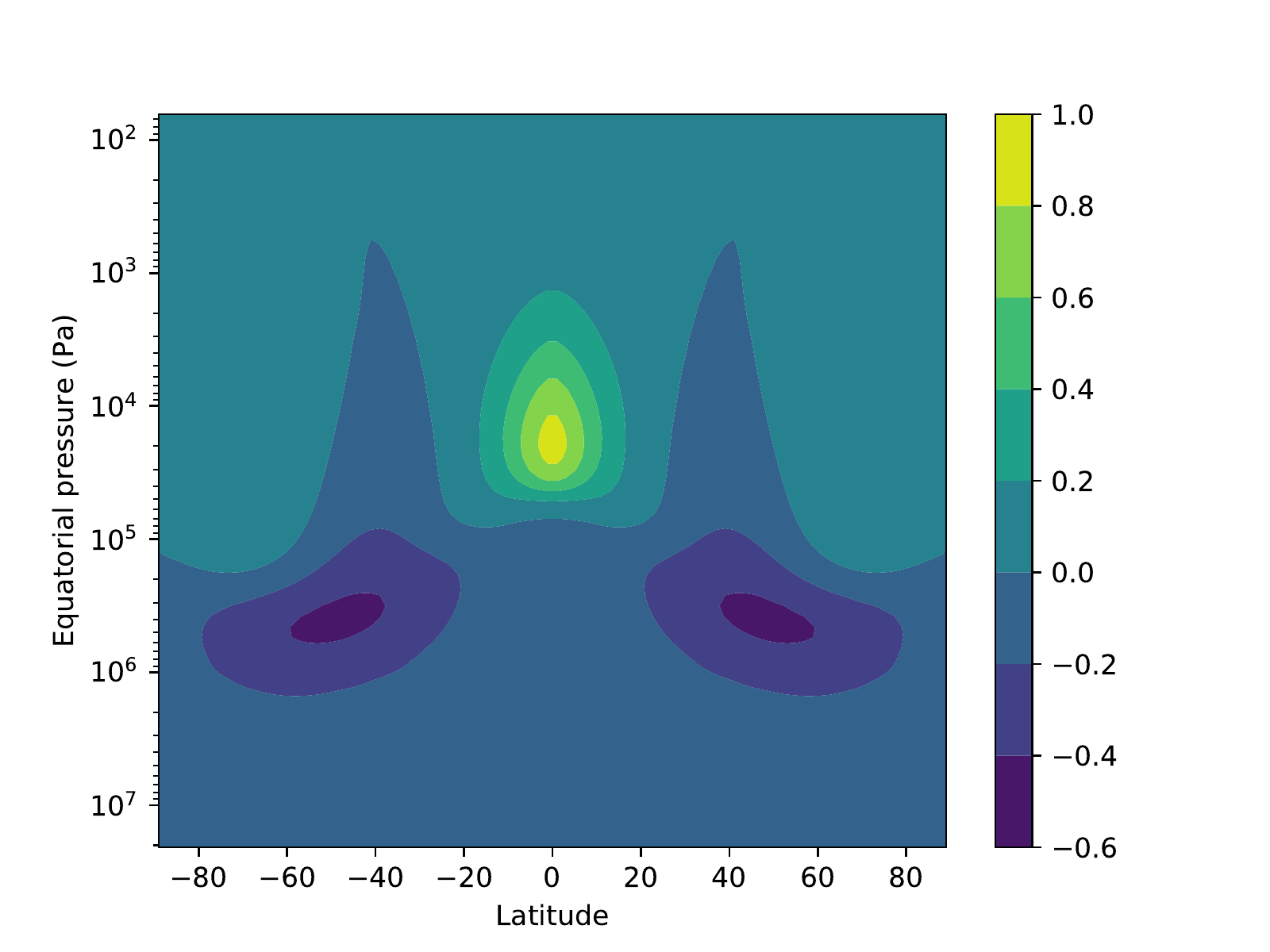}\label{fig:dt_100_50day}}
    \subfigure[]{\includegraphics[width=8.5cm,angle=0.0,origin=c]{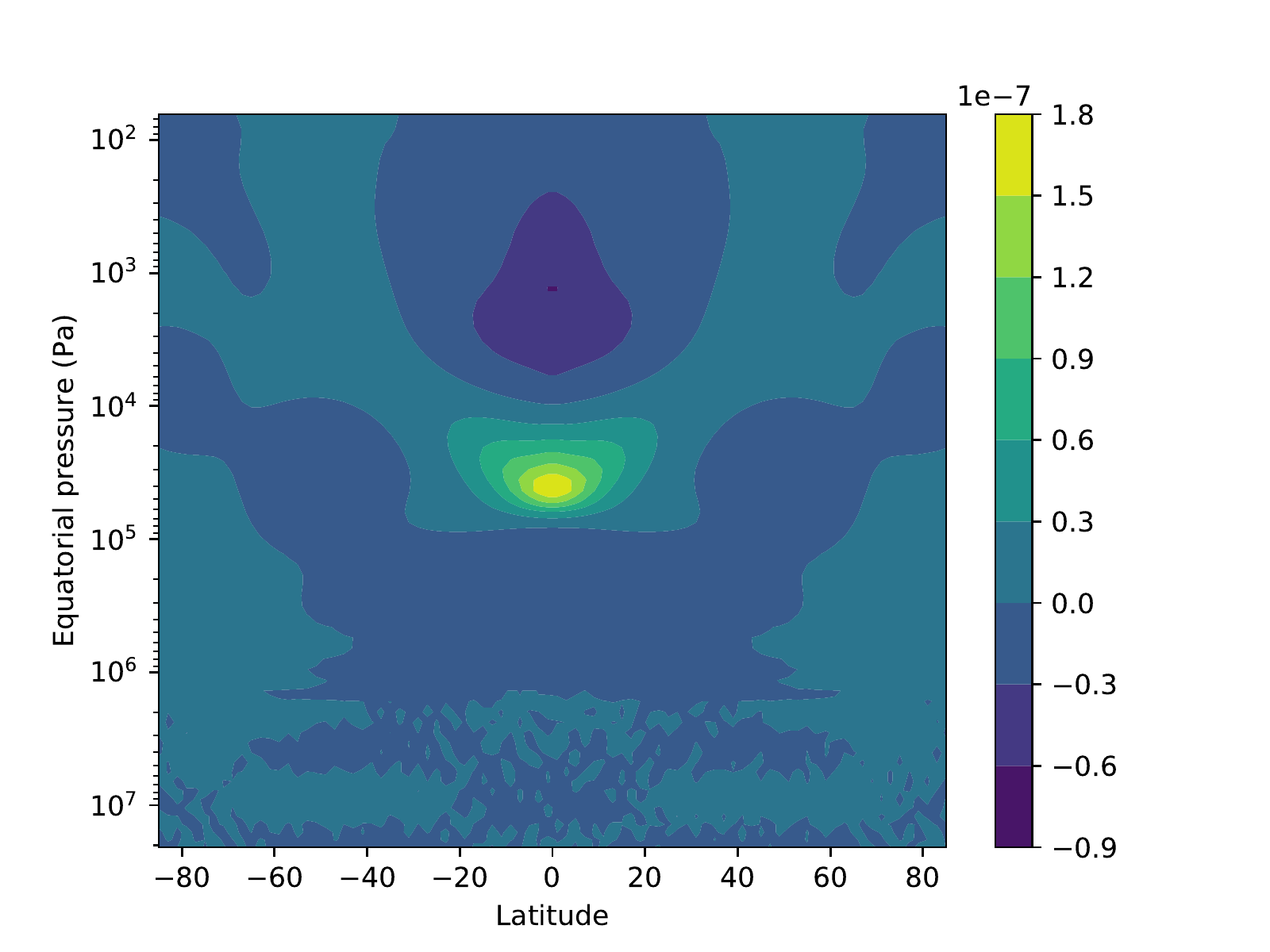}\label{fig:acc_dt_100_50day_ver}}
    \subfigure[]{\includegraphics[width=8.5cm,angle=0.0,origin=c]{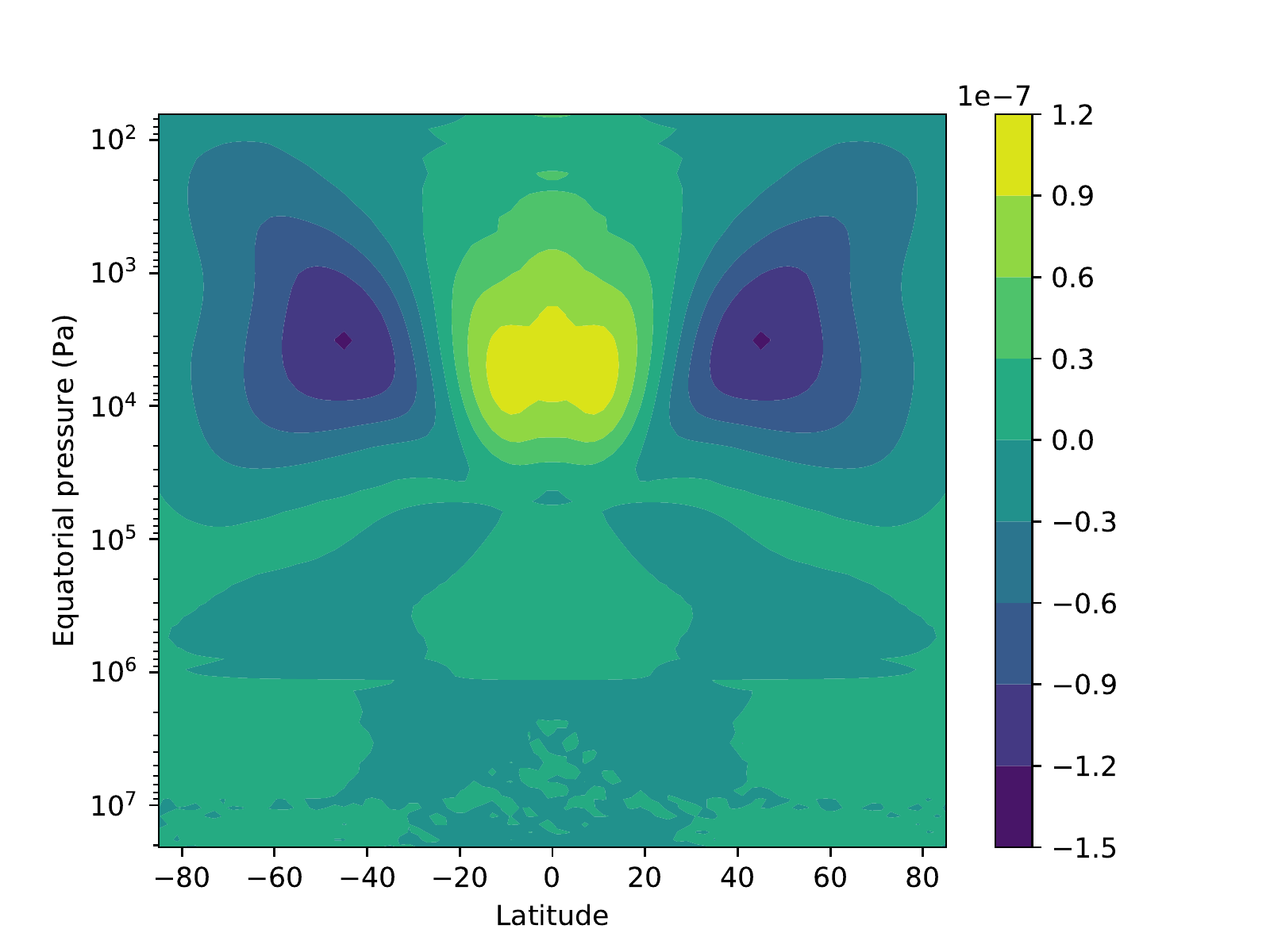}\label{fig:acc_dt_100_50day_mer}}
    \caption{(a) $(\overline{\rho}\,\overline{u})$ (kg.m$^{-2}$.s$^{-1}$) as a function of pressure and latitude after 50 days of simulation with $\Delta T_\mathrm{eq,top}= 100$\,K. (b) Vertical eddy acceleration (kg.m$^{-2}$.s$^{-2}$) for the same simulation. (c) Meridional eddy acceleration (kg.m$^{-2}$.s$^{-2}$) for the same simulation.}
    \label{fig:dt100_50days_tot}
    \end{center}
\end{figure}

Following \citet{Showman2011}, the meridional eddy accelerations, involving $v'$ and $u'$, should lead to momentum 
convergence at the equator from the MG steady state whereas the vertical component acts to decelerate the equatorial region. In Figure \ref{fig:dt100_50days_tot} we show the value of $(\overline{\rho}\,\overline{u})$ for $\Delta T_{eq,top} = 100$K after 50 days of simulations as well as the vertical and meridional accelerations. After 50 days, the jet extends from roughly $1$mbar to $1$bar with the maximum of $(\overline{\rho}\,\overline{u})$ around $0.2$bar. As in Figure 15 of \citet{Tsai2014}, we observe that the vertical accelerations are slowing down the upper part of the jet, while extending the jet to deeper pressures. The meridional accelerations on the other hand compensate the vertical component in agreement with both \citet{Tsai2014} and \citet{Showman2011}. It is interesting to note that the explanation for radius inflation of \citet{Tremblin2017} relies on the vertical wind in the deep atmosphere due to the equatorial jet, and that the spin up of the jet shows that the vertical accelerations are pushing the equatorial jet downwards. This seems to point towards a circulation in depth between the jet and the vertical velocities, that gets deeper with time.

\begin{figure*}  
    \begin{center}
    \subfigure[]{\includegraphics[width=8.5cm,angle=0.0,origin=c]{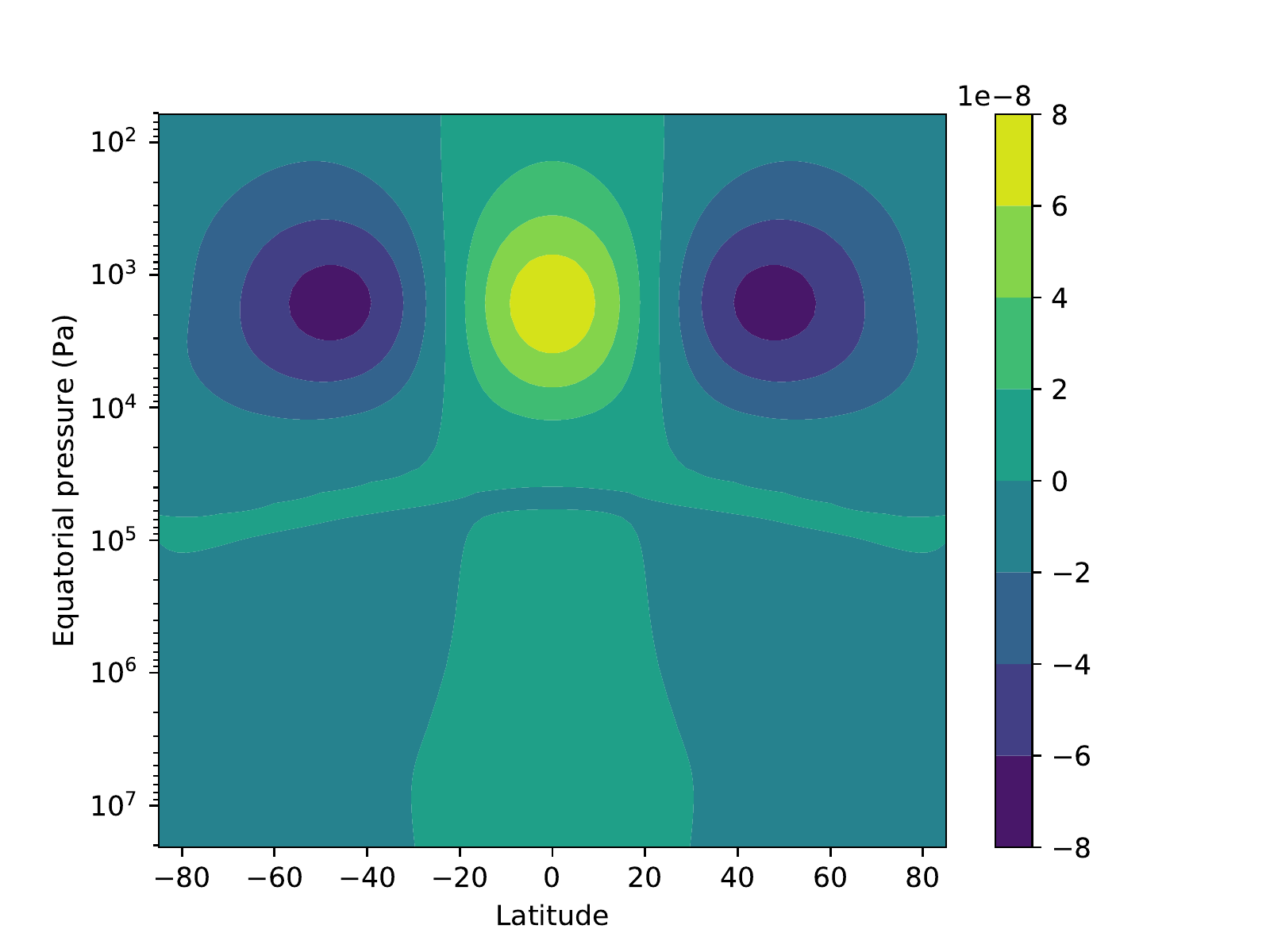}\label{fig:acc_dt100_1_mer}}
    \subfigure[]{\includegraphics[width=8.5cm,angle=0.0,origin=c]{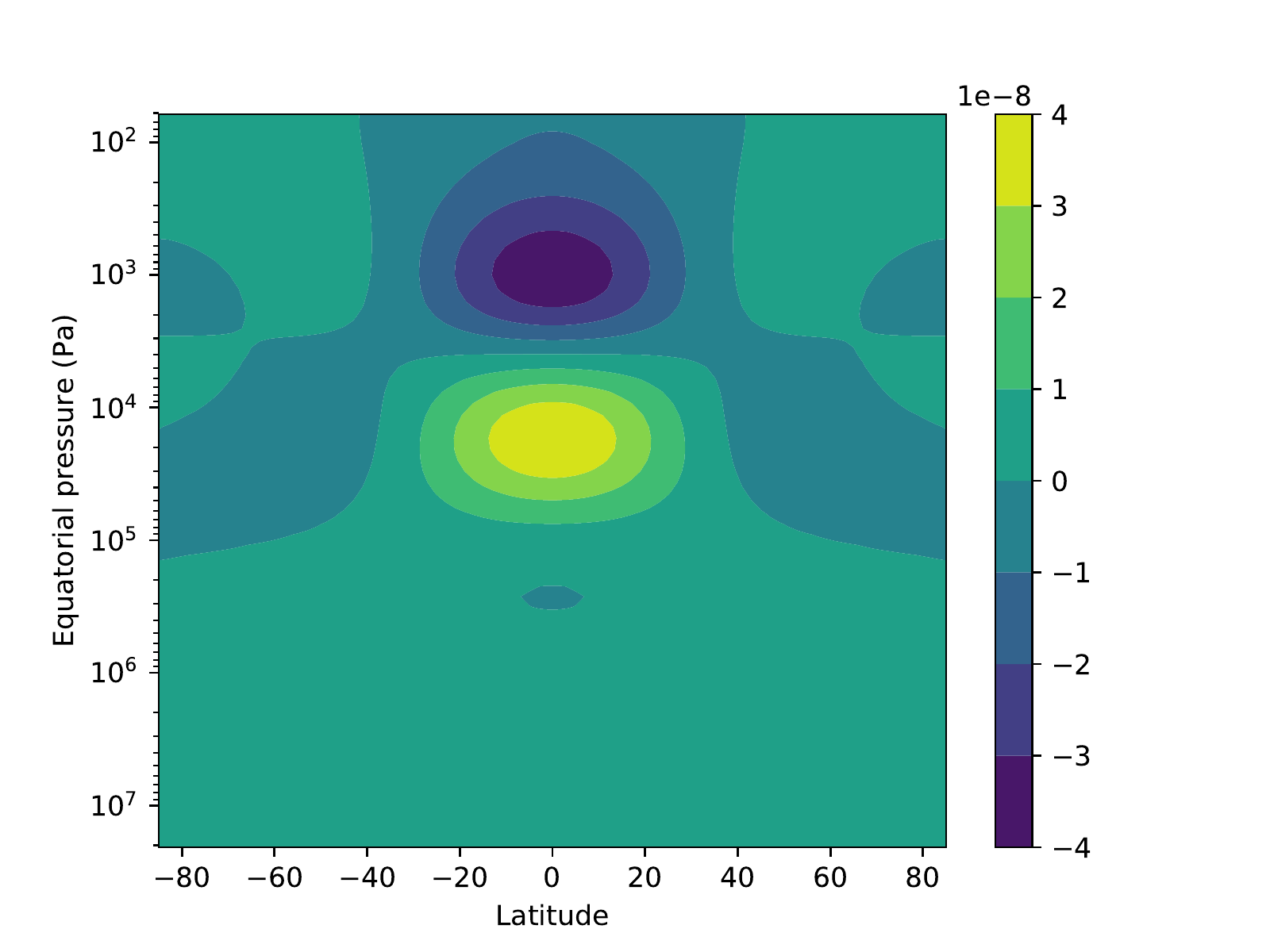}\label{fig:acc_dt100_1_ver}}
    \subfigure[]{\includegraphics[width=8.5cm,angle=0.0,origin=c]{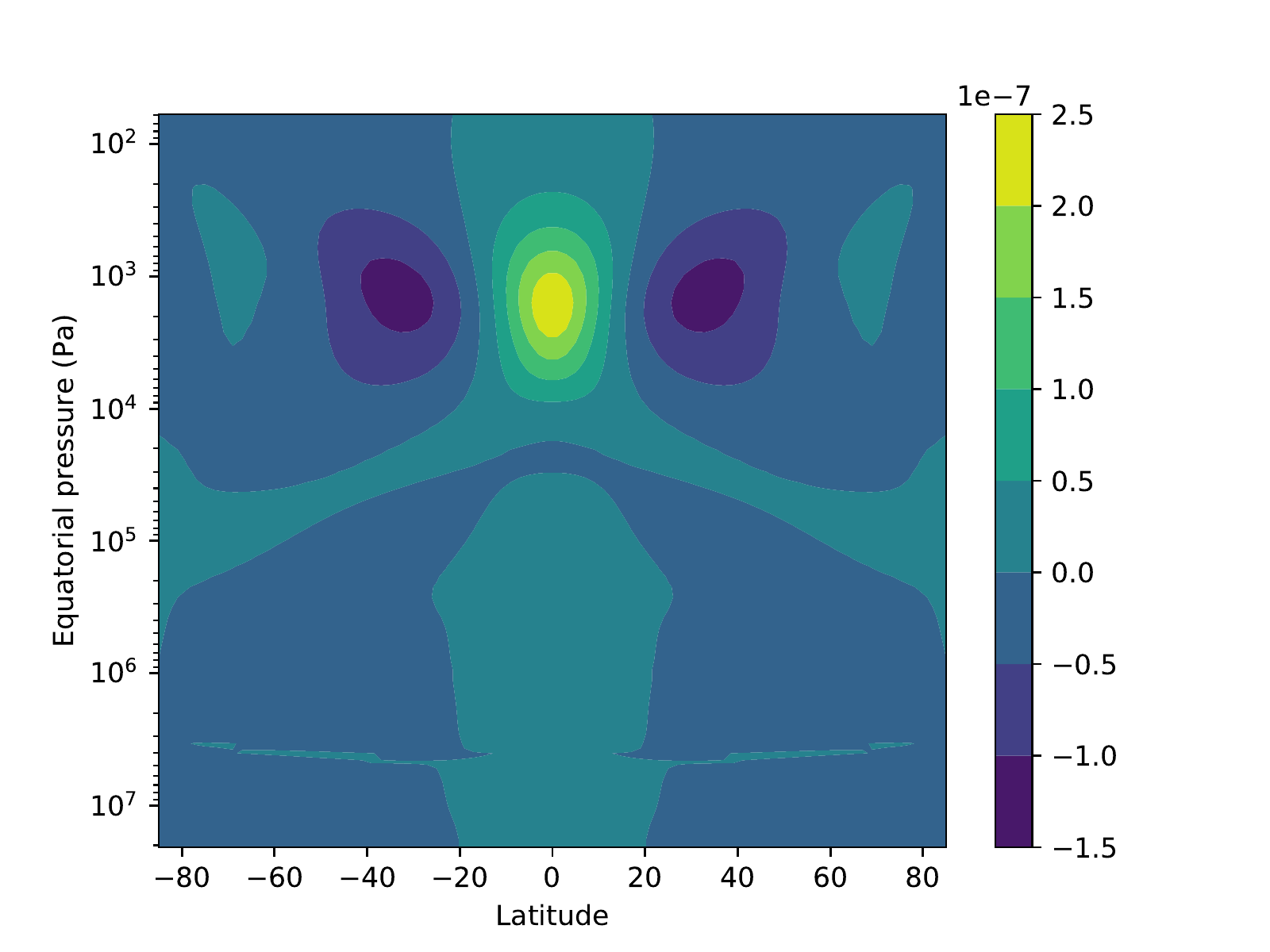}\label{fig:acc_dt100_5_mer}}
    \subfigure[]{\includegraphics[width=8.5cm,angle=0.0,origin=c]{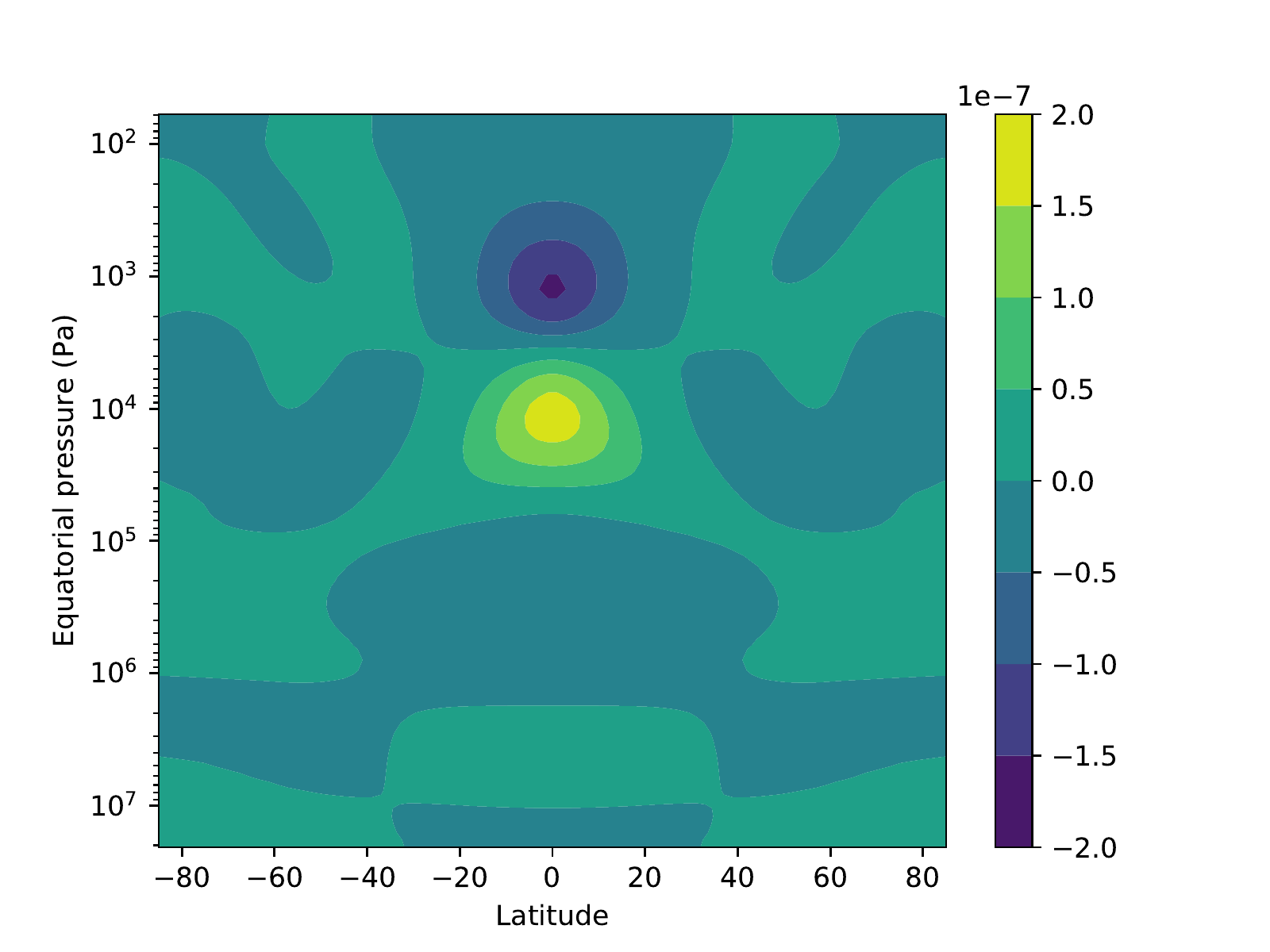}\label{fig:acc_dt100_5_ver}}
    \subfigure[]{\includegraphics[width=8.5cm,angle=0.0,origin=c]{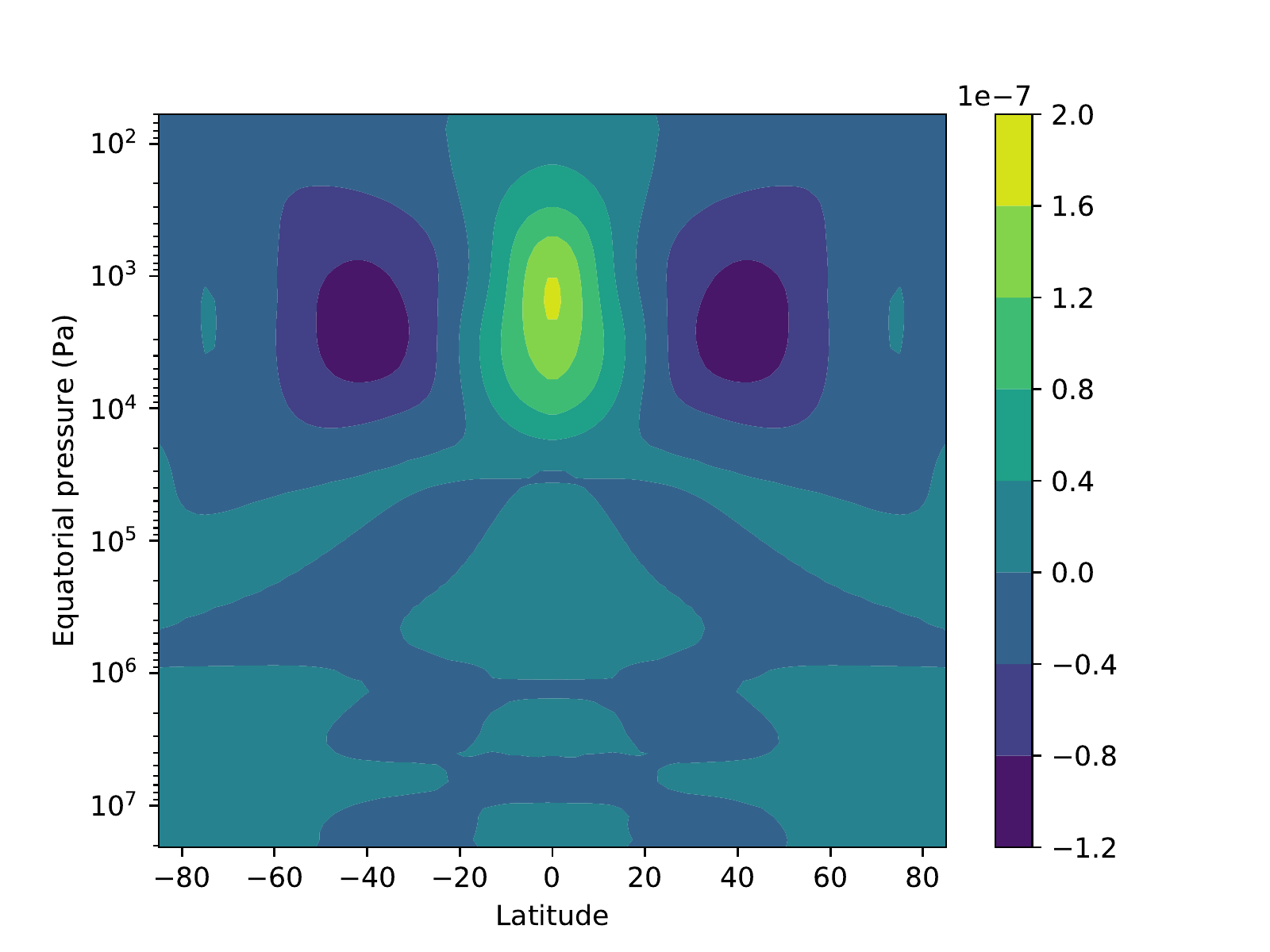}\label{fig:acc_dt100_20_mer}}
    \subfigure[]{\includegraphics[width=8.5cm,angle=0.0,origin=c]{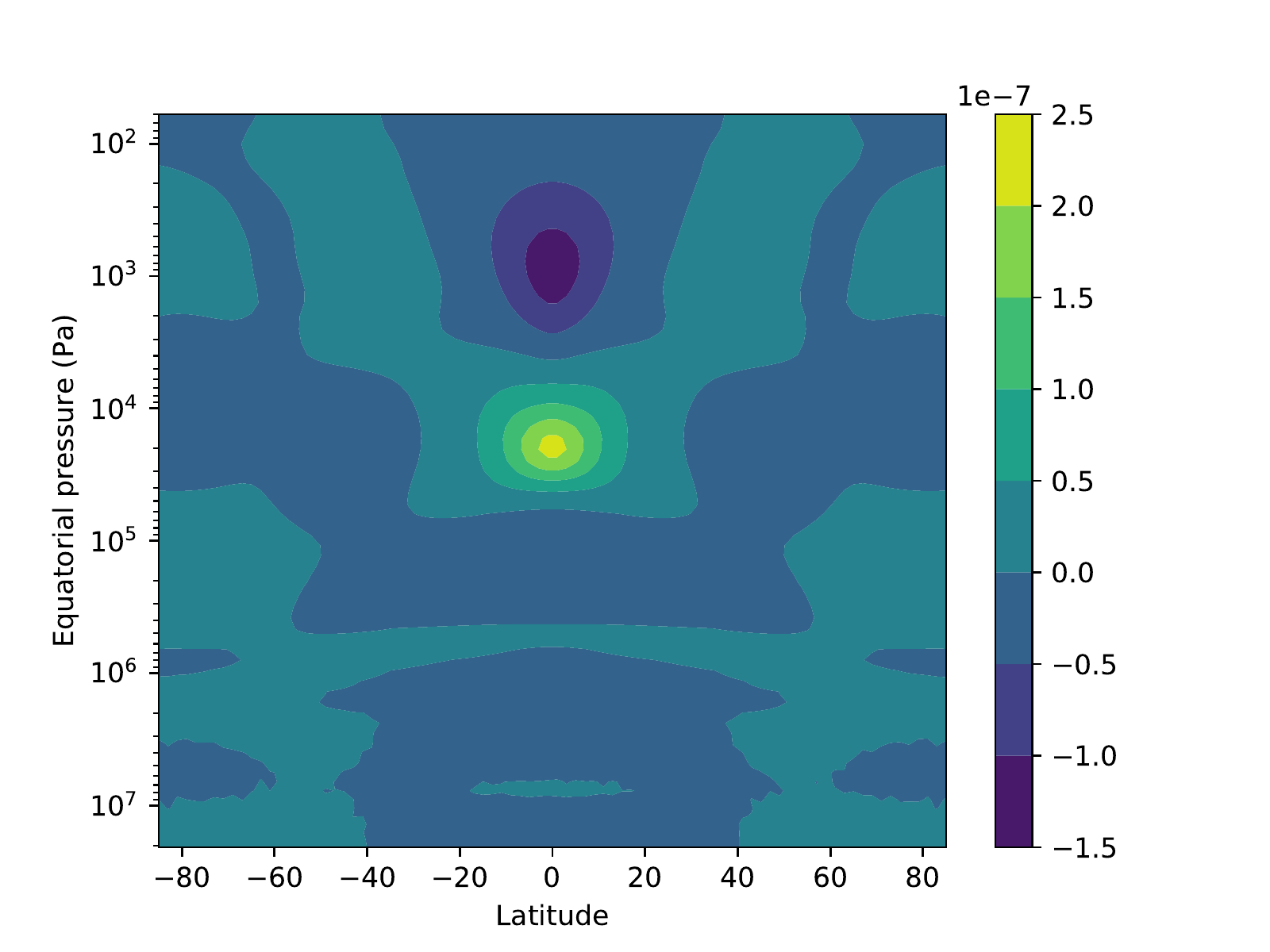}\label{fig:acc_dt100_20_ver}}
    \caption{Left column: meridional eddy accelerations as a function of pressure and latitude in the $\Delta T_\mathrm{eq,top} = 100$K case after 1 (top), 5 (middle) and 20 (bottom) days. Right column: same for vertical eddy acceleration. Units in kg.m$^{-2}$.s$^{-2}$}
    \label{fig:accelerations}
    \end{center}
\end{figure*}

We have also used these simulations to study the jet acceleration in more detail, during the earlier phases. Firstly, we observe that the jet sets up initially between $0.08$ and $0.1$ bar in about 15 days (not shown), and then extends upwards and downwards. We show in Figure \ref{fig:accelerations} the meridional and vertical accelerations after $1$, $5$ and $20$ days in the $\Delta T_\mathrm{eq,top}=100$K case. 

Figure \ref{fig:accelerations} shows that in the region where superrotation is the strongest (see Figure \ref{fig:dt100_50days_tot}), the vertical eddy acceleration always provides the maximum momentum convergence in the first 20 days, although the spatial extent of vertical acceleration decreases with time. Additionally, the location where the vertical motions accelerate the jet gets deeper with time, i.e. moves to higher pressures. We believe that this can be understood in the following way: the superrotation does not affect the mid--latitude eddy circulation, which keeps acting to converge eastward momentum at the equator. However, as radiation penetrates deeper and the jet extends, the vertical circulation is changed and the vertical winds carry momentum away from the jet. This inhibits the deposition of eastward momentum at the equator, which was accompanied by a deceleration of westward winds on the night side of the planet. Globally, it appears that both meridional and vertical accelerations set up the initial superrotation (with the vertical accelerations being dominant), which then tends to decrease and even change the sign of the vertical acceleration. Then, the global meridional motions act to sustain the jet once the vertical eddy acceleration is negative. 

A key question is whether one reaches a limiting level of the day--night temperature contrast, as a proxy for the radiative forcing, at which the superrotation would transition. Such a transition would occur once the non--linear terms become important and depend on the state of the atmosphere at that point. If the atmosphere is similar to Figure \ref{fig:MG_usual}, then superrotation would be favoured, but for a state such as \ref{fig:MG_komacek} the superrotation would be impeded. In our simulations, it seems that there is no threshold to superrotation, because the vertical accelerations spin-up the initial jet in all cases. The only limiting aspect is the time to reach a non linear superrotating state as the forcing gets lower.

\begin{figure}  
    \begin{center}
    \subfigure[]{\includegraphics[width=9.3cm,angle=0.0,origin=c]{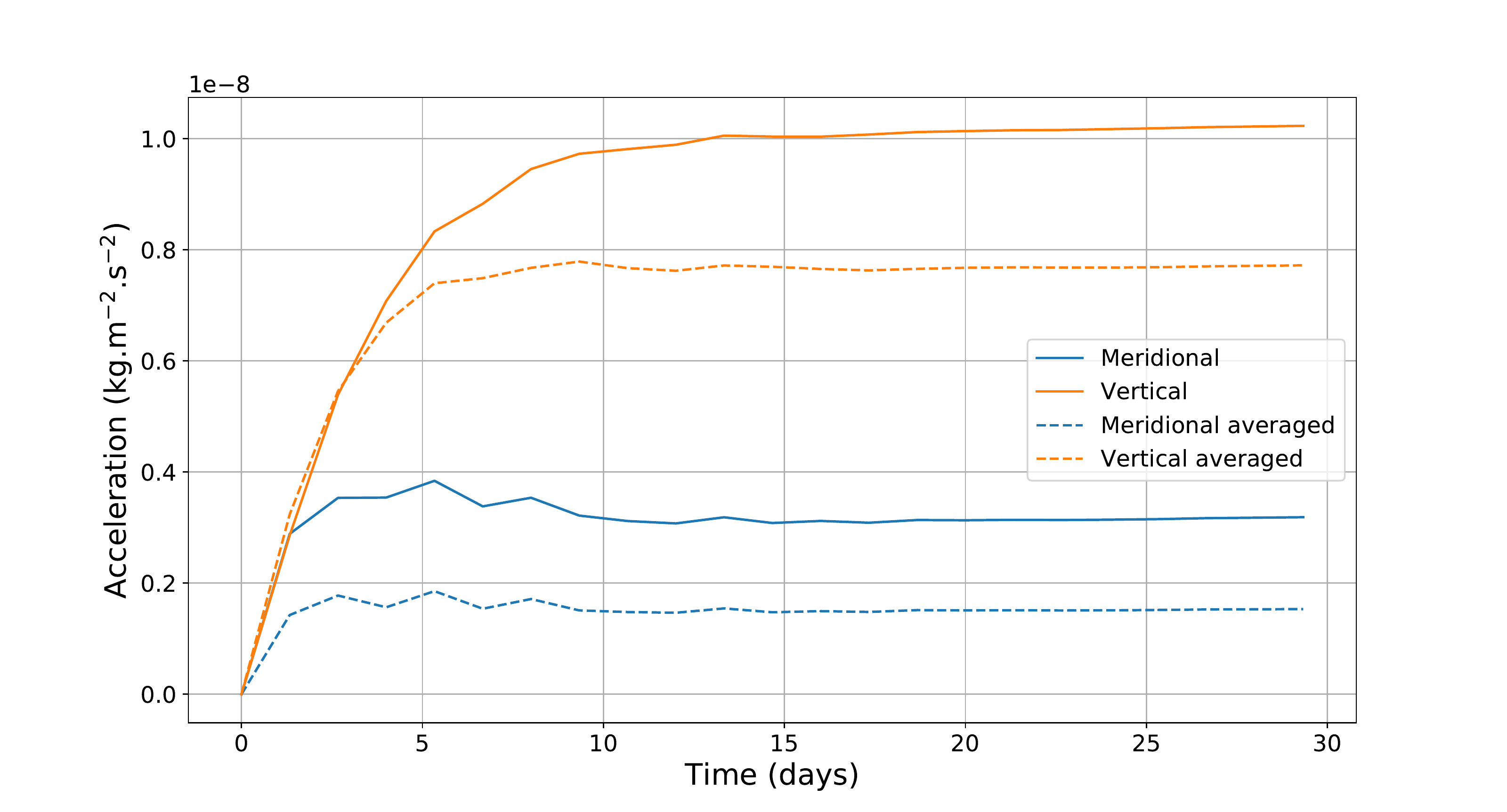}\label{fig:acc_low}}
    \subfigure[]{\includegraphics[width=9.3cm,angle=0.0,origin=c]{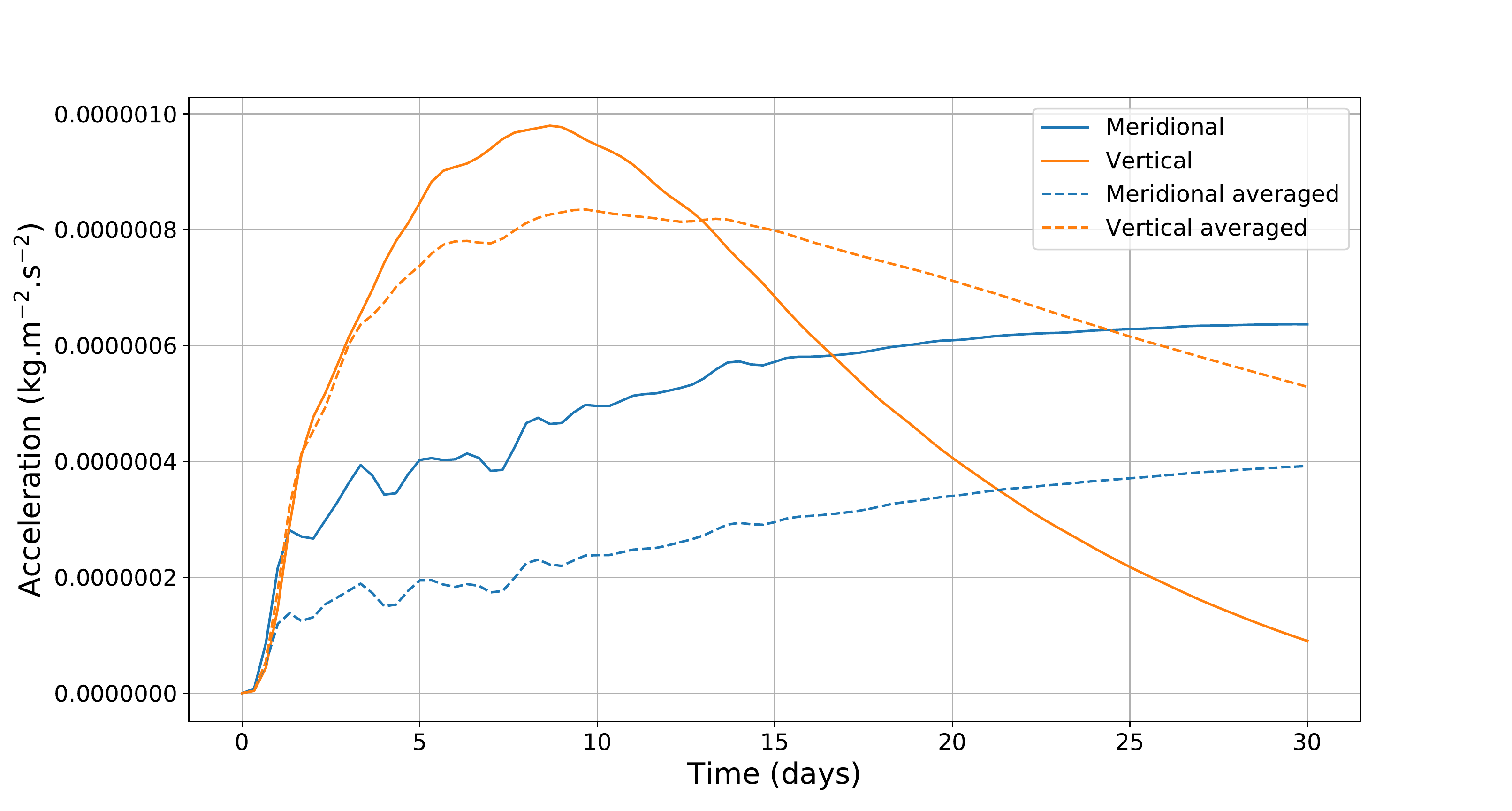}\label{fig:acc_high}}
    \caption{Eddy accelerations at the equator at the 80 mbar pressure level (plain lines) or averaged from 40mbar to 0.4bar (dashed lines)  as a function of time for (a): $\Delta T_\mathrm{eq,top} = 1$K, (b):  $\Delta T_\mathrm{eq,top} = 100$K}
    \label{fig:accelerations_lowhigh}
    \end{center}
\end{figure}

\begin{figure}  
    \begin{center}
    \subfigure[]{\includegraphics[width=8cm,height=5cm,angle=0.0,origin=c]{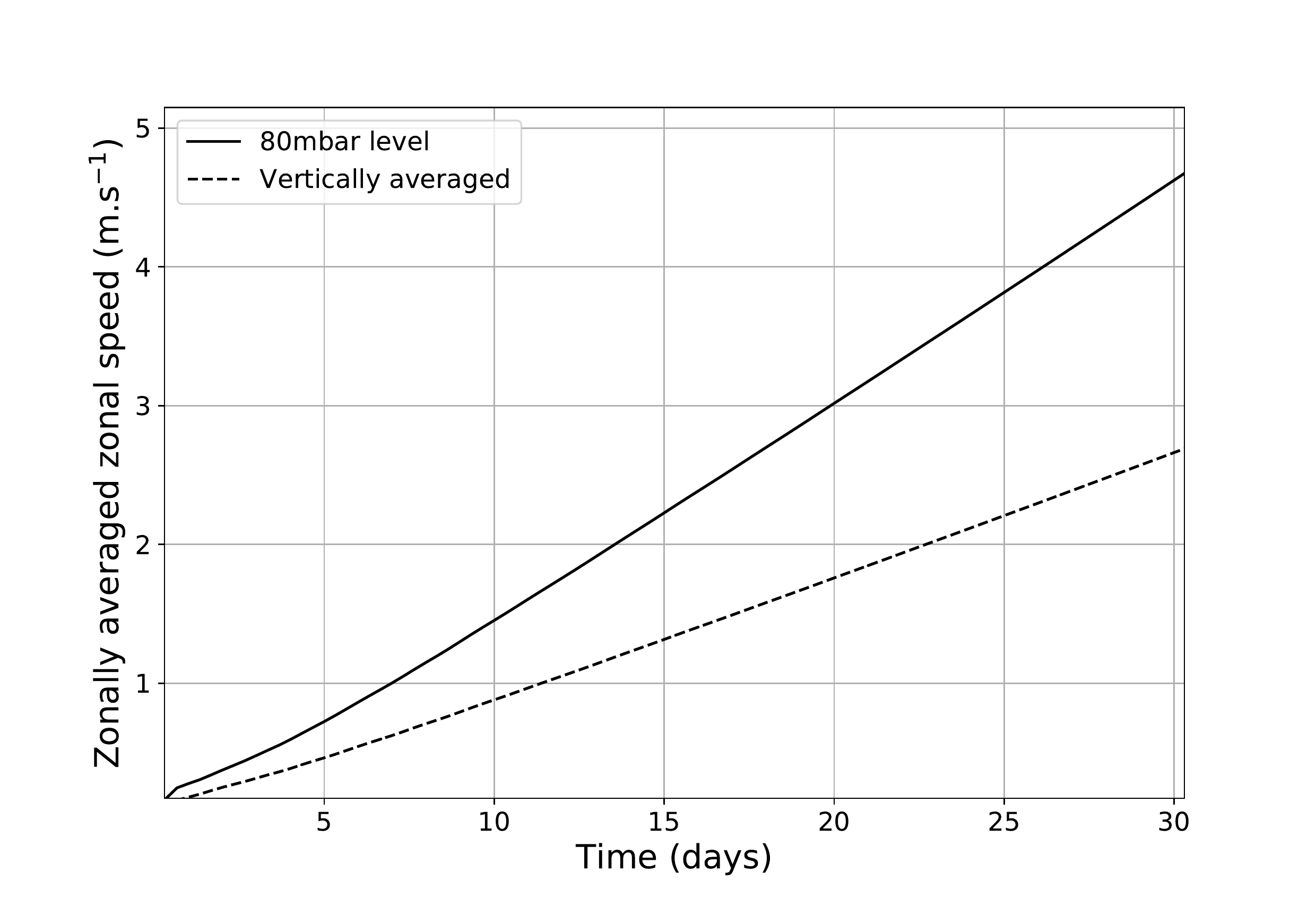}\label{fig:u_low}}
    \subfigure[]{\includegraphics[width=8cm,height=5cm,angle=0.0,origin=c]{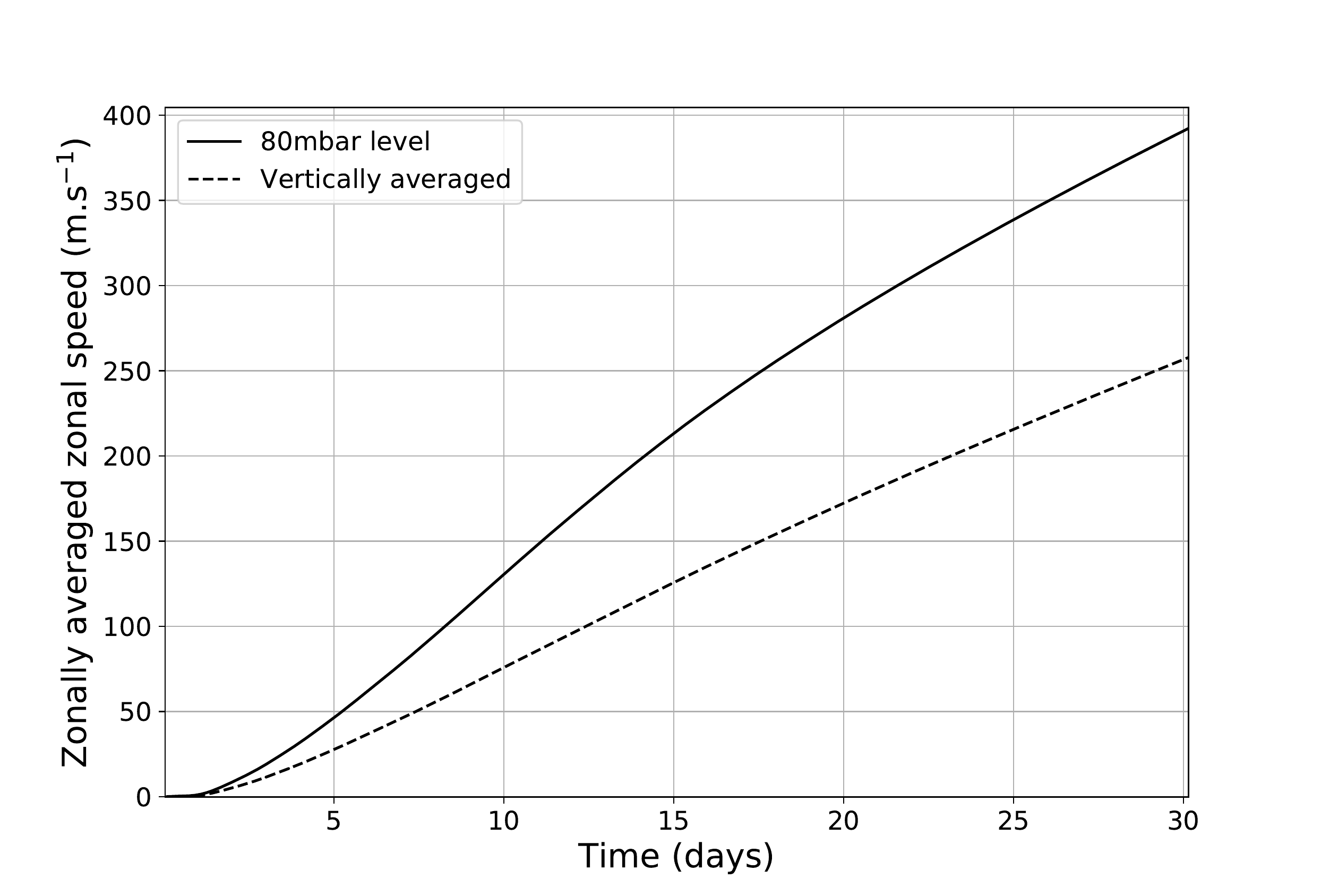}\label{fig:u_high}}
    \caption{Zonally averaged zonal speed at the 80 mbar pressure level (plain lines) or averaged from 40mbar to 0.4bar (dashed lines)  as a function of time for (a): $\Delta T_\mathrm{eq,top} = 1$K, (b):  $\Delta T_\mathrm{eq,top} = 100$K}
    \label{fig:u_lowhigh}
    \end{center}
\end{figure}

In the case of long drag and short radiative timescales  \citet{Komacek2016} have already noted that the meridional motion of the  linear steady state is opposite to what is necessary to drive superrotation, although they do observe that the non linear steady state is actually superrotating. We have explained this by considering the time dependent linear state in Section \ref{sssec:gcm_qual}. More precisely, Figure \ref{fig:accelerations_lowhigh} shows the meridional and vertical accelerations as a function of time for two simulations adopting $\Delta T_\mathrm{eq,top}= 1$K and $\Delta T_\mathrm{eq,top}= 100$K and the same dissipation. Figure \ref{fig:u_lowhigh} shows the evolution of the zonally averaged zonal wind for comparison. 


For both simulations, Figure \ref{fig:acc_low} and Figure \ref{fig:acc_high}, in the first two days meridional and vertical accelerations are both positive and of comparable magnitude. This was expected, as during the first days the atmosphere is comparable to the MG steady state explored by \citet{Showman2011}. The zonally averaged zonal speed, Figure \ref{fig:u_lowhigh}, is almost zero as expected from the propagation of waves with zonal wavenumber $m=1$. In the low forcing case, the vertical accelerations eventually dominate by almost an order of magnitude and the meridional terms can even lead to opposing the creation of a jet (for a given pressure level, not seen on the vertically averaged Figure \ref{fig:accelerations_lowhigh}). This is due to the fact that the atmosphere has reached the linear steady state of Figure \ref{fig:dt_100_r45.pdf}, associated with strong vertical accelerations but almost zero (or negative) meridional accelerations. For the high forcing case, although vertical accelerations contribute to the initiation of superrotation, they get smaller with time as the jet is being created, and eventually become negative (after $\sim$ 60 days, not shown). This confirms that the vertical accelerations initiate the jet but then tend to extend it to deeper pressure, while meridional momentum convergence allows for an equilibrated state.


Globally, we can now resolve the discrepancy of Figure 4 and Figure 5 of \citet{Komacek2016}, discussed throughout this paper (notably Section \ref{ssec:problem_komacek} and this section): nonlinearly superrotating atmospheres can have linear steady states that seem to oppose the triggering superrotation, in contradiction with \citet{Showman2011}. The study of \citet{Tsai2014}, in the limit of strong dissipation, does not offer explanation for the apparent paradox.  First, as we have shown, the treatment of the time dependent linear solution shows that the linear steady state is not relevant in the high forcing case. After 1 day, the non linear terms become dominant whereas the MG steady state would require linear effects to dominate for at least 10 days. The shape of the atmosphere after 1 day is again given by our Figure \ref{fig:MG_usual}: it is the usual chevron shape pattern of \citet{Showman2011}. Therefore, when non linear effects become dominant, they tend to accelerate the equator whereas if the linear steady state had been reached the deceleration by meridional motion could have been dominant: for adequate forcing in hot Jupiter conditions, the state of the atmosphere after 1 day always leads to meridional momentum convergence at the equator. Then, as seen in Figure \ref{fig:accelerations} and \ref{fig:accelerations_lowhigh}, the vertical accelerations also need to be taken into account: at the $80$mbar pressure range, vertical motion triggers the emergence of superrotation contrary to what is proposed by \citet{Showman2011}. Only when superrotation is settled (Figure \ref{fig:acc_dt_100_50day_ver}) do the vertical accelerations tend to decelerate the jet, and extend it deeper i.e. to higher pressures. 

Later on, once the superrotation is settled, the study of \citet{Tsai2014} applies to the slower evolution of the atmosphere, leading to the possible existence of a unique steady state. This steady state is permitted by both meridional and vertical accelerations, as we have seen throughout this work. This explains why superrotation is reached even when the dissipation is very low, although the initial acceleration is not included in the explanation of \citet{Tsai2014}. 

\section{Conclusion}
\label{sec:conclusions}

In this study we have explored the initial acceleration of superrotation in the context of a hot Jupiter atmosphere. We have also focused on an inherent discrepancy between the works of \citet{Showman2011} and \citet{Komacek2016}. \citet{Showman2011} propose that the superrotation is triggered by non linear accelerations around the linear steady state, that converge momentum to the equator. On the other hand,  \citet{Komacek2016} show that certain configurations that exhibit superrotation are also associated with momentum divergence at the equator in the linear steady state limit.

In order to resolve this apparent contradiction, we have studied the general form of the time dependent linear response of the atmosphere to a constant, asymmetrical heating. This response depends on the shape of the forcing, the global shape and frequency of the waves it generates and the decay rate of these waves. Our first conclusion, through the use of ECLIPS3D, is that changing  the longitudinal form of the forcing is not of prime importance in the qualitative understanding of superrotation, although quantitatively it does affect the results. The use of a Newtonian cooling scheme with a wavenumber 1 in longitude is therefore a reasonable approximation.

We have also obtained an equation for the frequency and decay rates of the propagating waves, as in \citet{Heng2014}. We could not solve this equation analytically for Rossby and gravity waves, and have therefore estimated the asymptotic
behaviour of the waves numerically. For Kelvin waves on the other hand, the analytical solution has been obtained. The estimated decay rates were reported in section \ref{sssec:waves_summary} and Figure \ref{fig:decay}.

From there, we have explained qualitatively the structure of the linear solutions with different drag and radiative timescales, as presented in Figure 5 of \citet{Komacek2016} and our Figure \ref{fig:MG_usual+komacek}. The zonal dependency had also been estimated by \citet{Wu2001} by other means. A major result of this present work is that in the limit of short times (compared to the damping rates), the waves present in the decomposition of the heating function contribute almost equally to the time dependent linear solution. This tends to create a Matsuno--Gill like circulation (Figure \ref{fig:MG_usual}) in the first day of the evolution of a hot Jupiter atmosphere in a GCM, although the actual linear steady state would be a "reverse" Matsuno-Gill, exhibiting eastward momentum divergence at the equator (Figure \ref{fig:MG_komacek}). With order of magnitude analysis, we have concluded that in simulations representative of hot Jupiters, the linear steady state could not be reached  but non linear terms were dominant after $\sim 1$ day, hence when the atmosphere resembles the Matsuno-Gill circulation. As a consequence, the equator is accelerated although the linear steady state would tend to decelerate the equator, resolving part of the discrepancy between \citet{Komacek2016} and \citet{Showman2011}.

Finally, we have considered the non linear accelerations during the spin up of superrotation from 3D GCM simulations with different contrasts in temperature between the day and night, or strengths of forcing, to assess the importance of the vertical accelerations. Once the jet is formed, the vertical acceleration tends to decelerate the upper part of the jet while extending it to deeper pressure. The meridional acceleration oppose this deceleration, and a steady state can be reached, as already shown by \citet{Tsai2014} and \citet{Showman2011}. On the other hand, during the acceleration, the vertical component contributes equally to the meridional component to form an initial superrotation. This is in disagreement with Figure 11 of \citet{Showman2011}, however, the data for this figure are averaged across the upper  atmosphere (above 30 mbar) where superrotation does not develop or is weaker. Numerically, it seems that as a jet is initiated, the vertical circulation is altered preventing the vertical deposition of eastward momentum at the equator, whereas the meridional circulation is roughly unchanged.

Overall, in this work, we have studied, on theoretical and semi analytical grounds the acceleration of superrotation. We have complemented previous studies to provide a coherent understanding of the initial acceleration of the equator of hot Jupiters. Combined with the works of \citet{Showman2011}, \citet{Tsai2014,Komacek2016,Hammond2018}, a somewhat complete picture of the initial phase of the atmospheric dynamics of simulated hot Jupiters can now be drawn. 

Our simulations suggest that there are regions of parameter space for which the linear steady state
does not accelerate superrotation, but the early spin-up from rest does.
This suggests that multiple long term nonlinear states might be possible \citep[as explore, e.g.,][]{Thrastarson2010,Liu2013},
depending on initial conditions, and it might be a track towards understanding peculiar observations such that of \citet{Dang2018}.

\begin{acknowledgements}
We wish to thank an anonymous referee for a very detailed report that greatly improved the argumentation of the paper. 
FD wishes to thank Antoine Venaille for his valuable insights on many questions regarding atmospheric dynamics. FD thanks the
European Research Council (ERC) for funding under the
H2020 research \& innovation programme (grant agreement
\#740651 NewWorlds). Some of the calculations for this work were performed using Met Office software. Additionally, some of calculations used the Dirac Complexity system, operated by the University of Leicester IT Services, which forms part of the STFC Dirac HPC Facility (www.Dirac.ac.uk ). This equipment is funded by BIS National E-Infrastructure capital grant ST/K000373/1 and  STFC Dirac Operations grant ST/K0003259/1. Dirac is part of the National E-Infrastructure. This research also made use of the ISCA High Performance Computing Service at the University of Exeter. NJM is part funded by a Leverhulme Trust Research Project Grant and partly supported by a Science and Technology Facilities Council Consolidated Grant (ST/R000395/1). This work is also partly supported by the ERC
grant 787361-COBOM. 
\end{acknowledgements}

\bibliography{bib_HJ}

\begin{thebibliography}{54}
\expandafter\ifx\csname natexlab\endcsname\relax\def\natexlab#1{#1}\fi

\bibitem[{Abramowitz \& Stegun(1965)}]{abramowitz1965}
Abramowitz, M. \& Stegun, I.~A. 1965, Handbook of mathematical functions: with
  formulas, graphs, and mathematical tables, Vol.~55 (Courier Corporation)

\bibitem[{Amundsen {et~al.}(2014)Amundsen, Baraffe, Tremblin,
  {et~al.}}]{Amundsen2014}
Amundsen, D., Baraffe, I., Tremblin, {et~al.} 2014, Astronomy and Astrophysics,
  564, A59

\bibitem[{{Amundsen} {et~al.}(2016){Amundsen}, {Mayne}, {Baraffe},
  {et~al.}}]{Amundsen2016}
{Amundsen}, D.~S., {Mayne}, N.~J., {Baraffe}, I., {et~al.} 2016, A \& A, 595,
  A36

\bibitem[{{Armstrong} {et~al.}(2016){Armstrong}, {de Mooij}, {Barstow},
  {Osborn}, {Blake}, \& {Saniee}}]{Armstrong2016}
{Armstrong}, D.~J., {de Mooij}, E., {Barstow}, J., {et~al.} 2016, Nature
  Astronomy, 1, 0004

\bibitem[{{Bakas} {et~al.}(2015){Bakas}, {Constantinou}, \&
  {Ioannou}}]{Bakas2015}
{Bakas}, N.~A., {Constantinou}, N.~C., \& {Ioannou}, P.~J. 2015, Journal of
  Atmospheric Sciences, 72, 1689

\bibitem[{{Bakas} \& {Ioannou}(2013)}]{Bakas2013}
{Bakas}, N.~A. \& {Ioannou}, P.~J. 2013, Journal of Atmospheric Sciences, 70,
  2251

\bibitem[{Baraffe {et~al.}(2010)Baraffe, Chabrier, \& Barman}]{Baraffe2010}
Baraffe, I., Chabrier, G., \& Barman, T. 2010, Reports on Progress in Physics,
  73, 16901

\bibitem[{{Bouchet} {et~al.}(2013){Bouchet}, {Nardini}, \&
  {Tangarife}}]{Bouchet2013}
{Bouchet}, F., {Nardini}, C., \& {Tangarife}, T. 2013, Journal of Statistical
  Physics, 153, 572

\bibitem[{{Charbonneau} {et~al.}(2002){Charbonneau}, {Brown}, {Noyes}, \&
  {Gilliland}}]{Charbonneau2002}
{Charbonneau}, D., {Brown}, T.~M., {Noyes}, R.~W., \& {Gilliland}, R.~L. 2002,
  The astrophysical journal, 568, 377

\bibitem[{{Cooper} \& {Showman}(2005)}]{Cooper2005}
{Cooper}, C.~S. \& {Showman}, A.~P. 2005, The Qstrophysical Journal letters,
  629, L45

\bibitem[{{Dang} {et~al.}(2018){Dang}, {Cowan}, {Schwartz}, {Rauscher},
  {Zhang}, {Knutson}, {Line}, {Dobbs-Dixon}, {Deming}, {Sundararajan},
  {Fortney}, \& {Zhao}}]{Dang2018}
{Dang}, L., {Cowan}, N.~B., {Schwartz}, J.~C., {et~al.} 2018, Nature Astronomy,
  2, 220

\bibitem[{{Debras} {et~al.}(2019){Debras}, {Mayne}, {Baraffe}, {Goffrey}, \&
  {Thuburn}}]{Debras2019}
{Debras}, F., {Mayne}, N., {Baraffe}, I., {Goffrey}, T., \& {Thuburn}, J. 2019,
  \aap, 631, A36

\bibitem[{Dobbs-Dixon \& Agol(2013)}]{Dobbs2013}
Dobbs-Dixon, I. \& Agol, E. 2013, Monthly Notices of the Royal Astronomical
  Society, 435, 3159

\bibitem[{{Drummond} {et~al.}(2018{\natexlab{a}}){Drummond}, {Mayne},
  {Baraffe}, {Tremblin}, {Manners}, {Amundsen}, {Goyal}, \&
  {Acreman}}]{Drummond2018}
{Drummond}, B., {Mayne}, N.~J., {Baraffe}, I., {et~al.} 2018{\natexlab{a}},
  Astronomy and Astrophysics, 612, A105

\bibitem[{{Drummond} {et~al.}(2018{\natexlab{b}}){Drummond}, {Mayne},
  {Manners}, {Baraffe}, {Goyal}, {Tremblin}, {Sing}, \&
  {Kohary}}]{Drummond2018c}
{Drummond}, B., {Mayne}, N.~J., {Manners}, J., {et~al.} 2018{\natexlab{b}},
  \apj, 869, 28

\bibitem[{{Drummond} {et~al.}(2018{\natexlab{c}}){Drummond}, {Mayne},
  {Manners}, {Carter}, {Boutle}, {Baraffe}, {H{\'e}brard}, {Tremblin}, {Sing},
  {Amundsen}, \& {Acreman}}]{Drummond2018b}
{Drummond}, B., {Mayne}, N.~J., {Manners}, J., {et~al.} 2018{\natexlab{c}}, The
  Astrophysical Journal, 855, L31

\bibitem[{{Drummond} {et~al.}(2016){Drummond}, {Tremblin}, {Baraffe},
  {et~al.}}]{Drummond2016}
{Drummond}, B., {Tremblin}, P., {Baraffe}, I., {et~al.} 2016, A \& A, 594, A69

\bibitem[{Gill(1980)}]{Gill80}
Gill, A.~E. 1980, Quarterly Journal of the Royal Meteorological Society, 106,
  447

\bibitem[{{Hammond} \& {Pierrehumbert}(2018)}]{Hammond2018}
{Hammond}, M. \& {Pierrehumbert}, R.~T. 2018, \apj, 869, 65

\bibitem[{{Hardiman} {et~al.}(2010){Hardiman}, {Andrews}, {White}, {Butchart},
  \& {Edmond}}]{Hardiman2010}
{Hardiman}, S.~C., {Andrews}, D.~G., {White}, A.~A., {Butchart}, N., \&
  {Edmond}, I. 2010, Journal of Atmospheric Sciences, 67, 1983

\bibitem[{{Helling} {et~al.}(2016){Helling}, {Lee}, {Dobbs-Dixon},
  {et~al.}}]{Helling2016}
{Helling}, C., {Lee}, G., {Dobbs-Dixon}, I., {et~al.} 2016, MNRAS, 460, 855

\bibitem[{Heng {et~al.}(2011)Heng, Menou, \& Phillipps}]{Heng2011}
Heng, K., Menou, K., \& Phillipps, P. 2011, MNRAS, 413, 2380

\bibitem[{{Heng} \& {Workman}(2014)}]{Heng2014}
{Heng}, K. \& {Workman}, J. 2014, \apjs, 213, 27

\bibitem[{{Herbert} {et~al.}(2019){Herbert}, {Caballero}, \&
  {Bouchet}}]{Herbert2019}
{Herbert}, C., {Caballero}, R., \& {Bouchet}, F. 2019, Journal of Atmospheric
  Sciences, in press

\bibitem[{{Hindle} {et~al.}(2019){Hindle}, {Bushby}, \& {Rogers}}]{Hindle2019}
{Hindle}, A.~W., {Bushby}, P.~J., \& {Rogers}, T.~M. 2019, \apjl, 872, L27

\bibitem[{{Iro} {et~al.}(2005){Iro}, {B{\'e}zard}, \& {Guillot}}]{iro2005}
{Iro}, N., {B{\'e}zard}, B., \& {Guillot}, T. 2005, Astrnonomy and
  Astrophysics, 436, 719

\bibitem[{{Knutson} {et~al.}(2007){Knutson}, {Charbonneau}, {Allen},
  {et~al.}}]{Knutson2007}
{Knutson}, H.~A., {Charbonneau}, D., {Allen}, L.~E., {et~al.} 2007, Nature,
  447, 183

\bibitem[{Komacek \& Showman(2016)}]{Komacek2016}
Komacek, T. \& Showman, A. 2016, The Astrophyiscal Journal, 821, 16

\bibitem[{{Lee} {et~al.}(2016){Lee}, {Dobbs-Dixon}, {Helling}, {Bognar}, \&
  {Woitke}}]{Lee2016}
{Lee}, G., {Dobbs-Dixon}, I., {Helling}, C., {Bognar}, K., \& {Woitke}, P.
  2016, A \& A, 594, A48

\bibitem[{{Lines} {et~al.}(2018{\natexlab{a}}){Lines}, {Manners}, {Mayne},
  {Goyal}, {Carter}, {Boutle}, {Lee}, {Helling}, {Drummond}, {Acreman}, \&
  {Sing}}]{Lines2018b}
{Lines}, S., {Manners}, J., {Mayne}, N.~J., {et~al.} 2018{\natexlab{a}},
  \mnras, 481, 194

\bibitem[{{Lines} {et~al.}(2018{\natexlab{b}}){Lines}, {Mayne}, {Boutle},
  {Manners}, {Lee}, {Helling}, {Drummond}, {Amundsen}, {Goyal}, {Acreman},
  {Tremblin}, \& {Kerslake}}]{Lines2018}
{Lines}, S., {Mayne}, N.~J., {Boutle}, I.~A., {et~al.} 2018{\natexlab{b}},
  \aap, 615, A97

\bibitem[{{Liu} \& {Showman}(2013)}]{Liu2013}
{Liu}, B. \& {Showman}, A.~P. 2013, \apj, 770, 42

\bibitem[{Matsuno(1966)}]{Matsuno}
Matsuno, T. 1966, Journal of the Meteorological Society of Japan, 44, 25

\bibitem[{Mayne {et~al.}(2014)Mayne, Baraffe, Acreman, {et~al.}}]{Mayne2014}
Mayne, N., Baraffe, I., Acreman, D., {et~al.} 2014, Geoscientific Model
  Development, 7, 3059

\bibitem[{{Mayne} {et~al.}(2014){Mayne}, {Baraffe}, {Acreman}, {Smith},
  {Browning}, {Sk{\aa}lid Amundsen}, {Wood}, {Thuburn}, \&
  {Jackson}}]{Mayne2014b}
{Mayne}, N.~J., {Baraffe}, I., {Acreman}, D.~M., {et~al.} 2014, Astronomy and
  Astrophysics, 561, A1

\bibitem[{{Mayne} {et~al.}(2017){Mayne}, {Debras}, {Baraffe}, {Thuburn},
  {Amundsen}, {Acreman}, {Smith}, {Browning}, {Manners}, \& {Wood}}]{Mayne2017}
{Mayne}, N.~J., {Debras}, F., {Baraffe}, I., {et~al.} 2017, A \& A, 604, A79

\bibitem[{Mayor \& Queloz(1995)}]{Mayor}
Mayor, M. \& Queloz, D. 1995, Nature, 378, 355

\bibitem[{{Perez-Becker} \& {Showman}(2013)}]{Perez2013}
{Perez-Becker}, D. \& {Showman}, A.~P. 2013, The Astrophysical Journal, 776,
  134

\bibitem[{Rauscher \& Menou(2012)}]{Rauscher2012}
Rauscher, E. \& Menou, K. 2012, The Astrophysical Journal, 750, 96

\bibitem[{{Roman} \& {Rauscher}(2019)}]{Roman2019}
{Roman}, M. \& {Rauscher}, E. 2019, \apj, 872, 1

\bibitem[{Showman {et~al.}(2008)Showman, Cooper, Fortney, \&
  Marley}]{Showman2008}
Showman, A., Cooper, C., Fortney, J., \& Marley, M. 2008, The Astrophyisical
  Journal, 682, 559

\bibitem[{{Showman} \& {Polvani}(2011)}]{Showman2011}
{Showman}, A.~P. \& {Polvani}, L.~M. 2011, The Astrophysical Journal, 738, 71

\bibitem[{{Sing} {et~al.}(2008){Sing}, {Vidal-Madjar}, {Lecavelier des Etangs},
  {et~al.}}]{Sing2008}
{Sing}, D.~K., {Vidal-Madjar}, A., {Lecavelier des Etangs}, A., {et~al.} 2008,
  The astrophysical journal, 686, 667

\bibitem[{{Snellen} {et~al.}(2008){Snellen}, {Albrecht}, {de Mooij}, \& {Le
  Poole}}]{Snellen2008}
{Snellen}, I.~A.~G., {Albrecht}, S., {de Mooij}, E.~J.~W., \& {Le Poole}, R.~S.
  2008, A \& A, 487, 357

\bibitem[{{Srinivasan} \& {Young}(2012)}]{Srivinasan2012}
{Srinivasan}, K. \& {Young}, W.~R. 2012, Journal of Atmospheric Sciences, 69,
  1633

\bibitem[{{Thrastarson} \& {Cho}(2010)}]{Thrastarson2010}
{Thrastarson}, H.~T. \& {Cho}, J. Y.-K. 2010, \apj, 716, 144

\bibitem[{Thuburn {et~al.}(2002)Thuburn, Wood, \& Staniforth}]{Thuburn}
Thuburn, J., Wood, N., \& Staniforth, A. 2002, Quarterly Journal of the Royal
  Meteorological Society, 128, 1771

\bibitem[{{Tobias} \& {Marston}(2013)}]{Tobias2013}
{Tobias}, S.~M. \& {Marston}, J.~B. 2013, Physical Review Letters, 110, 104502

\bibitem[{{Tremblin} {et~al.}(2017){Tremblin}, {Chabrier}, {Mayne}, {Amundsen},
  {Baraffe}, {Debras}, {Drummond}, {Manners}, \& {Fromang}}]{Tremblin2017}
{Tremblin}, P., {Chabrier}, G., {Mayne}, N.~J., {et~al.} 2017, \apj, 841, 30

\bibitem[{{Tsai} {et~al.}(2014){Tsai}, {Dobbs-Dixon}, \& {Gu}}]{Tsai2014}
{Tsai}, S.-M., {Dobbs-Dixon}, I., \& {Gu}, P.-G. 2014, The Astrophysical
  Journal, 793, 141

\bibitem[{{Tsai} {et~al.}(2018){Tsai}, {Kitzmann}, {Lyons}, {Mendon{\c{c}}a},
  {Grimm}, \& {Heng}}]{Tsai2017}
{Tsai}, S.-M., {Kitzmann}, D., {Lyons}, J.~R., {et~al.} 2018, \apj, 862, 31

\bibitem[{{Wu} {et~al.}(2000){Wu}, {Sarachik}, \& {Battisti}}]{Wu2000}
{Wu}, Z., {Sarachik}, E.~S., \& {Battisti}, D.~S. 2000, Journal of Atmospheric
  Sciences, 57, 2169

\bibitem[{{Wu} {et~al.}(2001){Wu}, {Sarachik}, \& {Battisti}}]{Wu2001}
{Wu}, Z., {Sarachik}, E.~S., \& {Battisti}, D.~S. 2001, Journal of Atmospheric
  Sciences, 58, 724

\bibitem[{{Zellem} {et~al.}(2014){Zellem}, {Lewis}, {Knutson},
  {et~al.}}]{Zellem2014}
{Zellem}, R.~T., {Lewis}, N.~K., {Knutson}, H.~A., {et~al.} 2014, The
  Astrophysical Journal, 790, 53

\end{thebibliography}

\appendix

\section{Non orthogonality in the case $\tau_\mathrm{rad} \neq \tau_\mathrm{drag}$}
\label{app:orthogonality}

In the case $\tau_\mathrm{rad} = \tau_\mathrm{drag}$, \citet{Matsuno} obtained 
a simple equation linking $u_{n,l}$ and $v_{n,l}$, the horizontal speeds 
of the eigenvector (n, l), from Eqs.\eqref{eq:matsuno_u} and \eqref{eq:matsuno_h}:

\begin{equation}
u_{n,l} = \dfrac{\omega_{n,l} y v_{n,l} + k \partial v_{n,l}/\partial y}{i (\omega_{n,l} - k)(\omega_{n,l} +k)}\,\, .
\label{eq:u_ortho}
\end{equation}
A similar equation is obtained for $h_{n,l}$.
As $v_{n,l} \propto \psi_n(y) = H_n(y) e^{-y^2/2}$, with $H_n$ the n$^{th}$ Hermite polynomial, 
the use of the recurrence relations of the Hermite polynomials: 

\begin{gather}
\dfrac{\mathrm{d} H_n(y)}{\mathrm{d} y} = 2 n H_{n-1}(y),\\
H_{n+1}(y) = 2y H_n(y) - 2n H_{n-1}(y),
\end{gather}
implies that the eigenvector (n,l) is simply written as:

\begin{equation}
\begin{pmatrix}
v_{n,l} \\
u_{n,l} \\
h_{n,l} \\
\end{pmatrix} = 
\begin{pmatrix}
\mathrm{i} (\omega_{n,l}^2-k^2) \psi_n(y) \\
\dfrac{1}{2} (\omega_{n,l}-k) \psi_{n+1}(y) + n (\omega_{n,l}+k) \psi_{n-1}(y) \\
\dfrac{1}{2} (\omega_{n,l}-k) \psi_{n+1}(y) - n (\omega_{n,l}+k) \psi_{n-1}(y)
\end{pmatrix} \,\, .
\end{equation}
The orthogonality of the eigenvectors is easily proven, and the completeness
is proved by use of the completeness of the Hermite polynomial functions. 

In the case $\tau_\mathrm{rad} \neq  \tau_\mathrm{drag}$, Eq.\eqref{eq:u_ortho}
is slightly changed to get: 

\begin{equation}
u_{n,l} = \dfrac{(\mathrm{i}\omega_{n,l}+ \tau_\mathrm{rad}^{-1}) y v_{n,l} + \mathrm{i} k \partial v_{n,l}/\partial y}
{k^2 + (\mathrm{i}\omega_{n,l}+ \tau_\mathrm{rad}^{-1})(\mathrm{i}\omega_{n,l}+ \tau_\mathrm{drag}^{-1})}, 
\label{eq:u_ortho_modified}
\end{equation}
and we recall that $v_{n,l} \propto H_n(c_{n,l}y) e^{-c_{n,l}^2 y^2/2}$ where 
\begin{equation}
c_{n,l}^4 = \dfrac{\mathrm{i} \omega_{n,l}+ \tau_\mathrm{rad}^{-1}}{\mathrm{i} \omega_{n,l}+ \tau_\mathrm{drag}^{-1}}\,\, .
\end{equation}

Therefore, the expression of the eigenvector (n,l) is: 
\begin{equation}
\begin{pmatrix}
v_{n,l} \\
u_{n,l} \\
h_{n,l} \\
\end{pmatrix} = 
\begin{pmatrix}
(k^2 + (\mathrm{i}\omega_{n,l}+ \tau_\mathrm{rad}^{-1})(\mathrm{i}\omega_{n,l}+ \tau_\mathrm{drag}^{-1}))\psi_n(c_{n,l}y) \\
\dfrac{1}{2} \left(\dfrac{\mathrm{i}\omega_{n,l} +  \tau_\mathrm{rad}^{-1}}{c_{n,l}} -\mathrm{i} k c_{n,l} \right) 
\psi_{n+1}(c_{n,l}y) + 
n \left(\dfrac{\mathrm{i}\omega_{n,l} +  \tau_\mathrm{rad}^{-1}}{c_{n,l}} +\mathrm{i} k c_{n,l} \right) 
\psi_{n-1}(c_{n,l}y) \\
\dfrac{1}{2} \left(c_{n,l}(\mathrm{i}\omega_{n,l} +  \tau_\mathrm{drag}^{-1}) -\dfrac{\mathrm{i} k} {c_{n,l}} \right) 
\psi_{n+1}(c_{n,l}y) - 
n \left(c_{n,l}(\mathrm{i}\omega_{n,l} +  \tau_\mathrm{drag}^{-1}) +\dfrac{\mathrm{i} k}{ c_{n,l}} \right) \psi_{n-1}(c_{n,l}y) 
\end{pmatrix} \,\, .
\label{eq:shape_degueu}
\end{equation}

Because of the dependency with $c_{n,l}$ in the parabolic cylinder function, the eigenvectors of Eq.\ref{eq:shape_degueu} are not orthogonal
anymore (the calculation is cumbersome but with no difficulty). 
On the other hand, it is easily shown that the set of eigenvectors (n,l) are linearly
independent (thanks to the Hermite polynomials). A rigorous proof would be needed to
assess that they form a complete set, allowing 
for a projection of the heating function onto these eigenvectors. 
A graphical representation of these waves is provided in Appendix \ref{app:waves}.

\section{Solutions to Eq.(40)}
\label{app:pol}
\subsection{The argument principle}

\begin{figure} [hb] 
    \begin{center}
    \subfigure{\includegraphics[height=8.8cm]{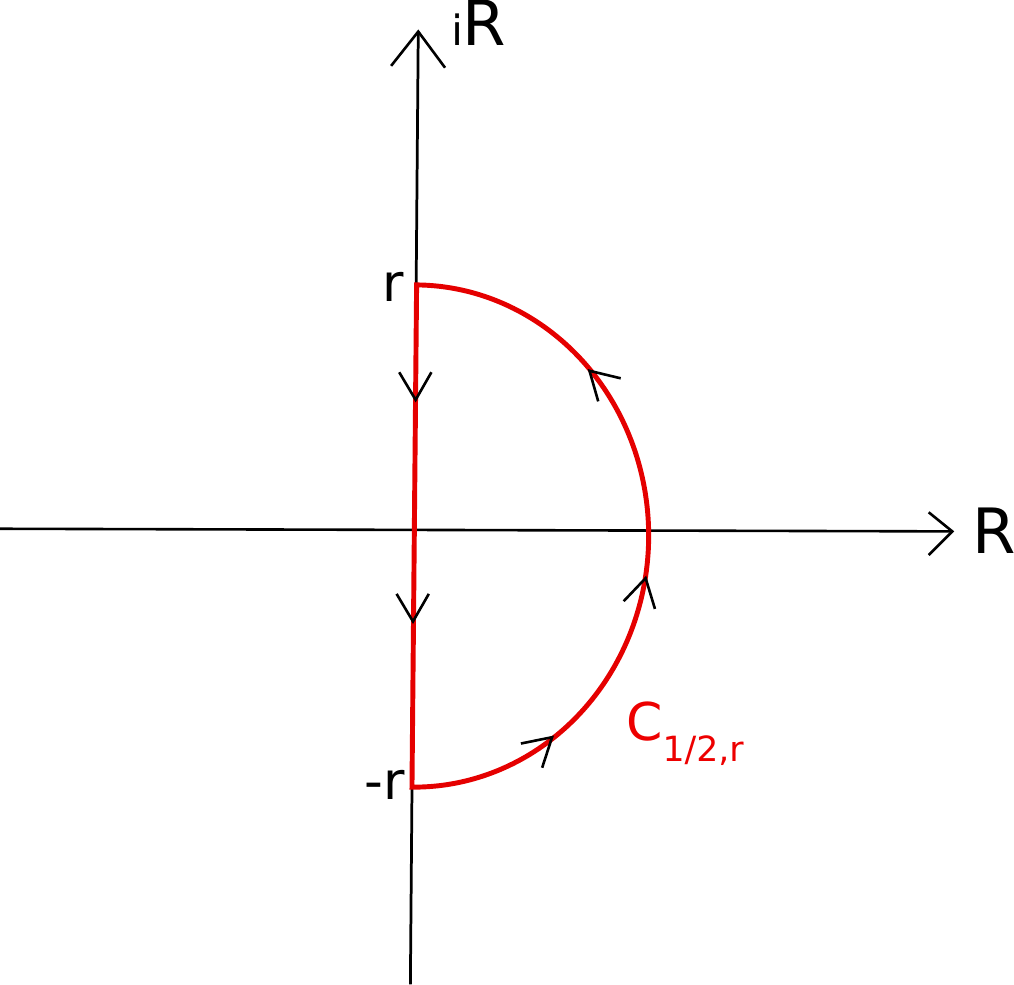}}
    \caption{Graphical representation of the $C_{1/2,r}$ contour, encompassing the complex numbers with positive real part when $r \rightarrow \infty$.}
    \label{fig:contour}
    \end{center}
\end{figure}

Left hand side of Eq.\eqref{eq:odd} may be confused with a second order polynomial, but the 
dependency of $c$ with $\omega$ actually leads to a polynomial of order 6. From Eq.\eqref{eq:odd} and Eq.\eqref{eq:defc4}, we obtain an equation for $X = c^2$ :

\begin{gather}
    -X^6 + \left( \dfrac{2 \mathrm{i} (2n +1) \Delta \tau}{k} \right) X^5  + (3 + 2 \mathrm{i} k \Delta \tau)X^4 - \left( \dfrac{4 \mathrm{i} (2n +1) \Delta \tau}{k} \right) X^3 \nonumber \\
    + \left(-3 + \dfrac{8 \mathrm{i} \Delta \tau^3}{k} - 4\mathrm{i}k \Delta \tau\right)X^2
    + \left( \dfrac{2 \mathrm{i} (2n +1) \Delta \tau}{k} \right)X + (1 + 2 \mathrm{i} k \Delta \tau) = 0 \ ,
    \label{eq:pol_c2}
\end{gather}
with $\Delta \tau =(\tau_\mathrm{drag}^{-1}-\tau_\mathrm{rad}^{-1})/2$. Determining the number of propagating waves therefore consists in determining the number of roots of Eq.\eqref{eq:pol_c2} with a positive real part, as explained in section \ref{sec:waves}. We will make use of the argument principle: denoting $P(X)$ the polynomial in $X$
of Eq.\eqref{eq:pol_c2}, the number $N$ of roots of $P$ in a domain $K$ is given by :

\begin{equation}
    N(\gamma) = \dfrac{1}{2 \mathrm{i} \pi} \oint_{\vec{\gamma}}{ \dfrac{P'(z)}{P(z)}\mathrm{d}z} \ ,
    \label{eq:principle}
\end{equation}
where $\vec{\gamma}$ is a positively oriented contour encompassing $K$. Denoting $C_{1/2,r}$ the semi--circle of radius $r$, centered on the origin and cut by the pure imaginary line (Figure \ref{fig:contour}), 
the number of zeros of $P$ with positive real part is given by Eq.\eqref{eq:principle} with $\gamma =  C_{1/2,r}$ in the limit $r \rightarrow \infty$. 

The right hand side of Eq.\eqref{eq:principle} can be estimated by decomposing the integral on the pure imaginary line and on the circle of radius $r$. Namely:

\begin{equation}
    2 \mathrm{i} \pi N(C_{1/2,r}) = \int_{r}^{-r} \dfrac{P'(\mathrm{i}t)}{P(\mathrm{i}t)}\mathrm{i}\mathrm{d}t +
    \int_{-\pi/2}^{\pi/2}\dfrac{P'(r\ \exp(\mathrm{i}\theta))}{P(r\ \exp(\mathrm{i}\theta))}r\mathrm{i}\exp(\mathrm{i}\theta)\mathrm{d}\theta \ .
    \label{eq:number_root}
\end{equation}

For large r, the second term of this expression can be calculated thanks to the asymptotical expansion of a polynomial of order 6: 

\begin{equation}
\dfrac{P'(r\ \exp(\mathrm{i}\theta))}{P(r\ \exp(\mathrm{i}\theta))} = 
\dfrac{6}{r\ \exp(\mathrm{i}\theta)} + \mathcal{O}\left(\dfrac{1}{r^2}\right) \ .
\end{equation}
Hence the second term in the right hand side of Eq.\eqref{eq:number_root} tends to $6 \mathrm{i} \pi$ when $r \rightarrow \infty$. 

The first term deserves further consideration. If the image of the imaginary numbers by $P$ does not cross a given half line $\mathbb{D}$  with origin at $z=0$, said otherwise $P(\mathrm{i\mathbb{R}}) \subset \mathbb{C} \setminus \mathbb{D}$, we can define a holomorphic function $\mathrm{Ln}$ such that $\mathrm{Ln}(\exp(z)) = \exp(\mathrm{Ln}(z)) = z$ and in that case: 

\begin{align}
    \int_{r}^{-r} \dfrac{P'(\mathrm{i}t)}{P(\mathrm{i}t)}\mathrm{i}\mathrm{d}t &= \int_{r}^{-r} \dfrac{\mathrm{d}}{\mathrm{d}t} \mathrm{Ln}(P(\mathrm{i}t))\mathrm{d}t  \nonumber \\ &= \mathrm{Ln}  \left(\dfrac{P(-\mathrm{i}r)}{P(\mathrm{i}r)}\right) \rightarrow_{r \rightarrow \infty} \mathrm{Ln}(1) = 0 \ ,
\end{align}
where the limit comes from the fact that $P$ is an even degree polynomial. In that case, our two expressions for the terms of the right hand side of Eq.\eqref{eq:number_root} would yield:
\begin{equation}
    N(C_{1/2,\infty}) = 3 \ ,
\end{equation}
confirming that Eq.\eqref{eq:polynomial_6} has exactly three roots of positive real part. We are therefore going to show that, except for $n=0$, the image of the imaginary numbers by $P$ is always included in $\mathbb{C} \setminus i \mathbb{R}^-$ or $\mathbb{C} \setminus i \mathbb{R}^+$, where $i \mathbb{R}^-$ and $i \mathbb{R}^+$ are the imaginary numbers with negative and positive imaginary part respectively. 

\subsection{Image of the polynomial}

With further calculation one can show that the polynomial $P$ from Eq.\eqref{eq:pol_c2} can be written: 

\begin{equation}
    P(\mathrm{i} Y) = (1+Y^2)^2 (Y^2-2 \alpha_n Y +1) + 2 \mathrm{i} k \Delta \tau (Y^4 -2 (2 \alpha_0^2-1)Y^2+1)\ ,
\end{equation}
for $Y \in \mathbb{R}$ and with $\alpha_n = (2n+1)\Delta \tau /k$. Hence we can write the imaginary part of $P(\mathrm{i}Y)$ as : 

\begin{equation}
    \Im (P(\mathrm{i}Y)) = 2 k \Delta \tau (Y^2-2 \alpha_0 Y +1)(Y^2+2 \alpha_0 Y +1) \ .
\end{equation}
There are two cases: 
\begin{itemize}
    \item if $|\alpha_0| =|\Delta \tau /k| < 1$, the imaginary part of $P(\mathrm{i}Y)$ has no roots for $Y \in \mathbb{R}$. Hence for all $n$, $P(\mathrm{i}\mathbb{R}) \subset \mathbb{C} \setminus i \mathbb{R}^-$ or $P(\mathrm{i}\mathbb{R}) \subset \mathbb{C} \setminus i \mathbb{R}^+$ depending on the sign of $\Delta \tau$, and we have the result: for all $n$, there are exactly three roots of positive real part of Eq.\eqref{eq:polynomial_6}, hence only three waves propagate.
    
    \item For $|\alpha_0| =|\Delta \tau /k| > 1$, the real part of the polynomial is:
\begin{equation}
    \Re (P(\mathrm{i}Y)) = (1+Y^2)^2 (Y^2-2 \alpha_n Y +1)\ ,
\end{equation}
which has two real roots:
\begin{equation}
     y_n^\pm = \alpha_n \pm \sqrt{\alpha_n^2-1}\ .
\end{equation}
Calculating the imaginary part of $P$ for $Y=y_n^\pm$ yields:
\begin{equation}
    \Im(P(\mathrm{i}y_n^\pm)) = (2 k \Delta \tau) (\alpha_n^2-\alpha_0^2) (4 {y_n^\pm}^2) \ .
\end{equation}
Hence, for $n>0$ the imaginary part of $P(\mathrm{i}y_n^\pm)$ is always of the same sign as $\Delta \tau$: for all $y$, $P(\mathrm{i}y)$ cannot be an imaginary number with a negative (resp. positive) imaginary part if $\Delta \tau$ is positive (resp. negative). Therefore $P(\mathrm{i}\mathbb{R}) \subset \mathbb{C} \setminus i \mathbb{R}^-$ (resp. $P(\mathrm{i}\mathbb{R}) \subset \mathbb{C} \setminus i \mathbb{R}^+$) and we have the same result: for $n>0$, there are exactly three roots of positive real part of Eq.\eqref{eq:polynomial_6}.
\end{itemize}


We therefore have confirmed that for $|\Delta \tau /k|<1$, only three waves propagate for all $n$ and that this results holds for $|\Delta \tau /k|>1$ and $n>0$. The case $|\Delta \tau /k|>1$ and $n=0$ is more complicated as when the real part of $P(\mathrm{i}y)$ cancels its imaginary part cancels as well and we can't apply the same argument as for $n>0$. However, it means that the polynomial $P$ for n=0, $P_0$, can be easily factorized:

\begin{equation}
    P_0(\mathrm{i}Y) = (Y^2-2 \alpha_0 Y +1)\bigg\{(1+Y^2)^2 + 2 \mathrm{i} k \Delta \tau (Y^2 + 2 \alpha_0 Y +1)\bigg\}\ ,
\end{equation}
which yields: 
\begin{align}
    P_0(Y) &= (-Y^2 + 2 \mathrm{i} \alpha_0 Y +1)\bigg\{(1-Y^2)^2 + 2 \mathrm{i} k \Delta \tau (-Y^2 - 2 \mathrm{i}\alpha_0 Y +1)\bigg\} \nonumber \\ 
    &\equiv (-Y^2 + 2 \mathrm{i} \alpha_0 Y +1) Q(Y) \ .
\end{align}
Looking for the roots of $P_0$ with positive real part is therefore equivalent to looking for the roots of $Q$ with positive real part. If we can show that $Q(\mathrm{i}\mathbb{R})\subset \mathbb{C} \setminus i \mathbb{R}^-$ or $Q(\mathrm{i}\mathbb{R})\subset \mathbb{C} \setminus i \mathbb{R}^+$, then we can use the argument principle on $Q$, a polynomial of order 4, and Eq.\eqref{eq:number_root} will ensure that $P$ has only 2 roots with positive real part. This is easily proven by looking at the real part of $Q$: 

\begin{equation}
    \Re(Q(\mathrm{i}Y)) = (1+Y^2)^2 > 0\ ,
\end{equation}
hence for all $y \in \mathbb{R}$ the real part of $Q(\mathrm{i}y)$ never cancels hence $Q(\mathrm{i}\mathbb{R})\subset \mathbb{C} \setminus i \mathbb{R}$: $P$ has only two roots of positive real part and only two waves can propagate. 

It might seem surprising to have a different behaviour for $n=0$, but this was already obtained by \citet{Matsuno} where only two of the three solutions of the $n=0$ case were actual solutions of the linearised equations of motion. When considering that $\tau_\mathrm{drag} \neq \tau_\mathrm{rad}$, this degeneracy is removed when $|\Delta \tau /k|>1$ where only two roots of the polynomial have positive real part. These findings have been tested and confirmed numerically.

\newpage

\section{Waves when $\tau_\mathrm{rad},\tau_\mathrm{drag} \ne 0$}
\label{app:waves}
\begin{figure}  
    \begin{center}
    \subfigure[]{\includegraphics[width=10cm,angle=0.0,origin=c]{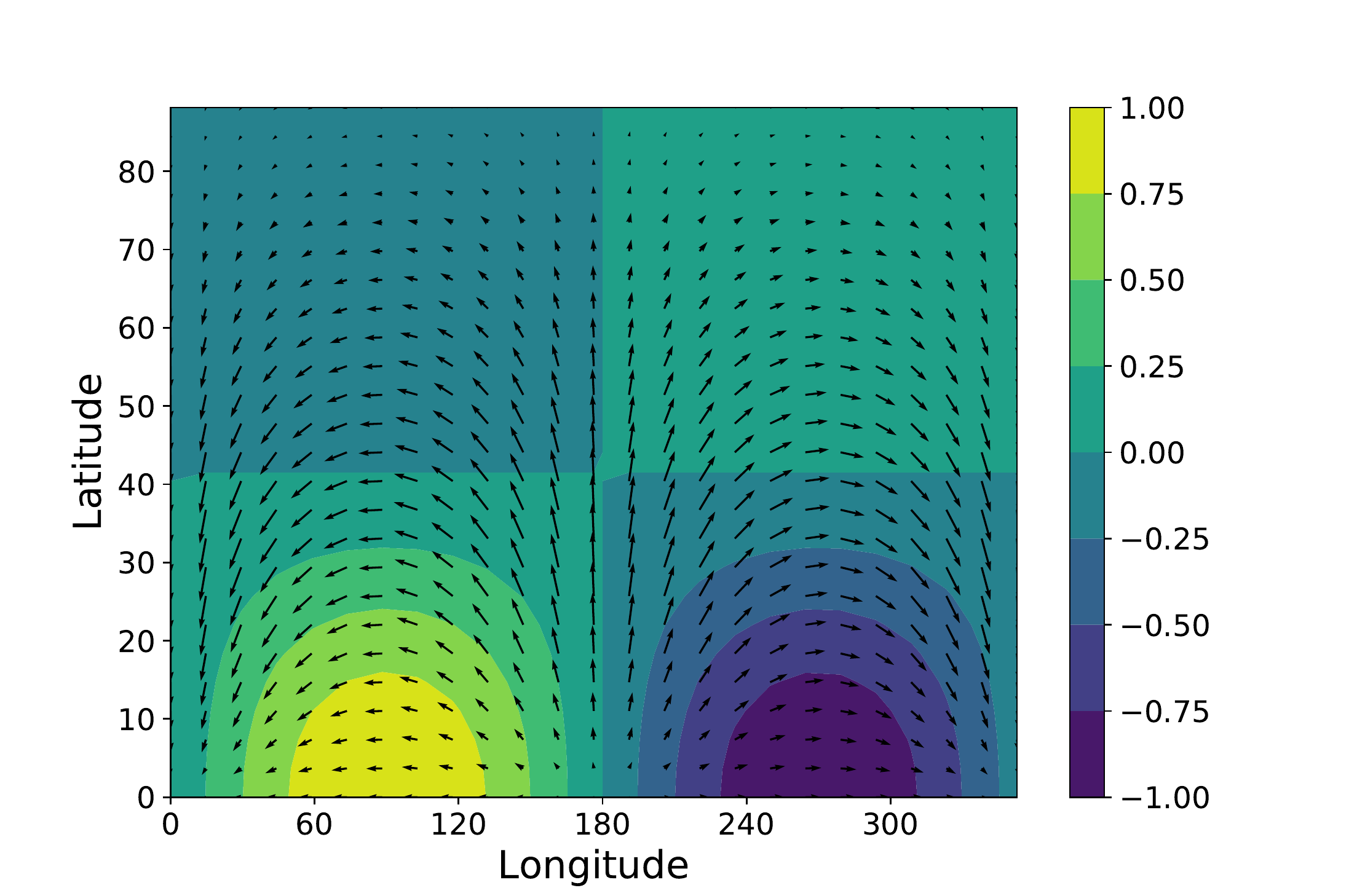}}
    \subfigure[]{\includegraphics[width=10cm,height=6.5cm,angle=0.0,origin=c]{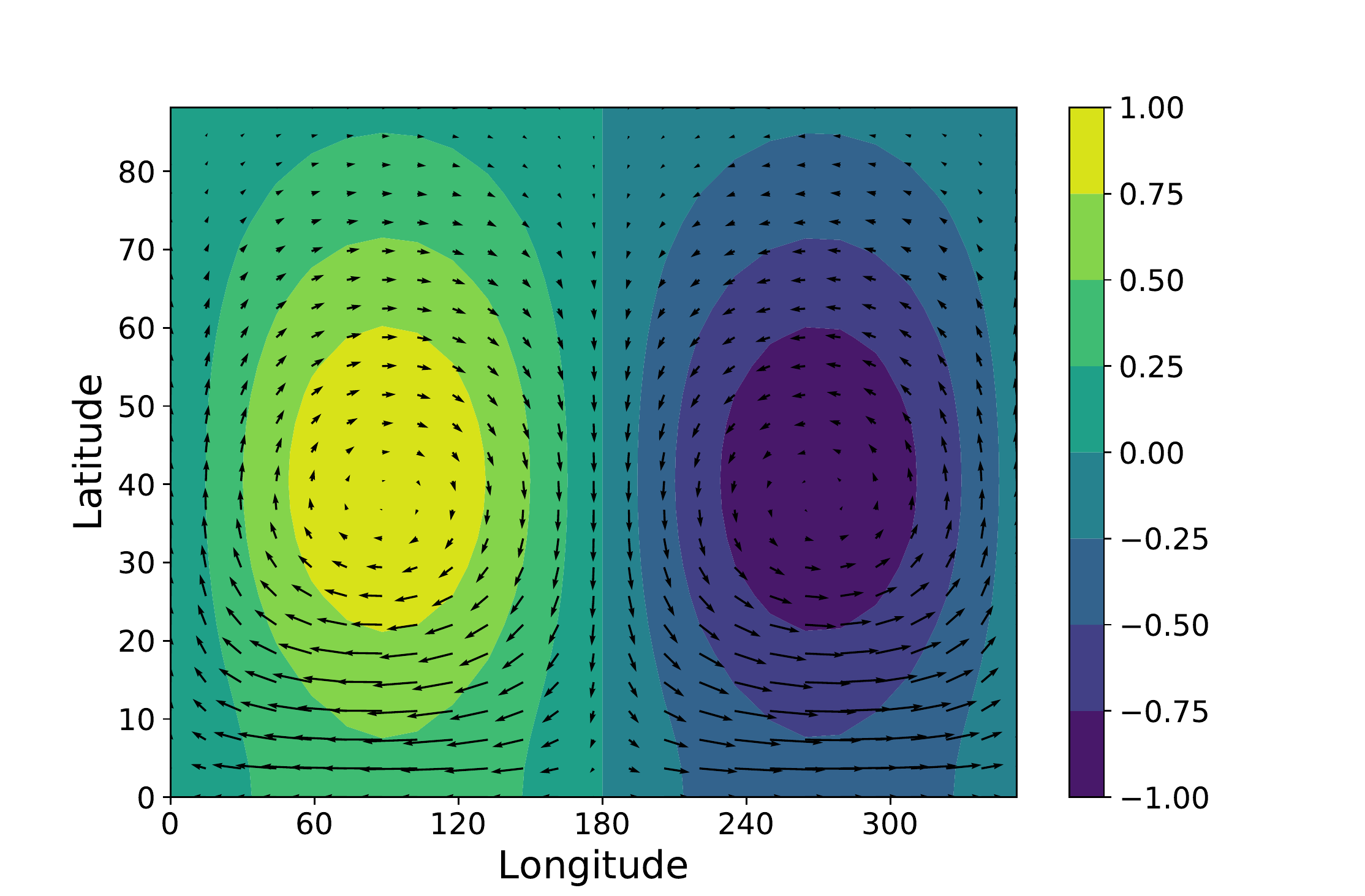}} 
    \subfigure[]{\includegraphics[width=10cm,angle=0.0,origin=c]{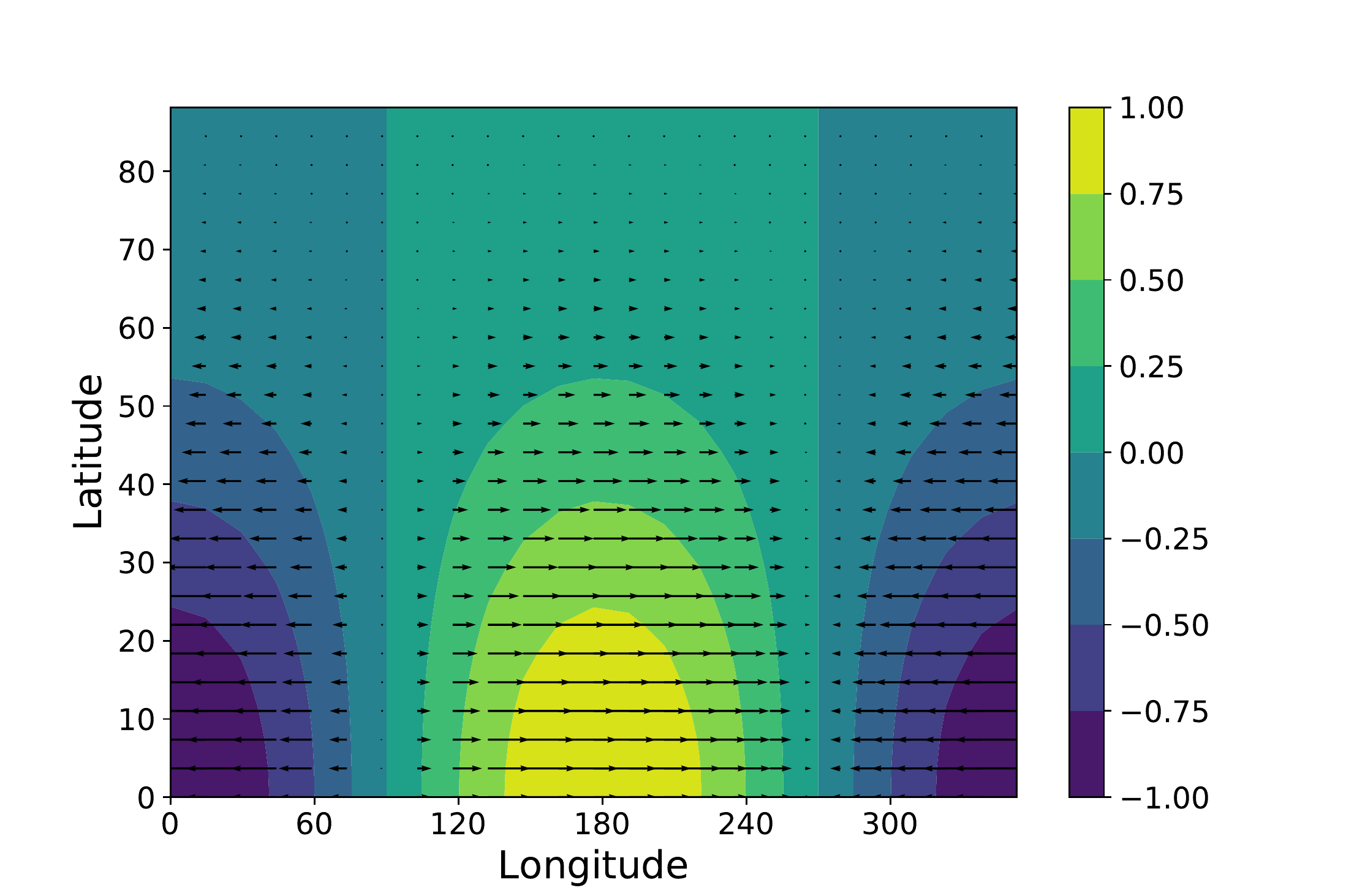}}
    \caption{Temperature (colors) and winds (arows), both in arbitrary unit, as a function of longitude and latitude for the shallow water solutions of 
    Eqs. \eqref{eq:polynomial_6} and \eqref{eq:shape_degueu} considering $\tau_\mathrm{rad}=\tau_\mathrm{drag}=2.8$ which is $\sim 10^5$\ s in dimensional units. (a) Westward propagating gravity wave (b) Rossby wave and (c) Kelvin wave.  } 
    \label{fig:waves_eq}
    \end{center}
\end{figure}

This appendix intends to show the change in the structure of the waves when $\tau_\mathrm{rad},\tau_\mathrm{drag} \ne 0$ from the numerical solutions of equations \ref{eq:polynomial_6} and \ref{eq:shape_degueu}. Figure \ref{fig:waves_eq} shows a Rossby, gravity and Kelvin wave when $\tau_\mathrm{rad} = \tau_\mathrm{drag} = 0.35$ which corresponds to $\sim 10^5$\ s in dimensional units. Their shape is similar to the waves studied by \citet{Matsuno}, as the only difference is in the decay timescale which is not zero but equal to $\tau_\mathrm{drag}$. 

\begin{figure}  
    \begin{center}
    \subfigure[]{\includegraphics[width=10cm,angle=0.0,origin=c]{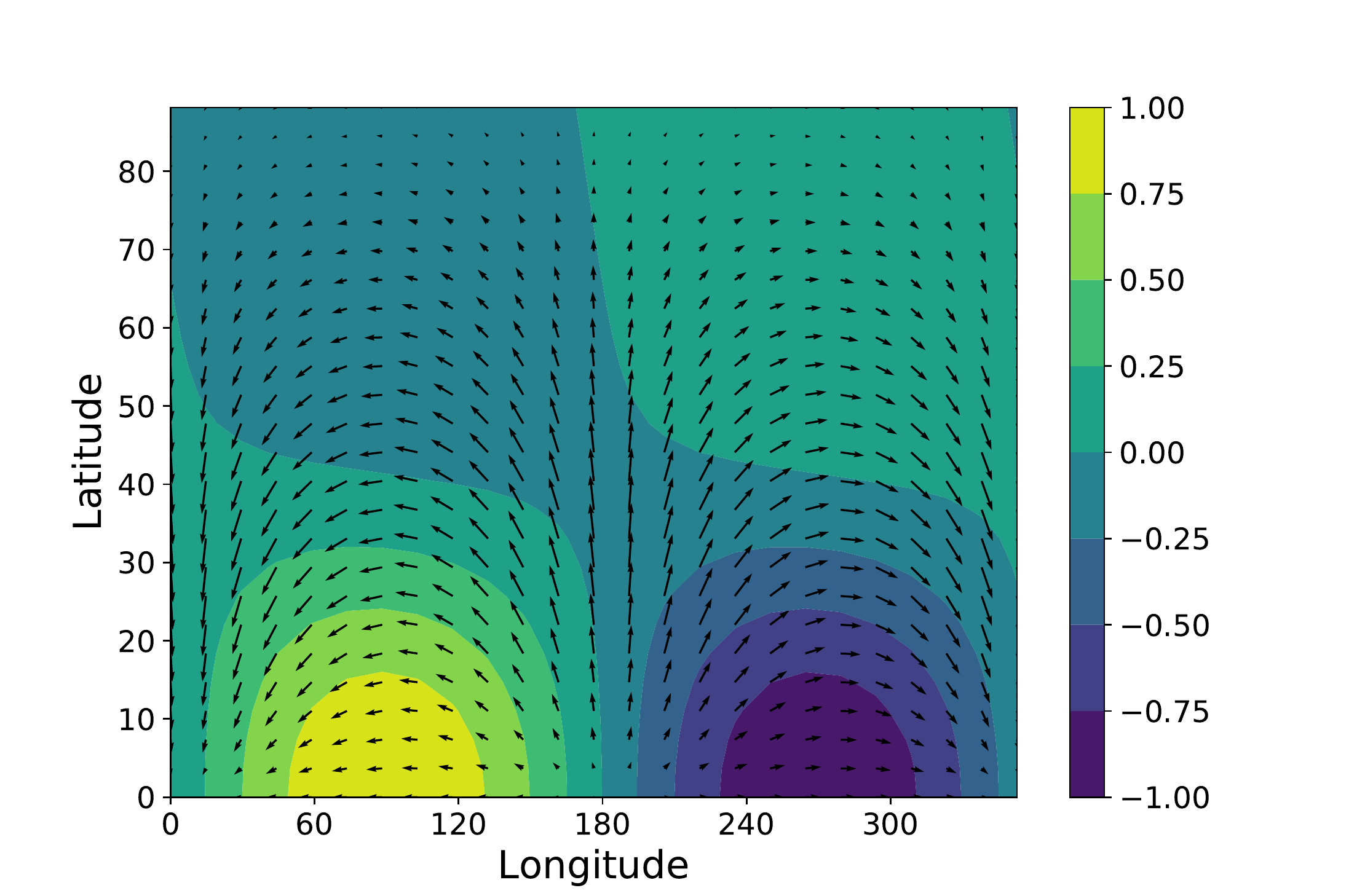}}
    \subfigure[]{\includegraphics[width=10cm,height=6.5cm,angle=0.0,origin=c]{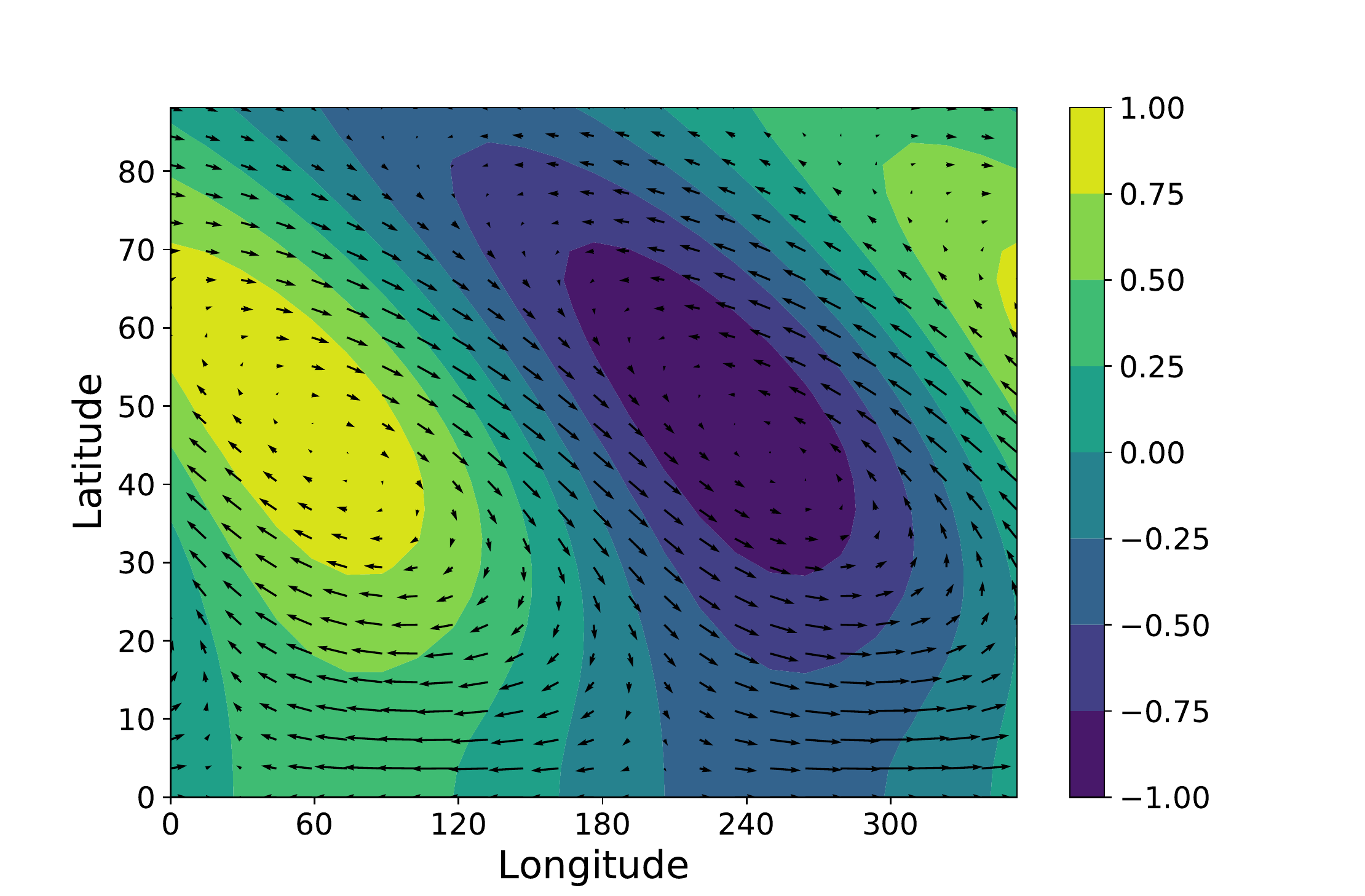}} 
    \subfigure[]{\includegraphics[width=10cm,angle=0.0,origin=c]{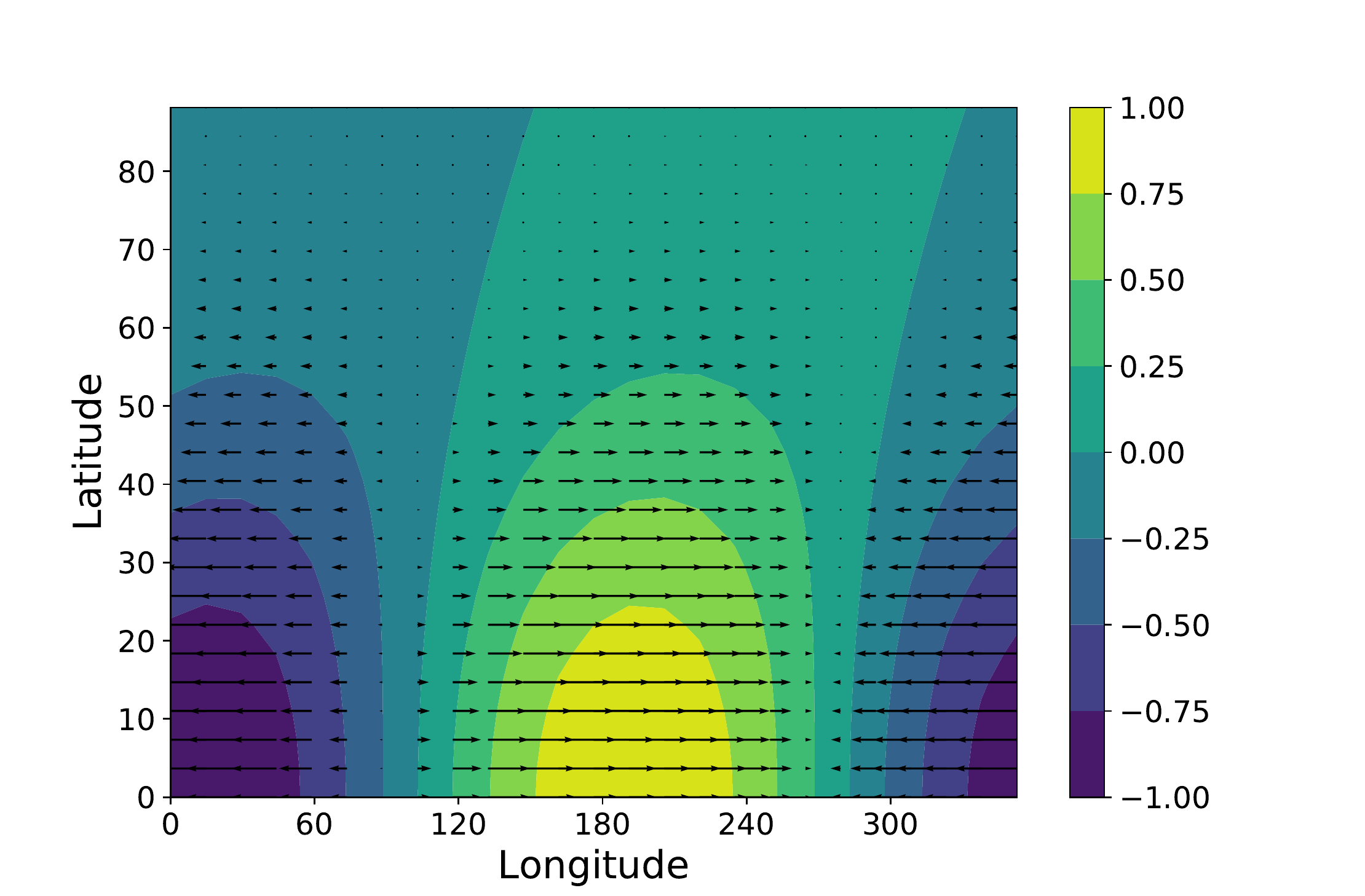}}
    \caption{Temperature (colors) and winds (arows), both in arbitrary unit, as a function of longitude and latitude for the shallow water solutions of 
    Eqs. \eqref{eq:polynomial_6} and \eqref{eq:shape_degueu} considering $\tau_\mathrm{rad}=2.8$ and $\tau_\mathrm{drag} = 28$. (a) Westward propagating gravity wave (b) Rossby wave and (c) Kelvin wave.  } 
    \label{fig:waves_rad}
    \end{center}
\end{figure}

\begin{figure}  
    \begin{center}
    \subfigure[]{\includegraphics[width=10cm,angle=0.0,origin=c]{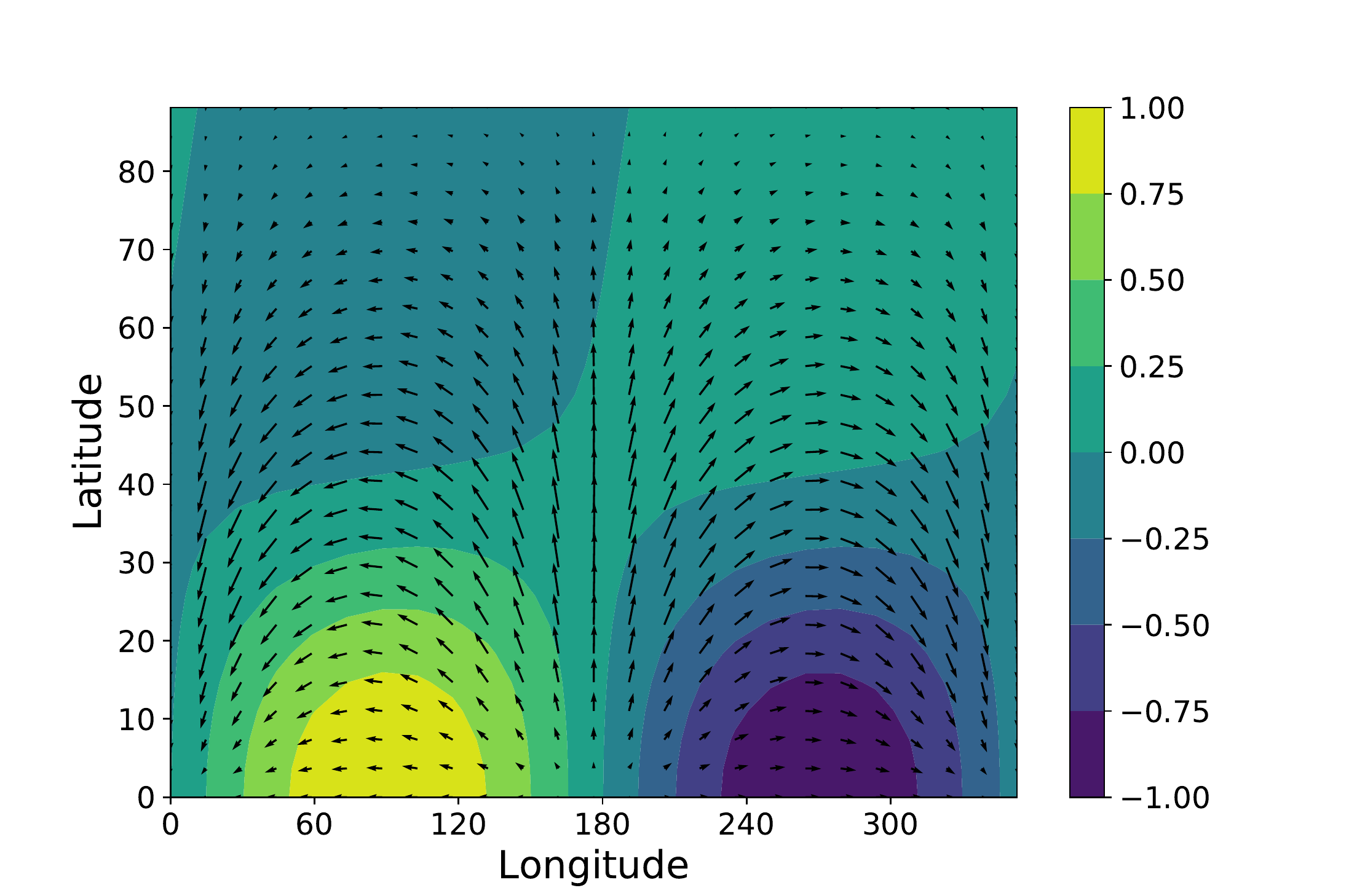}}
    \subfigure[]{\includegraphics[width=10cm,height=6.5cm,angle=0.0,origin=c]{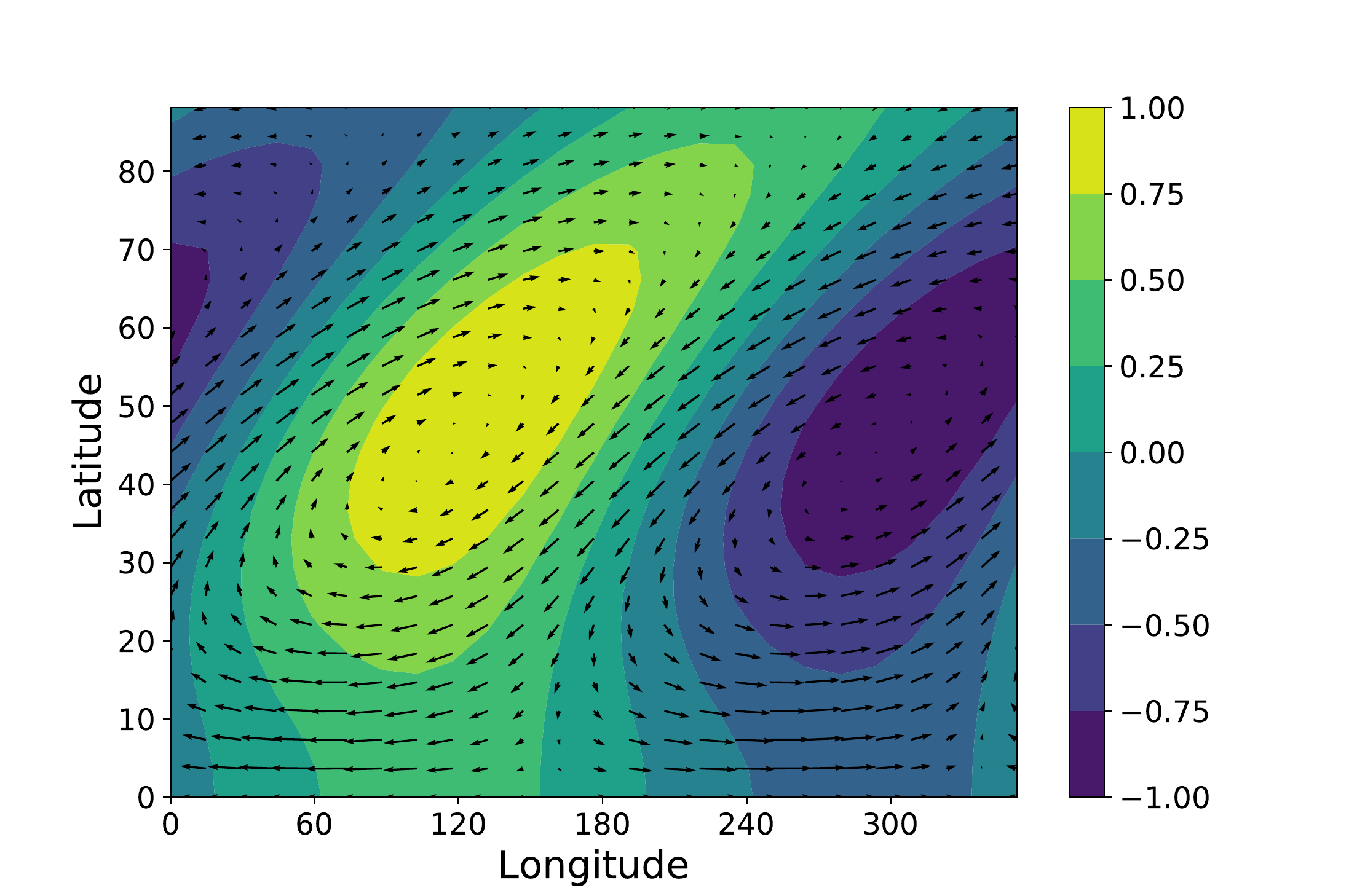}} 
    \subfigure[]{\includegraphics[width=10cm,angle=0.0,origin=c]{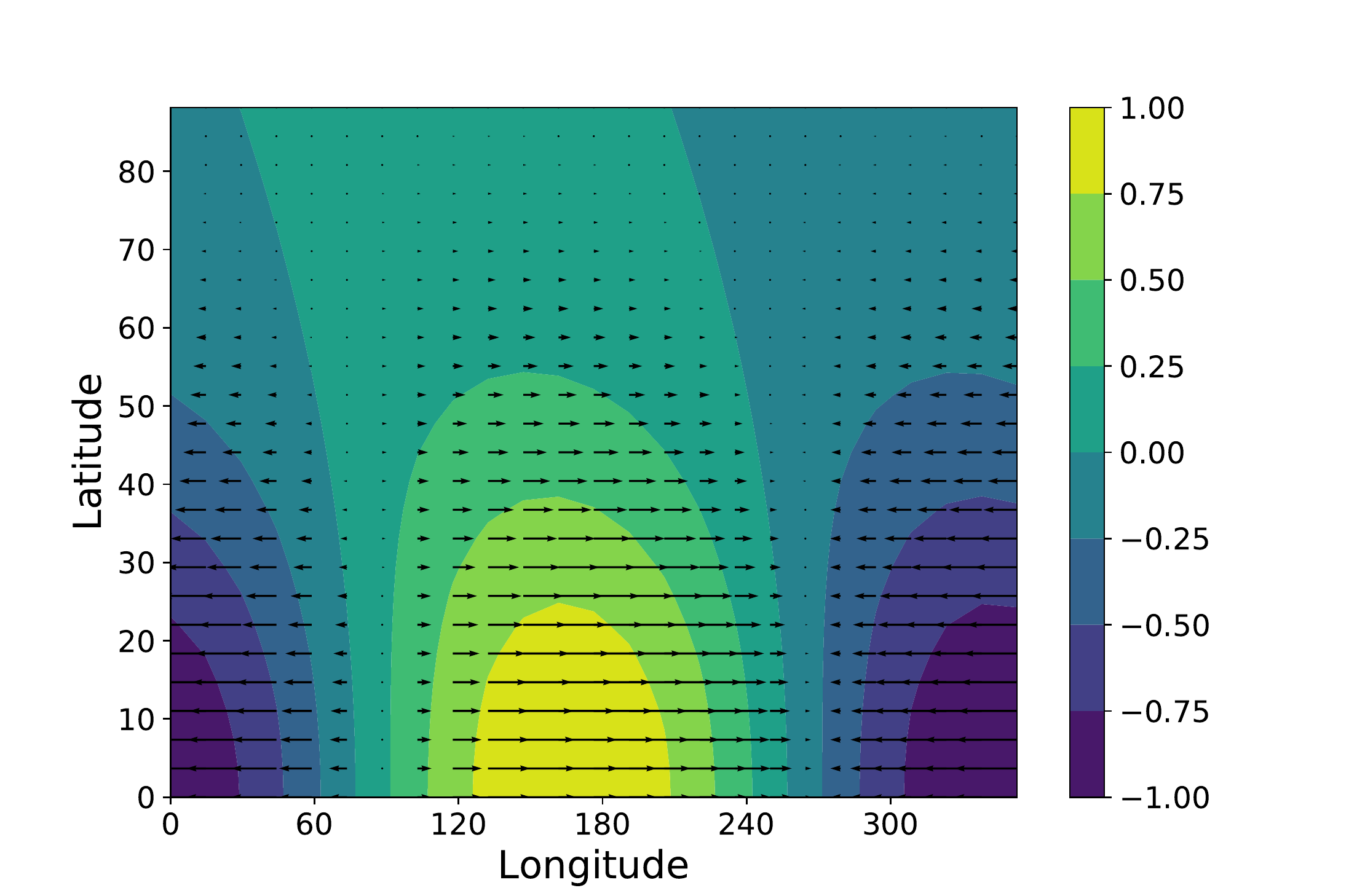}}
    \caption{Temperature (colors) and winds (arows), both in arbitrary unit, as a function of longitude and latitude for the shallow water solutions of 
    Eqs. \eqref{eq:polynomial_6} and \eqref{eq:shape_degueu} considering $\tau_\mathrm{rad}=28$ and $\tau_\mathrm{drag} = 2.8$. (a) Westward propagating gravity wave (b) Rossby wave and (c) Kelvin wave. } 
    \label{fig:waves_drag}
    \end{center}
\end{figure}

Then, Figure \ref{fig:waves_rad} and \ref{fig:waves_drag} shows the same waves when the drag or radiative timescales has been multiplied by ten, respectively. The shape of the waves is almost unaltered for gravity and Kelvin waves, apart from a tilt of the perturbation with latitude.  Rossby waves on the other hand are more affected by the change in the drag and radiative timescale.

\begin{figure}  
    \begin{center}
    \subfigure[]{\includegraphics[width=10cm,angle=0.0,origin=c]{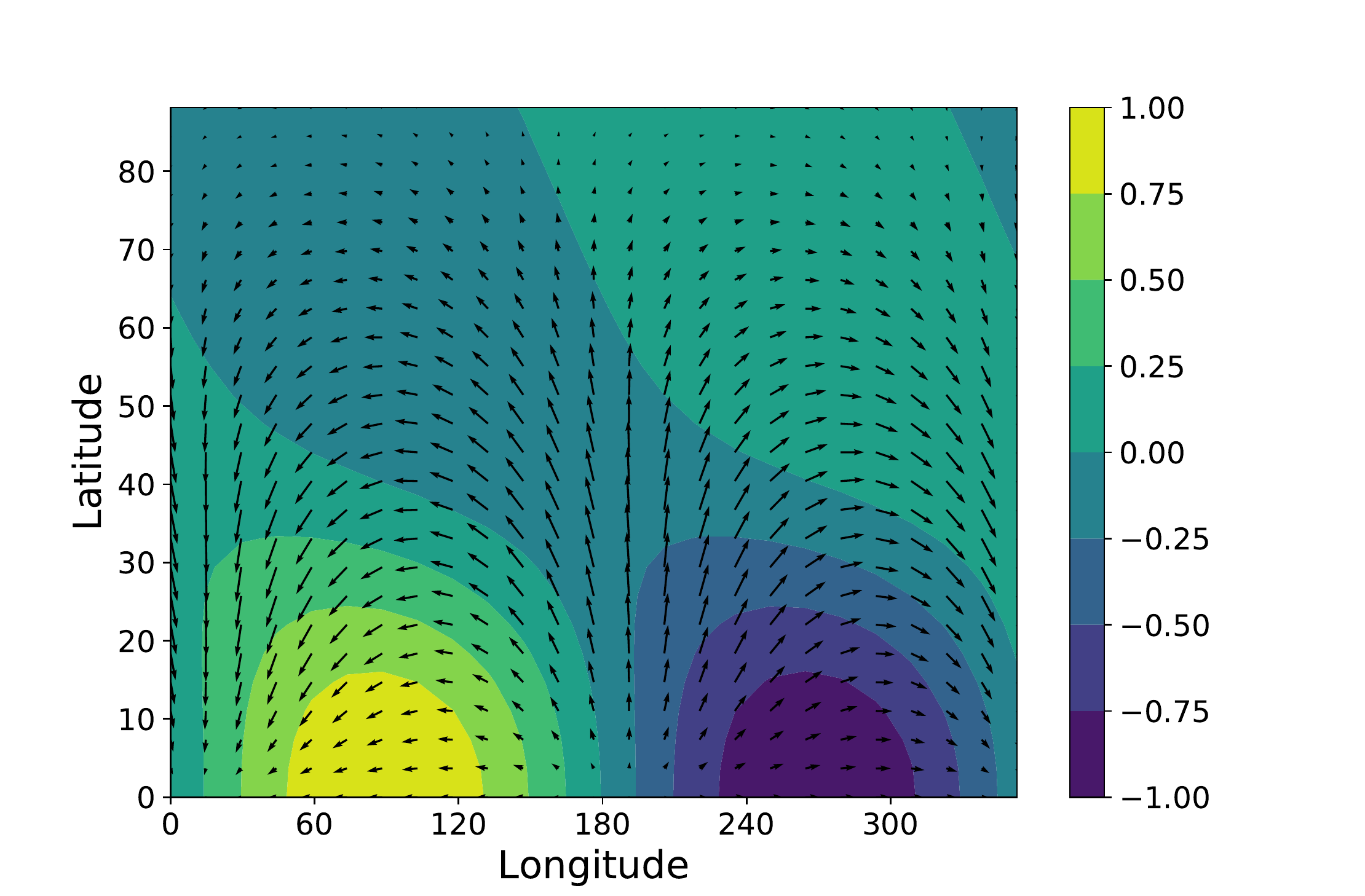}}
    \subfigure[]{\includegraphics[width=10cm,height=6.5cm,angle=0.0,origin=c]{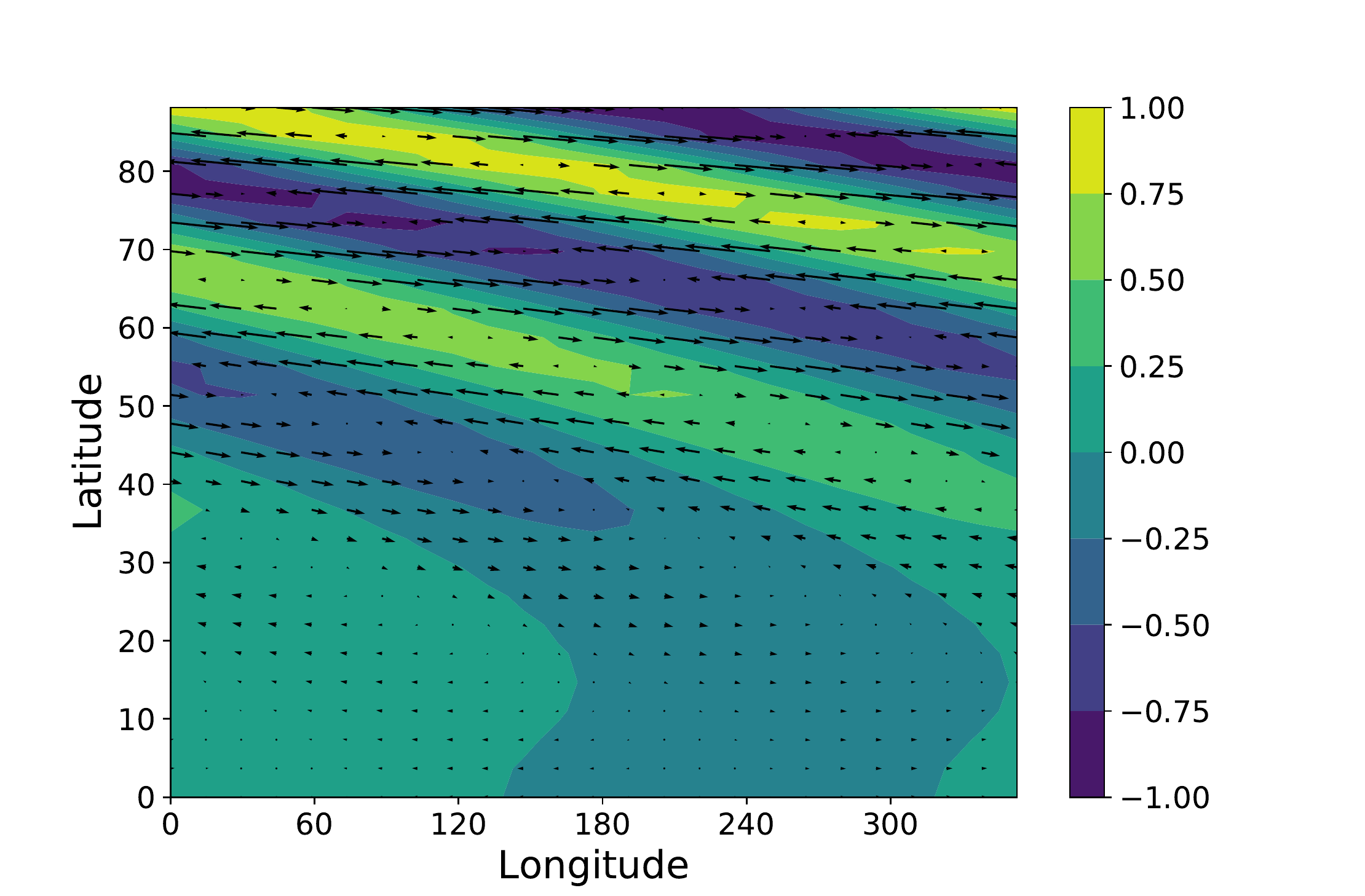}} 
    \subfigure[]{\includegraphics[width=10cm,angle=0.0,origin=c]{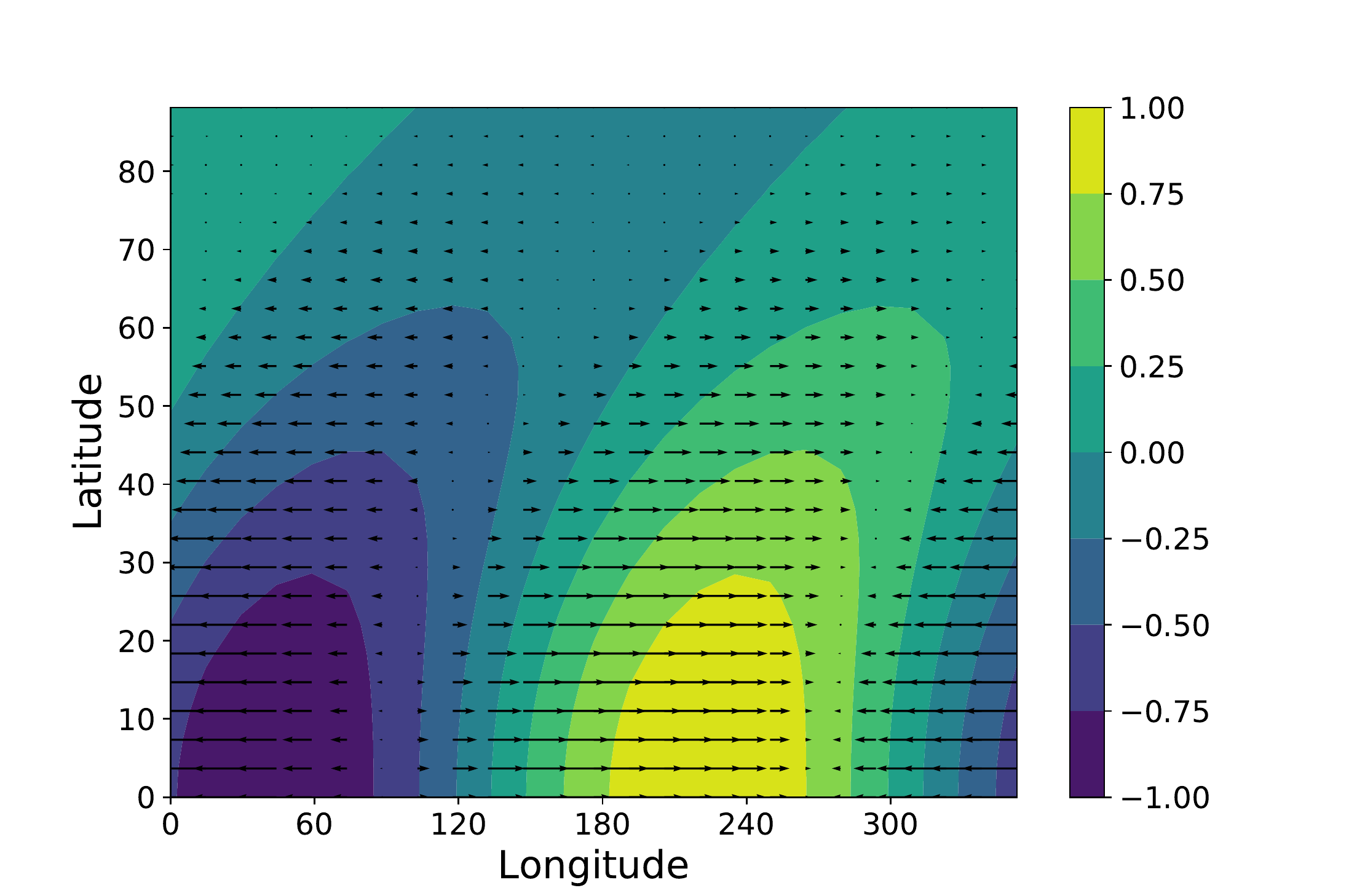}}
    \caption{Temperature (colors) and winds (arows), both in arbitrary unit, as a function of longitude and latitude for the shallow water solutions of 
    Eqs. \eqref{eq:polynomial_6} and \eqref{eq:shape_degueu} considering $\tau_\mathrm{rad}=1$ ($\sim 3.5 \times 10^4$ \ s in dimensional units) and $\tau_\mathrm{drag} = 28$ ($\sim 10^6$ \ s in dimensional units). (a) Westward propagating gravity wave (b) Rossby wave and (c) Kelvin wave. } 
    \label{fig:waves_bonus}
    \end{center}
\end{figure}

Finally,  Figure \ref{fig:waves_bonus} shows the same waves when  $\tau_\mathrm{rad} = 1$ and $\tau_\mathrm{drag} = 28$, which is about $3.5 \times 10^4$ \ s and $10^6$\ s respectively. The tilt in the gravity and Kelvin modes is amplified compared to Figure \ref{fig:waves_rad}, and the shape of the Rossby wave is extremely altered. Because the majority of the energy of the Rossby wave is in very high latitudes, where the equatorial $\beta$--plane frameworks breaks, we have excluded these modes in our semi-analytical treatment. Physical Rossby waves are nonetheless recovered in 3D by ECLIPS3D with such timescales, see section \ref{ssec:eclips3d_confirm}.

\newpage

\newpage

\section{Numerical development of the linear steady state}
\label{app:figure}
\begin{figure*}  
    \begin{center}
    \subfigure[]{\includegraphics[width=8.5cm,angle=0.0,origin=c]{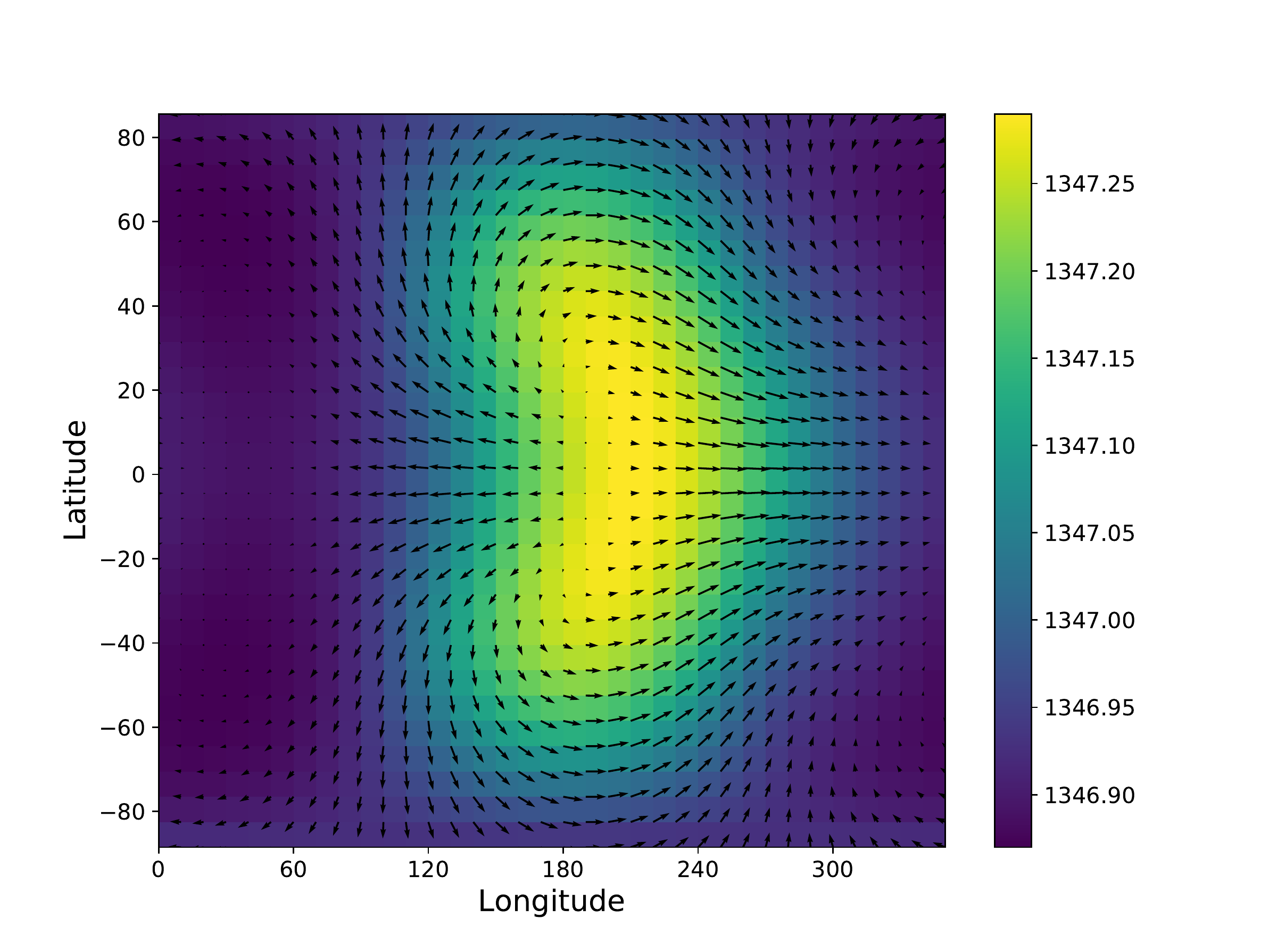}}
    \subfigure[]{\includegraphics[width=8.5cm,height=6.5cm,angle=0.0,origin=c]{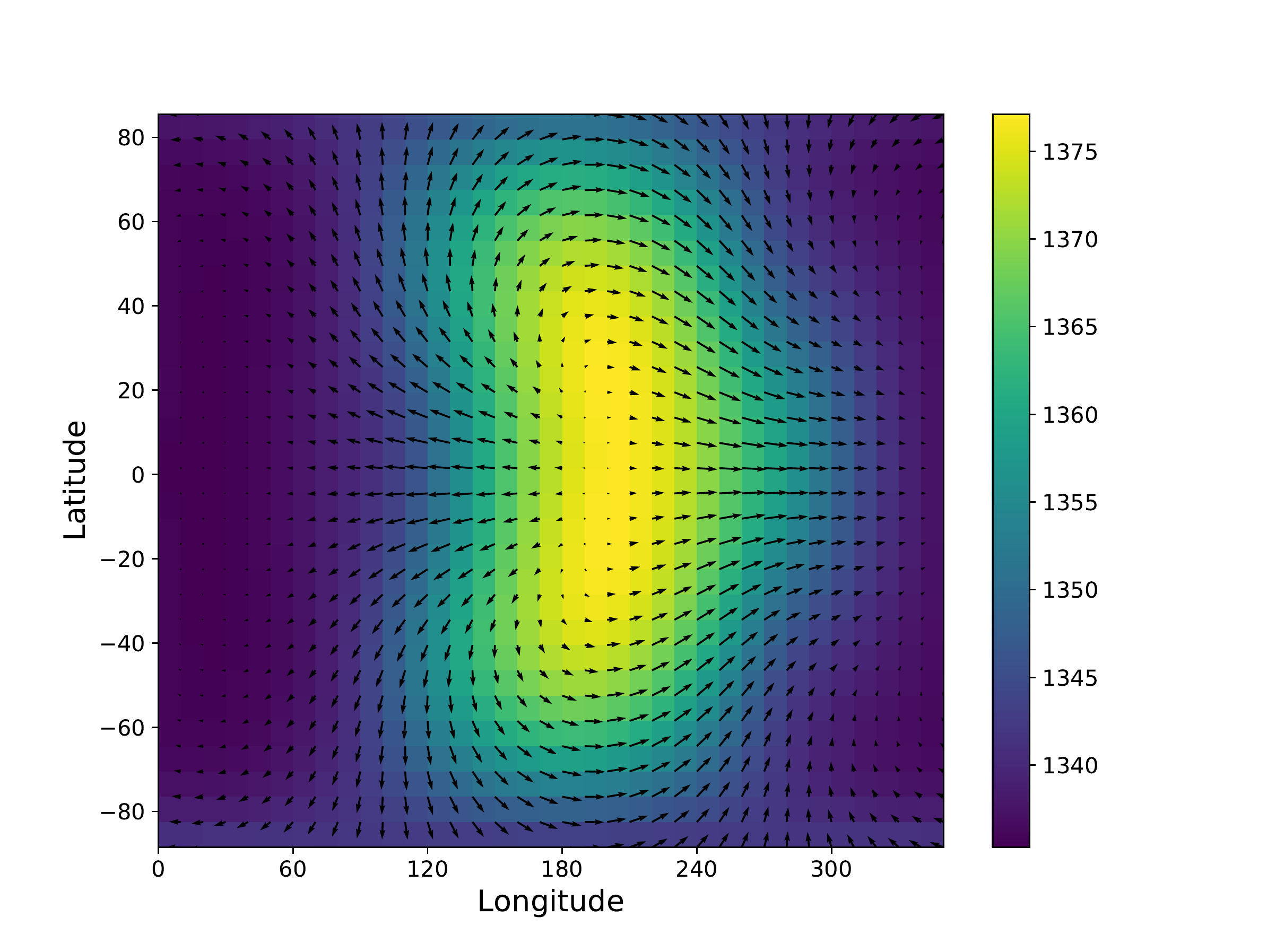}} \\
    \subfigure[]{\includegraphics[width=8.5cm,angle=0.0,origin=c]{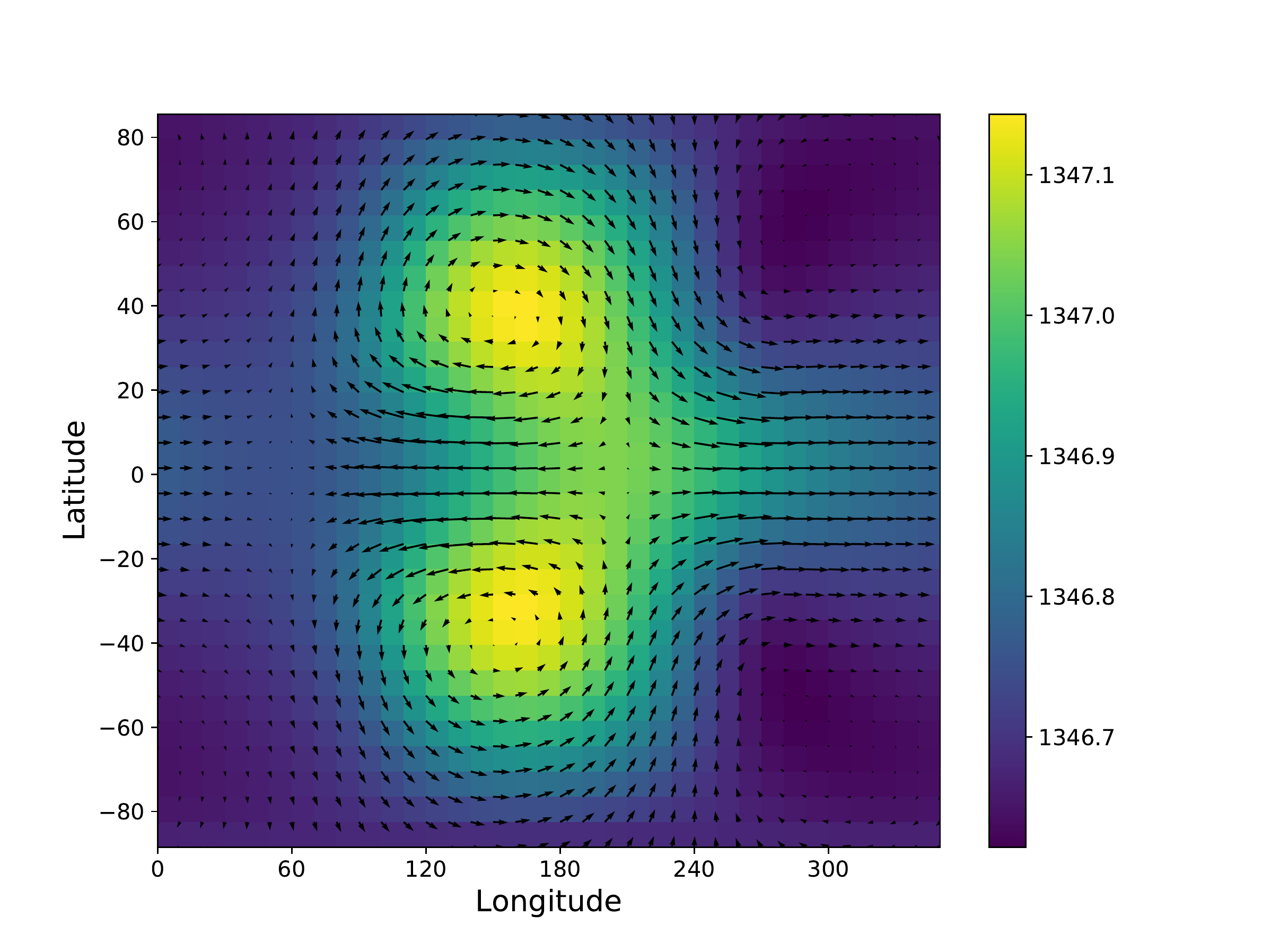}}
    \subfigure[]{\includegraphics[width=8.5cm,height=6.5cm,angle=0.0,origin=c]{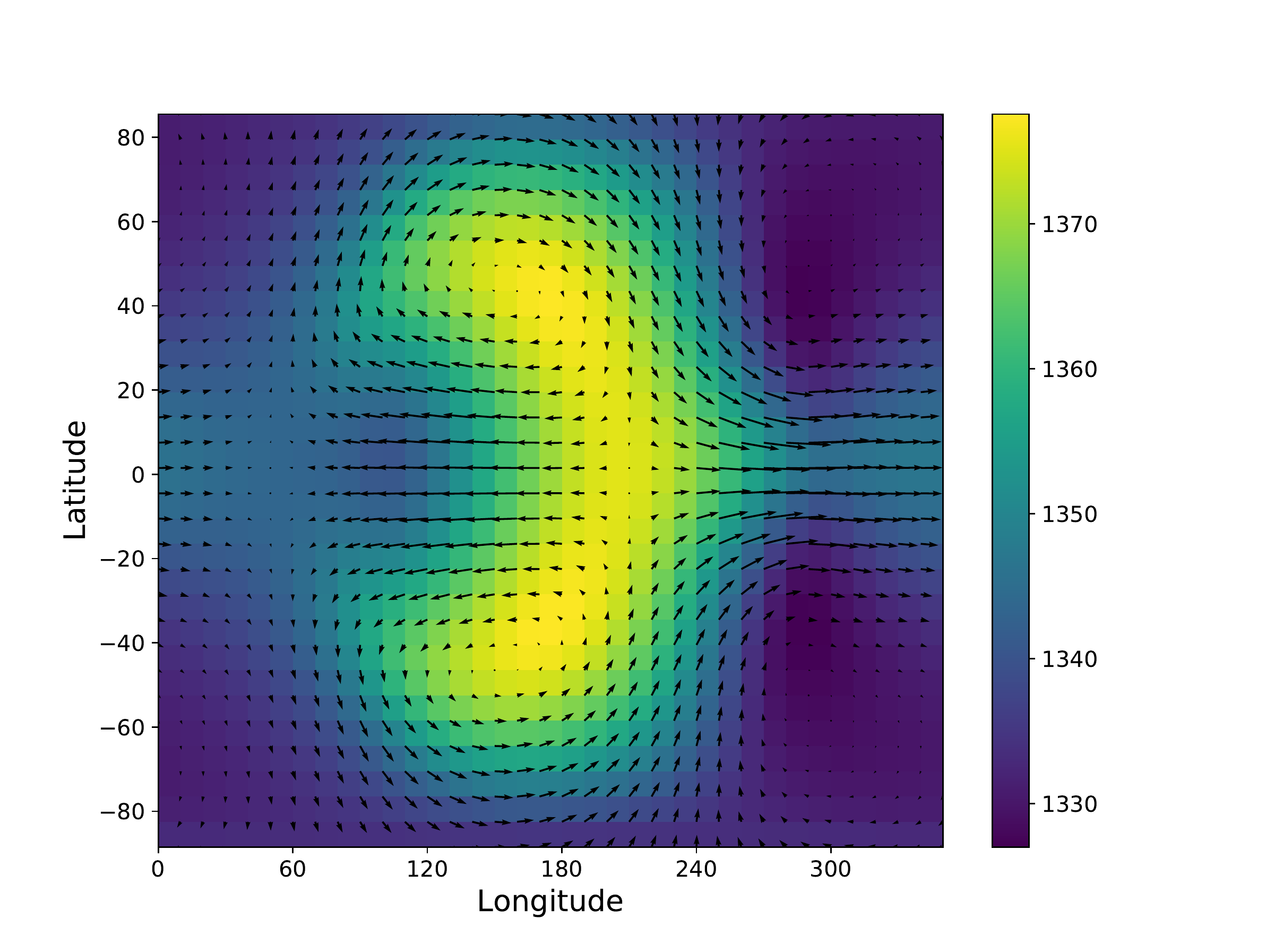}} \\
    \subfigure[]{\includegraphics[width=8.5cm,angle=0.0,origin=c]{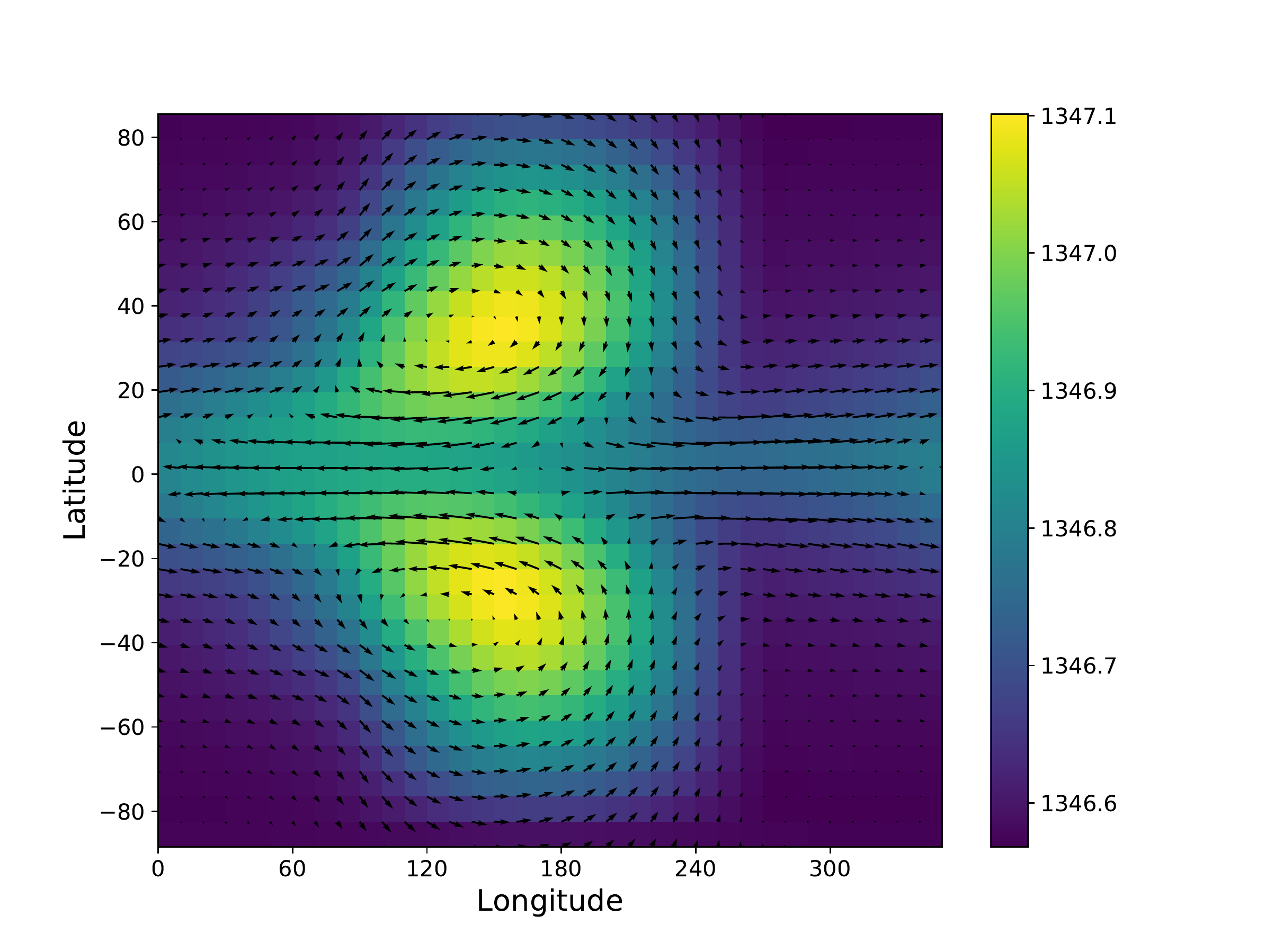}}
    \subfigure[]{\includegraphics[width=8.5cm,height=6.5cm,angle=0.0,origin=c]{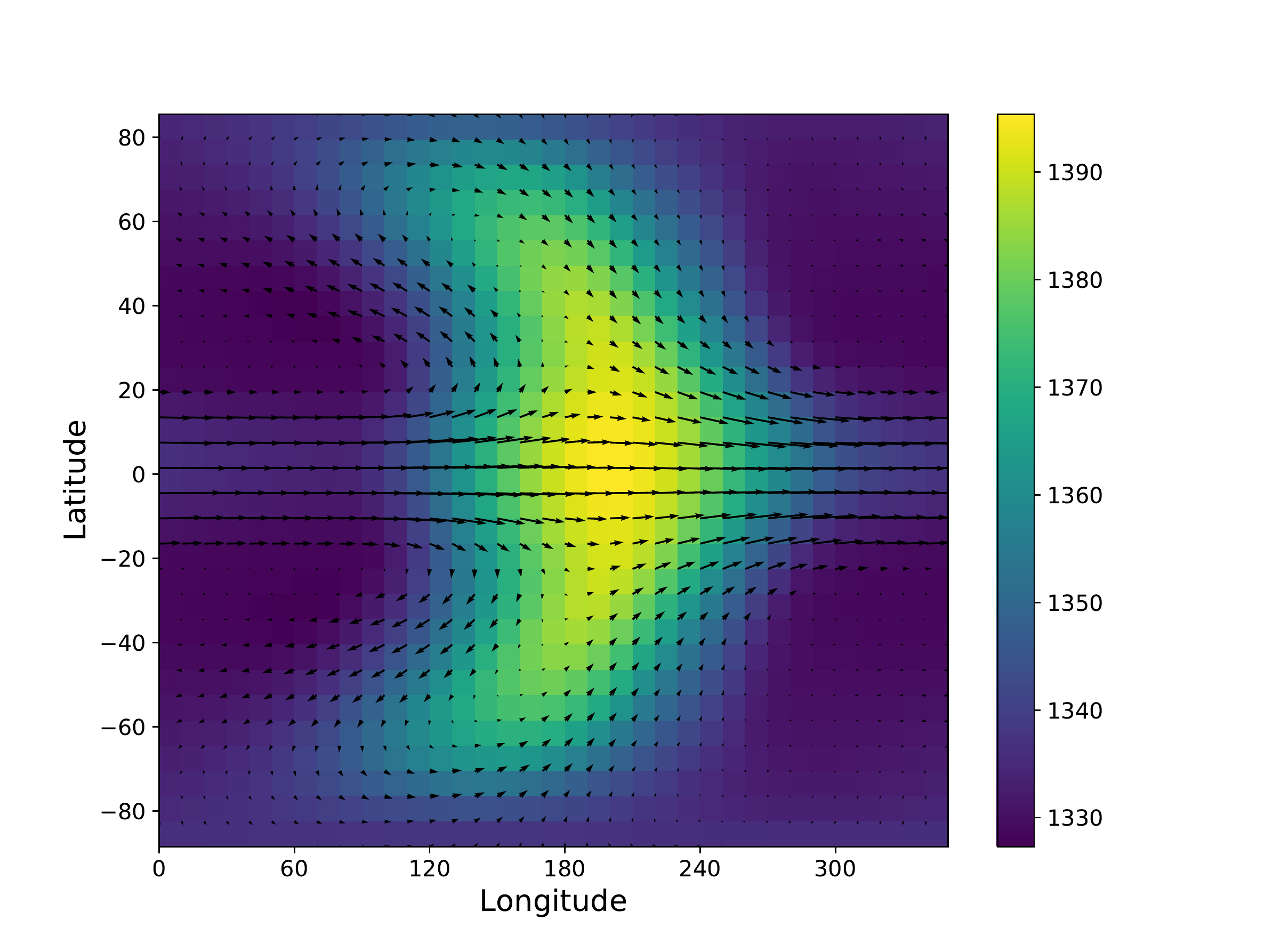}} \\
    \caption{Temperature (colors) and winds at a height corresponding to $80$mbar pressure at $t=0$. $\tau_\mathrm{drag} = 10^6$\ s and $\tau_\mathrm{rad}$ follows the prescription of \citet{iro2005} for all simulations. The left column has $\Delta T_\mathrm{eq,top} = 1$K while the right column has $\Delta T_\mathrm{eq,top} = 100$K. Finally, the top panel shows the atmosphere after 1 day of evolution, middle panel after 3 days and bottom panel after 50 days. Maximum speeds are (a) 1 m.s$^{-1}$, (b) 100 m.s$^{-1}$ (c) 2 m.s$^{-1}$ (d) 300 m.s$^{-1}$ (e) 5 m.s$^{-1}$ (f) 1000 m.s$^{-1}$.} 
    \label{fig:superrot_6}
    \end{center}
\end{figure*}

In this appendix, we show the development of the linear steady state for a simulation with $\Delta T_\mathrm{eq,top} = 1K$, $\tau_\mathrm{drag} = 10^6$\ s and $\tau_\mathrm{rad}$ following the prescription of \citet{iro2005} and the same simulation with $\Delta T_\mathrm{eq,top} = 100K$. 
Our estimates show that such a drag timescale should lead to a linear steady circulation similar to Figure \ref{fig:MG_komacek} after $\sim 20$ days of evolution in the low forcing case. Whereas the time to depart from the linear evolution in the highly forcing case should be about $\sim 1$ day. This is clearly recovered in Figure \ref{fig:superrot_6}. 

Figure \ref{fig:superrot_6} shows the temperature and winds of both simulations after 1, 3 and 50 days of evolution. We see that the low-forcing simulation resembles Figure \ref{fig:MG_komacek} after 50 days while the highly forced simulation is superrotating.

After 1 day of simulation a Matsuno-Gill like circulation is recovered in both cases. The maximum speed in the high forcing case is a hundred times the maximum speed of the low forcing case. After 3 days already, the maximum speed of the high forcing case is more than 100 times that of the low forced case. After 50 days, there is a factor 200 difference between the two cases, highlighting the influence of non linear terms.

In the low forcing case, the propagation and dissipation of mid latitude Rossby waves shifts the circulation obtained after 1 day towards a reverse Matsuno-Gill state after 50 days, whereas in the highly-forced case the eddies from the circulation after 1 day accelerate the equator towards a superrotating state.  Notably, we see on the middle panel of Figure 
\ref{fig:superrot_6} that the winds in the low-forced simulation are slowly shifted westward by the mid-latitude Rossby waves, whereas the highly forced simulation leads to a shrink of the westward winds that disappear after $\sim 30$ days.

\end{document}